\documentclass[]{aa}  

\usepackage{graphicx}
\usepackage{txfonts}
\usepackage{lipsum}
\usepackage[caption=false]{subfig}
\usepackage{lscape}             
\usepackage{placeins}           
\usepackage[]{hyperref}

\usepackage{textgreek}
\usepackage{amsmath}
\usepackage{stmaryrd}
\SetSymbolFont{stmry}{bold}{U}{stmry}{m}{n}

\usepackage{chemformula}
\usepackage{chemmacros}

\usepackage{xspace}

\newcommand{\fc}{\textsc{FastChem}\xspace}
\newcommand{\fcc}{\textsc{FastChem Cond}\xspace}

\begin{document} 

   \title{FastChem 4:\\ New chemical elements and improved convergence behaviour}

   \author{D. Kitzmann
          \inst{1,2},
          J. W. Stock,
          \and
          A.B.C. Patzer\inst{3}
          }

   \institute{
    Space Research and Planetary Sciences, Physics Institute, University of Bern, Gesellschaftsstrasse 6, 3012 Bern, Switzerland\\
   \email{daniel.kitzmann@unibe.ch}
   \and
    Center for Space and Habitability, University of Bern, Gesellschaftsstrasse. 6, 3012 Bern, Switzerland
   \and
     Institut f\"ur Physik und Astronomie (IfPA), Technische Universit\"at Berlin (TUB), Hardenbergstr. 36, 10623 Berlin, Germany
    }

   \date{Received May 18, 2026; accepted November 4, 2026}

  \abstract{
  Chemical equilibrium calculations are a key ingredient for modelling and interpreting spectroscopic observations of (exo)planets, brown dwarfs, cool stars, and protoplanetary disks. As these applications increasingly probe non-solar elemental abundances and previously underrepresented elements, equilibrium chemistry solvers must be both numerically robust and capable of handling complex chemical systems. Here we present \fc 4, a major update to the open-source \fc equilibrium chemistry code. We extend the gas-phase solver with a multidimensional Newton-method convergence accelerator that mitigates the slow convergence previously encountered for strongly non-solar elemental abundances. We further reformulate the gas-phase equations in logarithmic element densities, removing the dependence on quadruple-precision arithmetic and allowing \fc to be applied at low temperatures on any platform supporting double precision. 
  The condensate solver is upgraded with adaptive Levenberg--Marquardt regularisation, a perturbed-Hessian fallback for nearly singular Jacobians, and a combined gas--condensate Newton solver. These changes lead to a strong increase in computational performance and stability. The thermochemical data is expanded using thermochemical data from the NIST-JANAF tables and the Barin compilation, and now comprises 800 gas-phase molecules and ions and 511 condensates spanning 44 elements, of which about 290 are newly added liquid and solid species. We apply the updated code to a wide pressure--temperature grid for both solar and carbon-rich ($\mathrm{C/O}=2$) elemental compositions. The resulting grids reproduce the classical solar-composition condensation sequence and reveal the marked shifts that occur under carbon-rich conditions, including reduced condensates such as \ch{CaS}, \ch{AlN}, \ch{MgS}, and graphite. We also find that silicon monoxide is stable as a condensate over a limited pressure--temperature range, consistent with recent JWST observations of brown dwarfs. \fc 4 is released under the GPLv3 licence, together with a pre-compiled Python package.}


   \maketitle
   \nolinenumbers
%

\section{Introduction}
 
The interpretation of spectroscopic observations and the modelling of planetary and stellar atmospheres rely critically on an accurate description of atmospheric chemistry. Assuming chemical equilibrium when computing gas-phase compositions and condensate formation significantly simplifies this problem. Although this assumption is not universally valid, it remains reasonable in a wide range of scenarios. In many other cases, however, chemical kinetics, including vertical and horizontal transport as well as photochemical processes, must be taken into account to obtain a more realistic estimate of the chemical composition \citep[e.g.,][]{Venot2012A&A...546A..43V, Moses2011ApJ...737...15M, Rimmer2016ApJS..224....9R, Tsai2021ApJ...923..264T}.
 
However, such non-equilibrium models require detailed knowledge of the full reaction network connecting all relevant chemical species. In practice, these networks are often incomplete, and many reaction rate coefficients remain poorly constrained. As a consequence, the number of chemical species that can be included in complex kinetic models is frequently limited by missing chemical pathways or unknown rates. Equilibrium chemistry models, in contrast, can readily handle hundreds or even thousands of species involving dozens of elements, as they do not depend on explicit reaction rates but only on equilibrium constants. Thus, despite its known limitations in certain environments, chemical equilibrium calculations remain a crucial component of atmospheric modelling. Results of such calculations can also provide a fundamental benchmark that forms the basis for extending models to include kinetic effects. For example, without a reference equilibrium state, it is not possible to determine whether a system is in or out of equilibrium, nor to quantify the degree of deviation from equilibrium.
 
\fc is an open-source code\footnote{\url{https://github.com/NewStrangeWorlds/FastChem}} designed to compute chemical equilibrium abundances of gas-phase species and condensates in primarily astrophysical and planetary environments \citep{Stock2018MNRAS.479..865S, Stock2022MNRAS.517.4070S, Kitzmann2024MNRAS.527.7263K}. While several other equilibrium chemistry codes are available in astrophysics, such as GGchem \citep{Woitke2018A&A...614A...1W}, CONDOR \citep{Lodders1993E&PSL.117..125L}, TEA \citep{Blecic2016ApJS..225....4B}, or the NASA CEA code \citep{Gordon1994, McBride96}, \fc is among the fastest open-source implementations. It is distributed as a pre-compiled Python package (\textsc{pyfastchem}\footnote{\url{https://pypi.org/project/pyfastchem}}), making it straightforward to deploy across different platforms.
 
The \fc code has been successfully adopted for a wide range of astrophysical systems, including modelling the atmospheres of brown dwarfs and exoplanets \citep[e.g.,][]{Grasser2025A&A...698A.252G, Petz2025AJ....169..267P, Scott2025MNRAS.540.1909S} and interpreting observations of these objects using data from the \textit{James Webb} Space Telescope (JWST) and ground-based facilities \citep{Picos2025A&A...703A..65P, Fisher2026MNRAS.545S2187F, Verma2025AJ....170...69V, Bachmann2025A&A...700A.105B, deRegt2025A&A...696A.225D}. It has also been applied to scenarios involving outgassing from magma oceans and the formation of secondary atmospheres \citep{Ito2025ApJ...987..174I, Hakim2026MNRAS.546ag133H, Bower2025ApJ...995...59B}. Its applications even extend beyond astrophysics: \fc has been used to study combustion reaction networks in industrial contexts and has also been adapted into a smartphone application that enables chemical equilibrium calculations on mobile devices.
 
The first two versions of \fc, presented by \citet{Stock2018MNRAS.479..865S} and \citet{Stock2022MNRAS.517.4070S}, were restricted to gas-phase chemistry. \citet{Kitzmann2024MNRAS.527.7263K} introduced condensation in the third major version, \textsc{FastChem 3}, also referred to as \fcc. As discussed by \citet{Kitzmann2024MNRAS.527.7263K}, the phase rule implies that not all potentially stable condensates can coexist simultaneously: the number of stable condensates is limited to the number of elements minus one. In \fcc, we therefore adopted an approach based on the work of \citet{Leal2016AdWR...96..405L}, in which the set of stable condensates emerges naturally as a direct outcome of the algorithm, without the need for iterative addition or removal of condensates. Other standard codes in the field, such as NASA CEA or GGChem, need to iteratively add condensates to and remove them from the system of equations to find the set of stable condensates. In contrast, \fc, includes all potentially stable condensates in its system of equations from the start, and the stable set is simply an outcome of its solution.
 
In previous releases of \fc, the thermochemical data and species selection were focused on elements more abundant than vanadium under solar elemental abundances. However, the recent detection of several rare-earth, transition, and heavy metals, such as yttrium (Y), scandium (Sc), and barium (Ba), in the atmospheres of ultra-hot Jupiters \citep[e.g.,][]{Hoeijmakers2019A&A...627A.165H, Borsato2023A&A...673A.158B, Prinoth2024A&A...685A..60P} has rendered this restriction insufficient for accurately characterising the chemistry of these objects.
 
Moreover, as \fc has increasingly been applied to arbitrary element selections and strongly non-solar elemental abundance distributions, convergence issues have occasionally emerged when using the numerical algorithms implemented in earlier versions. These issues arise from strong non-linear coupling between elements that simultaneously participate in the formation of the same molecules and have comparable abundances. Addressing these limitations is therefore a key motivation of the present work.
 
In this paper, we first improve the gas-phase chemistry solver by introducing a convergence accelerator based on the Newton method. We then expand the chemical database by adding additional elements together with their associated gas-phase species and condensates. Finally, we use this extended database to investigate chemical equilibrium sequences for all included elements over a wide range of temperatures and pressures.

\section{Improvements to the FastChem code}

\begin{figure}
    {
    \centering
    \includegraphics[width=\linewidth,clip]{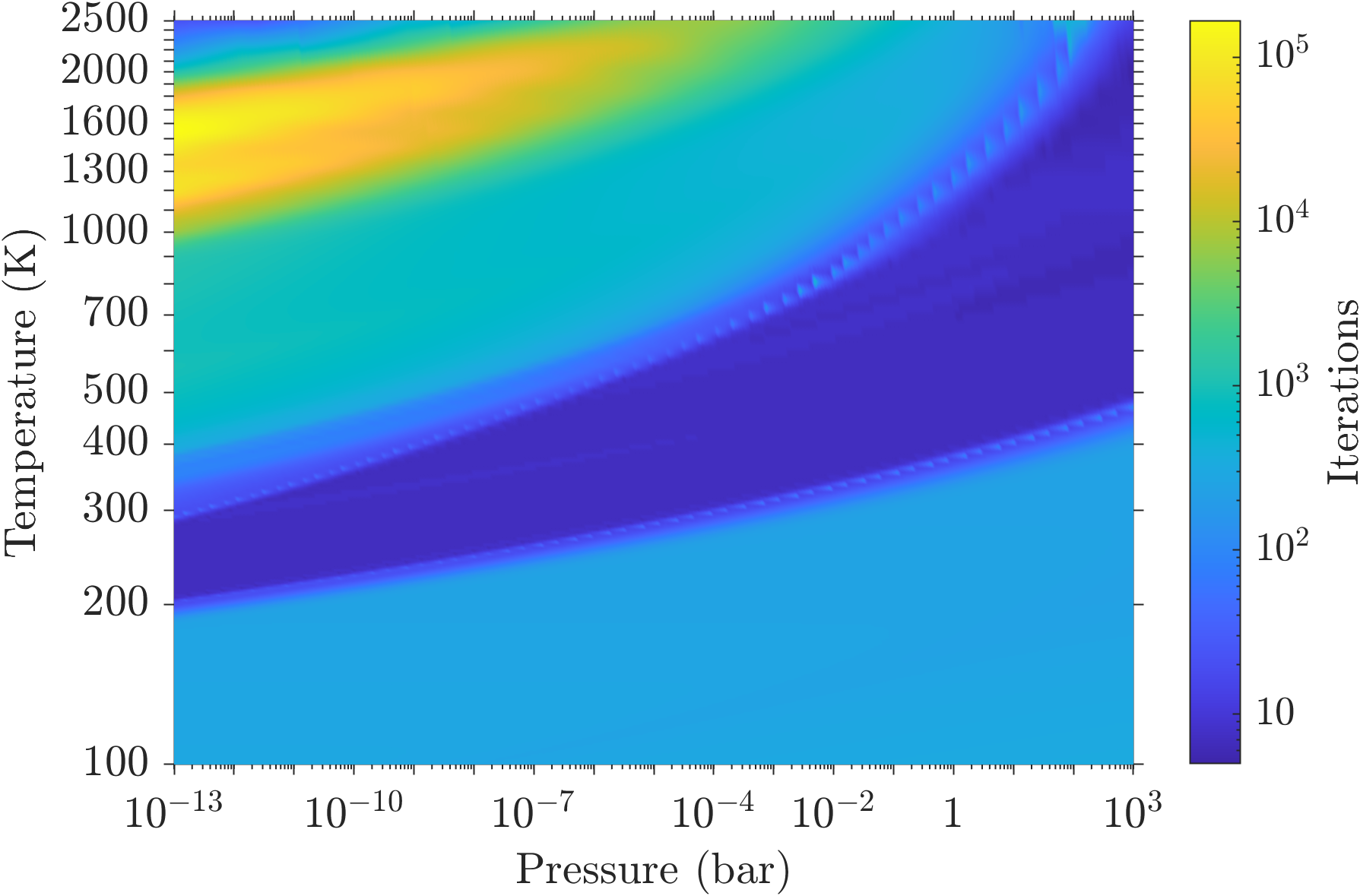}
    \caption{Number of iterations necessary for convergence using the previous version of \fc \citep{Stock2022MNRAS.517.4070S, Kitzmann2024MNRAS.527.7263K}. The pure gas-phase calculations were performed over a wide $p$-$T$ plane using solar elemental abundances but in particular with a C/O ratio of 1.}
    \label{fig:iterations_old}
    }
\end{figure}

When performing \fc calculations over a wide parameter range in temperature and pressure, the number of iterations required for convergence can become extremely large under certain conditions. This increase is strongly influenced by the chosen elemental composition, particularly when competing key elements have similar abundances. This effect is illustrated in Fig.~\ref{fig:iterations_old}, which shows the number of iterations needed across the pressure-temperature ($p$--$T$) range used by \citet{Stock2018MNRAS.479..865S}. In contrast to those calculations, we adopted solar elemental abundances but set the carbon-to-oxygen (C/O) ratio equal to one. To remain comparable with the earlier calculations, only gas-phase species were considered, thus neglecting condensation.
 
\fc typically converges within tens to a few hundred iterations across most of the $p$--$T$ range (see Fig.~\ref{fig:iterations_old}). However, in specific regions, particularly at higher temperatures ($\sim$1500\,K) and very low pressures (below $10^{-7}$\,bar), the number of required iterations increases significantly. In some cases, more than 100\,000 iterations are needed to reach convergence. Similar convergence challenges occur for other specific elemental compositions as well (not shown), although these issues are generally confined to smaller regions of the parameter range.

This behaviour is a consequence of splitting the multidimensional problem into a series of one-dimensional problems that are solved iteratively \citep{Stock2018MNRAS.479..865S, Stock2022MNRAS.517.4070S}. In this scheme, the information on which elements are bound in which molecules is only exchanged between the one-dimensional problems once per global iteration. While this approach is sufficient in the vast majority of cases, elements with similar abundances that are bound in the same molecules can be tightly coupled. In such cases, the separated iteration scheme sometimes requires a very large number of global iterations to converge.
 
Two additional numerical issues have emerged in some applications of \fc. First, at low temperatures the atomic number densities can become so small that they underflow standard double-precision arithmetic. Previous versions of \fc therefore relied on quadruple precision, which is not natively supported on some computational platforms (for example, Apple ARM processors), leading to convergence problems below roughly 300\,K. Second, for strongly non-solar elemental abundances, \fc may experience convergence issues in the coupled gas--condensate system, particularly at low temperatures where most elements are condensed out.
 
We address these issues in three ways. First, we introduce a multidimensional Newton method as a convergence accelerator for the gas-phase solver (Sect.~\ref{sec:numeric_update}). Second, we reformulate the gas-phase equations in logarithmic element densities, removing the dependence on quadruple-precision arithmetic (Sect.~\ref{sec:log_densities}). Third, we upgrade the condensate solver with adaptive Levenberg--Marquardt regularisation and add a combined gas--condensate Newton solver as a fallback (Sect.~\ref{sec:condensate_changes}). A small number of additional code-level improvements are described in Sect.~\ref{sec:further_improvements}.
 
\subsection{Multidimensional Newton method for gas-phase calculations}
\label{sec:numeric_update}
 
We followed the terminology and notation used by \citet{Kitzmann2024MNRAS.527.7263K}. \fc employs the law of mass action together with the element conservation equations. 
 
The conservation equation can be written for each chemical element $j \in \mathcal E$ as:
\begin{equation}
    N_j = n_j +  \sum_{i \in \mathcal S \setminus \mathcal E} \nu_{ij} n_i + \sum_{c \in \mathcal C} \nu_{cj} n_c \ .
    \label{eq:gas_element_conservation}
\end{equation}
The set $\mathcal{S}$ includes all gas-phase species (atoms/ions/molecules), the set $\mathcal{C}$ all condensates, while $\mathcal{E}$ contains only the elements. Specifically, $n_j$ refers to the number density of element $j$ in its atomic form. The number densities of the condensates $n_c$ are calculated separately \citep[see][]{Kitzmann2024MNRAS.527.7263K} and are assumed to be known for the gas-phase calculations described below.
The corresponding stoichiometric coefficients are given by $\nu_{ij}$ and $\nu_{cj}$. 
 
The law of mass action yields the number density of any gas-phase species $n_i$, $i \in \mathcal S \setminus \mathcal E$, as a function of the temperature dependent equilibrium constant $K_i$:
\begin{equation}
    n_{i} = K_i(T) \prod_{l \in \mathcal E} {n_l}^{\nu_{il}}  \ , \qquad\qquad i \in \mathcal S \setminus \mathcal E \ .
    \label{eq:mass_action_molecules}
\end{equation}
 
By replacing the unknown number densities $n_i$ in the equation~(\ref{eq:gas_element_conservation}) with the corresponding laws of mass action, we obtain a non-linear system of equations for the element densities $n_j$ only
\begin{equation}
    N_j - \sum_{c \in \mathcal C} \nu_{cj} n_c = n_j +  \sum_{i \in \mathcal S \setminus \mathcal E} \nu_{ij} K_i(T) \prod_{l \in \mathcal E} {n_{l}}^{\nu_{il}} \ , \qquad\quad j \in \mathcal E \ .
    \label{eq:system_of_eq}
\end{equation}
 
In the original computational scheme of \fc, this non-linear system is split into separate equations for each element, solving them iteratively \citep{Stock2018MNRAS.479..865S, Stock2022MNRAS.517.4070S}. However, because element densities are updated only after each iteration step, the algorithm cannot fully account for the competing interdependence of key chemical elements. Even if \fc ultimately converges in all cases with this numerical scheme, the computational speed can become extremely slow when this effect dominates the iteration behaviour.

\subsubsection{System of equations}
In the following, we briefly describe the implemented multidimensional Newton method based on the derivations outlined by \citet{Kitzmann2024MNRAS.527.7263K}. Specifically, calculations are performed for $\ln n_{j}$ to prevent the number densities $n_{j}$ from becoming negative or zero during the numerical procedure.
 
For a Newton step $(k)$ a first-order Taylor series expansion of Eq. \eqref{eq:gas_element_conservation} is done with respect to $\ln n_{j}$. Assuming that $n_c$ are considered constant during the calculation of the gas phase, this yields
\begin{equation}
  b_j = \sum_{l\in \mathcal E} \left( n_{j}^{(k)} \delta_{jl} + \sum_{i \in \mathcal S \setminus \mathcal E} \nu_{ij} \nu_{il} n_i^{(k)} \right) \Delta \ln n_{l}
  =
  \sum_{l\in \mathcal E} \left(\boldsymbol{\mathcal J}\right)_{jl} \Delta \ln n_{l} \ ,
  \label{eq:element_cons_matrix}
\end{equation}
with the components of the Jacobian matrix $\boldsymbol{\mathcal J}$ given by
\begin{equation}
  \left(\boldsymbol{\mathcal J}\right)_{jl} = n_j^{(k)} \delta_{jl} + \sum_{i \in \mathcal S \setminus \mathcal E} \nu_{ij} \nu_{il} n_i^{(k)} \ .
  \label{eq:system5}
\end{equation}
The final system of equations then has the general form
\begin{equation}
   \mathbf{b}=\boldsymbol{\mathcal J} \cdot \mathbf{\Delta \ln n}\ ,
  \label{eq:final_system_log_full}
\end{equation}
with the components of the vector on the left-hand side:
\begin{equation}
    b_j := N_j - n_j^{(k)} - \sum_{i \in \mathcal S \setminus \mathcal E} \nu_{ij} n_i^{(k)} - \sum_{c \in \mathcal C} \nu_{cj} n_c \ .
    \label{eq:rhs_element_conserv}
\end{equation}
 
In contrast to the Jacobian matrix derived by \citet{Kitzmann2024MNRAS.527.7263K} for the condensate system, $\boldsymbol{\mathcal J}$ here is generally well-conditioned and invertible. Consequently, we solve the system of equations using an LU decomposition with partial pivoting by default. Only in rare cases the system becomes ill-conditioned and has to be solved by the more robust approach discussed in Sect. \ref{sec:condensate_changes}.

\subsubsection{Implementation in FastChem}
 
Since the standard \fc iteration typically ensures rapid convergence to the solution of the system of equations \citep{Stock2018MNRAS.479..865S, Stock2022MNRAS.517.4070S}, we chose not to implement the Newton method as the default solver. Each Newton step requires setting up and solving a linear system with the full Jacobian matrix, which would increase the computational time significantly in these cases without improving stability. Moreover, applying the Newton method to a system that is still far from its solution can reduce both performance and stability.

Instead, the Newton method has been included as a convergence accelerator, used only if \fc fails to converge within a user-specified number of iterations. By default, this limit is set to 400 iterations, an empirical and conservative choice that ensures the Newton method is only used when needed. If more iterations are needed, \fc assembles the equation system (\ref{eq:system5}) for all elements that have not converged so far which is typically a small subset of all elements considered.
 
This application of the Newton method is not used here as a replacement for the standard \fc iteration. Instead, it is interleaved between iterations to improve convergence speed. Since the Newton method is only invoked after several \fc iterations, the initial solution used in a Newton step is usually inside the convergence radius, allowing Newton corrections to significantly improve the number densities $n_j$ used in the next standard \fc iteration step. The Newton method directly accounts for the non-linear coupling between elements through its Jacobian matrix $\boldsymbol{\mathcal J}$, resulting typically in a substantial convergence acceleration. In practice, only a few calls of the Newton method are needed to achieve convergence. 
 
\subsubsection{Improved convergence behaviour}
 
\begin{figure*}[t!]
{
  \centering
  \includegraphics[width=0.45\linewidth,clip]{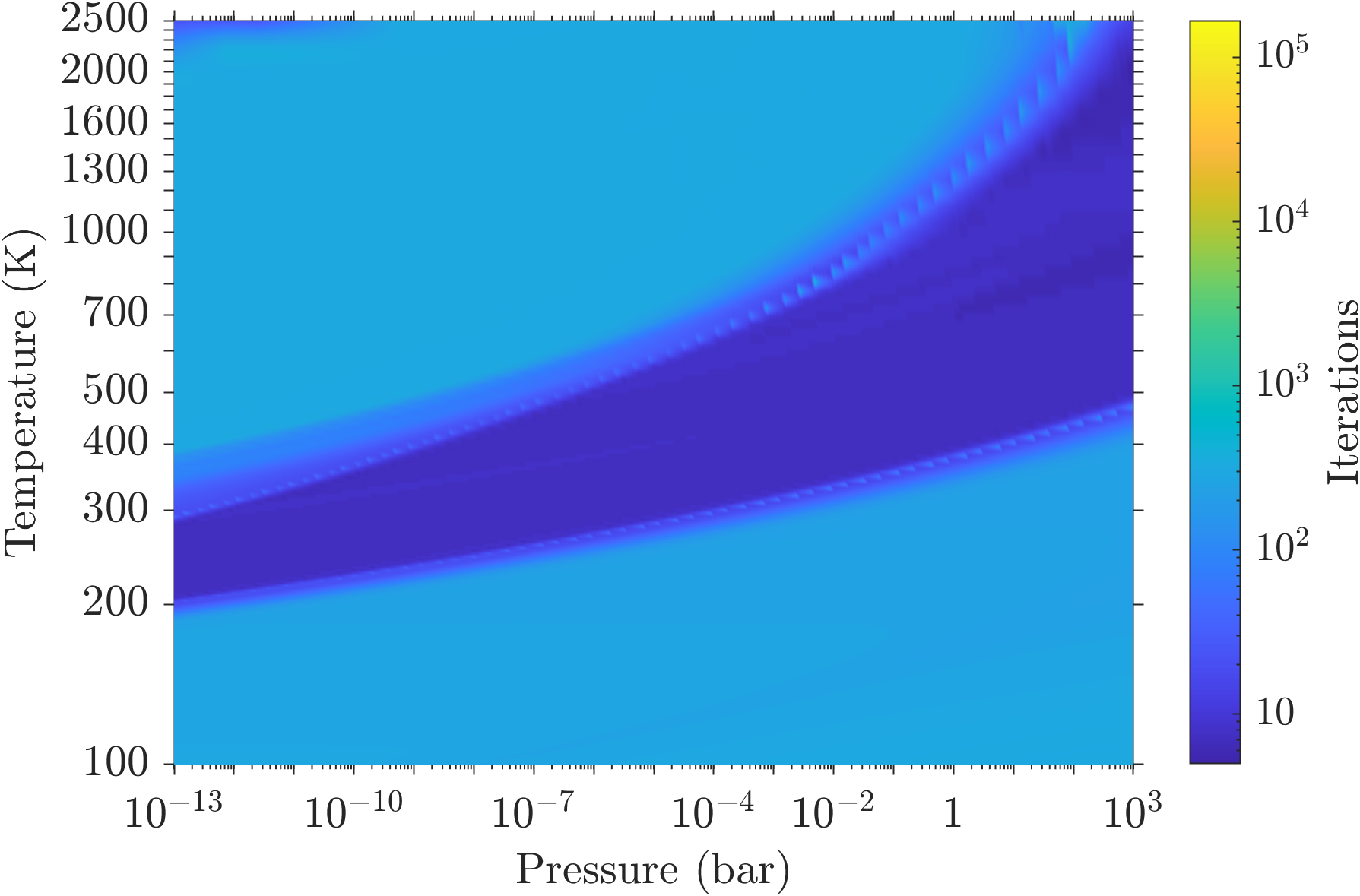} \includegraphics[width=0.45\linewidth,clip]{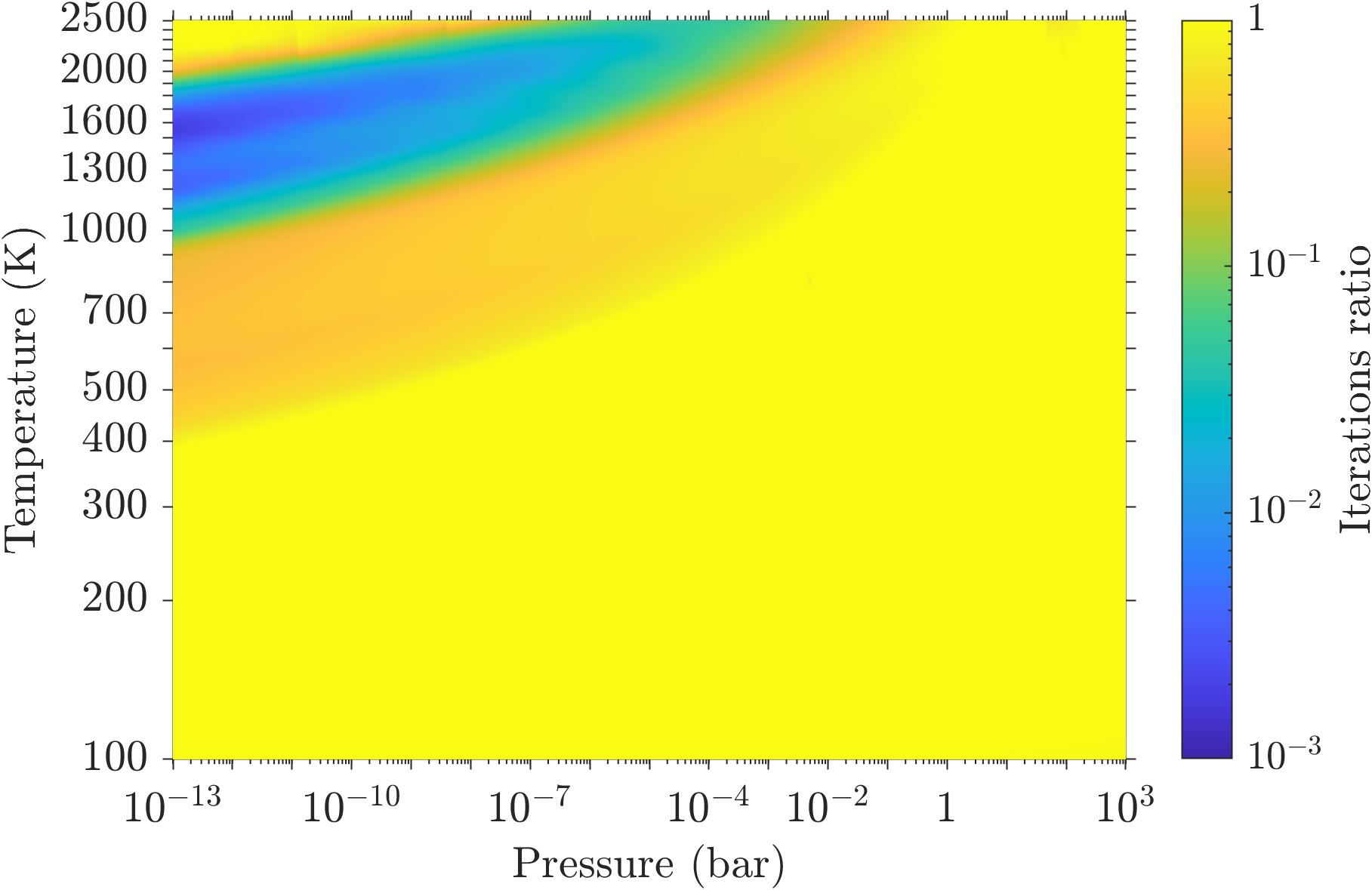}
   \caption{Improved convergence behaviour as implemented in \fc. The left panel depicts the number of iterations and employs the same colour scale as in Fig.~\ref{fig:iterations_old} to allow for direct comparison. The right panel shows the ratio of iterations between the revised and original numerical schemes.}
   \label{fig:iterations_new}
}
\end{figure*}
 
To evaluate the performance of the improved code, only gas-phase chemistry calculations were used.
To do so, we computed the chemical gas-phase composition on the same $p$-$T$ grid shown in Fig.~\ref{fig:iterations_old}. Using the same colour scale for direct comparison, the left panel of Fig.~\ref{fig:iterations_new} shows the number of iterations required for \fc to converge. The right panel of Fig.~\ref{fig:iterations_new} displays the ratio of iterations between the revised and original numerical schemes. For this test, the Newton accelerator was activated after 300 standard iterations.
 
These results demonstrate a significant improvement by including the new solver. While the original \fc approach required over 100\,000 iterations in certain regions of the parameter space, the revised scheme requires just a few hundred iterations. For the most problematic cases, the required number of iterations is reduced by a factor of almost 550. For instance, the most demanding case in Fig.~\ref{fig:iterations_old} converged in nearly 168\,000 iterations, whereas the same calculation is completed in only 309 steps applying the revised approach. Thereby, the Newton method was used just nine times.
 
This clearly highlights the superior performance of including the Newton method in the standard \fc scheme. However, the new approach is only necessary in a small regime of the parameter space. In most cases, the standard \fc iteration scheme performs well, as indicated by an iteration ratio between the two methods of approximately one.

\subsection{Introduction of log-based element densities}
\label{sec:log_densities}
 
As described by \citet{Stock2022MNRAS.517.4070S}, the element conservation equation used in the gas-phase calculations of \fc takes the form
\begin{equation}
\hat{\epsilon}_j \, n_\mathrm{g} =
n_j + \sum_{\substack{i \in \mathcal{S} \setminus \mathcal{E}}}
\bigl[\nu_{ij} + \hat{\epsilon}_j\,\sigma_i\bigr]\, n_i
+ \bar{n}_j + n_{j,\mathrm{min}} \ ,
\label{eq:elem_conserv}
\end{equation}
where $\sigma_i = 1 - \sum_{j} \nu_{ij}$, $\bar{n}_j$ accounts for species containing elements more abundant than $j$, and $n_{j,\mathrm{min}}$ accounts for species containing less abundant elements (\citet{Stock2022MNRAS.517.4070S}, Eqs.~19–20). In the former \fc algorithm, $\bar{n}_j$ and $n_{j,\mathrm{min}}$ are treated as known quantities while solving for $n_j$. The use of this formulation ensures that the abundances of all gas-phase species are consistent with the user-supplied gas pressure.
 
Substituting the mass-action law (Eq.~\ref{eq:mass_action_molecules}) for the molecular densities $n_i$, Eq.~\ref{eq:elem_conserv} can be written as
\begin{equation}
\hat{\epsilon}_j \, n_\mathrm{g} - \bar{n}_j - n_{j,\mathrm{min}} =
n_j +
\sum_{k=1}^{N_j} n_j^k
\underset{\hat{\epsilon}_i=\hat{\epsilon}_j}{\underset{\nu_{ij}=k}{\sum_{i\in \mathcal{S}\setminus\mathcal{E}}}}
\kappa_{ij} K_i
\underset{l\neq j}{\prod_{l\in\mathcal{E}}}n_l^{\nu_{il}},
\label{eq:master3}
\end{equation}
with $\kappa_{ij} = \nu_{ij} + \hat{\epsilon}_j\,\sigma_i > 0$.
As shown by \citet{Stock2022MNRAS.517.4070S}, this expression can be reduced to the polynomial equation
\begin{equation}
P_j(n_j):=\sum_{k=0}^{N_j}A_{jk}n_j^k=0,
\label{eq:defPj}
\end{equation}
which is solved by \fc independently for each element.
 
A key numerical issue arises at low temperatures, where the atomic number densities $n_j$ can become very small. This can lead to numerical underflow, effectively driving $n_j$ to zero. Since in the \fc formalism the atomic densities $n_j$ serve as the basis for computing the molecular abundances $n_i$ through the law of mass action, a vanishing $n_j$ would violate element conservation by forcing all species containing element $j$ to have zero abundance. To avoid this behaviour, previous versions of \fc internally used quadruple-precision variables with an extended numerical range.
 
Although this approach mitigates underflow problems to some degree, it also introduces portability issues. For example, the current generation of Apple ARM processors does not provide native hardware support for quadruple precision. On such platforms, \fc effectively runs in double precision, which can lead to convergence problems at temperatures below roughly 300\,K due to the aforementioned underflow issues.
 
To overcome these limitations, we redesigned \fc to operate with logarithmic element densities rather than linear ones. We therefore define the logarithmic element number densities
\begin{equation}
y_j := \ln n_j, \qquad j \in \mathcal{E} \ ,
\label{eq:yj_def}
\end{equation}
which serve as the primary unknowns. In this formulation, the mass-action law becomes
\begin{equation}
\ln n_i = \ln K_i + \sum_{j \in \mathcal{E}} \nu_{ij} \, y_j,
\label{eq:mass_action_log}
\end{equation}
while the element conservation equation (Eq.~\ref{eq:master3}) can be written as
\begin{equation}
\hat{\epsilon}_j \, n_\mathrm{g} - \bar{n}_j - n_{j,\mathrm{min}} = 
\mathrm{e}^{y_j} +
\sum_{i} \kappa_{ij}
\exp\,\Bigl(\ln K_i + \sum_{l \neq j} \nu_{il}\,y_l + \nu_{ij}\,y_j\Bigr).
\label{eq:conserv_exp}
\end{equation}
 
The right-hand side is thus expressed as a sum of exponentials of the form $S = \sum_k a_k \mathrm{e}^{x_k}$. Such expressions can be evaluated numerically in a stable manner using the log-sum-exp identity
\begin{equation}
\ln S =
x_{\max} +
\ln \left(\sum_k a_k\, \mathrm{e}^{x_k - x_{\max}}\right),
\qquad x_{\max} = \max_k x_k ,
\label{eq:lse}
\end{equation}
which prevents numerical overflow and underflow. The shifted exponents $x_k - x_{\max} \leq 0$ ensure that $\mathrm{e}^{x_k - x_{\max}} \in (0,1]$, such that terms negligible relative to the dominant contribution naturally evaluate to zero without explicit underflow.
 
Structurally, Eq.~\ref{eq:lse} can be cast into the same general form as Eq.~\ref{eq:defPj}, albeit with modified coefficients. We therefore reformulated all internal gas-phase solver functions described by \citet{Stock2018MNRAS.479..865S} to operate on logarithmic element densities $y_j$. The multidimensional Newton method introduced above, as well as the Newton method used for the condensate system \citep{Kitzmann2024MNRAS.527.7263K}, already operated in log-space and therefore required no modification.
 
Using standard double-precision variables, the logarithmic element densities and mass-action constants now remain well within the floating-point range for any physically plausible scenario, effectively eliminating underflow and overflow issues. With the new logarithmic formulation, \fc can, without ions, therefore be applied at temperatures as low as 1\,K without encountering numerical instabilities. Furthermore, since only double precision is required, the platform-specific limitations associated with quadruple precision have been removed. Support for quadruple-precision calculations has therefore been discontinued in this version of \fc, simplifying the code base and resolving portability issues on platforms without native quadruple-precision support.

\subsection{Changes to the condensate system}
\label{sec:condensate_changes}
 
As described by \citet{Kitzmann2024MNRAS.527.7263K}, the Jacobian of the condensate system can become ill-conditioned, posing a significant numerical challenge. Previous versions of \fc therefore employed strongly damped Newton steps to prevent divergence, together with a singular-value-decomposition fallback for nearly singular Jacobians.
 
In the present version, more sophisticated strategies have been implemented to address these issues. The Newton method used for the condensate system has been upgraded with an adaptive Levenberg--Marquardt regularisation, which introduces adaptive damping into the Jacobian matrix in nearly singular cases. If convergence still fails, \fc falls back to a perturbed Hessian approximation of the Jacobian, which helps remove singularities.
 
In addition, previous versions of \fc occasionally experienced long convergence times for coupled gas-phase and condensate systems with strongly non-solar elemental compositions, such as those representing an evaporating rocky planetary mantle \citep[e.g.][]{Ito2025ApJ...987..174I}. To address this issue, \fc now includes a combined Newton solver that simultaneously treats both the gas-phase and condensed species. This joint Newton method combines the multidimensional Newton method described in Sect.~\ref{sec:numeric_update} with the condensate solver presented by \citet{Kitzmann2024MNRAS.527.7263K}. The combined solver is only used if \fc fails to converge after 1000 iterations by default, as activating it too early can be detrimental when the system is still outside the convergence radius of the coupled solution.
 
Together, these improvements significantly enhance the numerical robustness of \fc. Moreover, the transition to double-precision variables and the improved convergence behaviour of the coupled gas--condensate solver considerably increase computational performance. For example, the protoplanetary disk calculation presented in \citet{Kitzmann2024MNRAS.527.7263K} originally required approximately 60 minutes of runtime on a 16-core processor, because of the very low temperatures (about 10\,K) in the outer disk regions. With the new version of \fc, the same calculation completes in only about 18 seconds.

\subsection{Further FastChem improvements and additions}
\label{sec:further_improvements}
 
Several further improvements are included in this version of \fc. Three new interface functions have been added to help users navigate and analyse the \fc output. The first, \texttt{convertToHillNotation(str formula)}, converts a given chemical formula into the Hill system \citep{Hil00} used by the standard \fc input files for gas-phase species (see Table~\ref{table:gas-phase_species}). It automatically handles special cases such as isomers, returning, for example, the \fc Hill notation for both \ch{HCN} and \ch{HNC}. The other two functions, \texttt{getGasSpeciesStoichiometry(int index)} and \texttt{getCondSpeciesStoichiometry(int index)}, return the stoichiometric vector of a given gas-phase or condensed species, respectively. They allow users to identify, for instance, all species containing exactly two carbon atoms.
 
In addition, a release version of \fc is now automatically compiled on GitHub for a wide range of platforms prior to being uploaded as a package to the Python Package Index (PyPI). As a result, users no longer need to compile \fc themselves: on most platforms, \texttt{pip install pyfastchem} downloads a pre-compiled version of \fc, eliminating the need for any user-side compilation or compiler installation.
 
A full description of all \fc interface functions can be found in the documentation hosted on GitHub\footnote{\url{https://newstrangeworlds.github.io/FastChem/}}.

\section{Addition of new elements and thermochemistry data overview}
 
\subsection{Additional chemical elements}
 
\renewcommand{\arraystretch}{1.1} 
\begin{table}[t]
  \caption{Adopted element abundances for the solar photosphere based on data from \citet{Asplund2021A&A...653A.141A}.}
  \label{table:abundances_asplund}
  \centering
  \footnotesize
  \begin{tabular}{llllll}
  \hline\hline
  \multicolumn{2}{c}{Element}                 & $x_j$ &  \multicolumn{2}{c}{Element}                 & $x_j$\\
  \hline  
  H  & Hydrogen    & 12.0   &Zn           & Zinc        & 4.56\\
  He & Helium      & 10.914 &F            & Fluorine    & 4.40\\
  O  & Oxygen      & 8.69   &Cu           & Copper      & 4.18\\
  C  & Carbon      & 8.46   &V            & Vanadium    & 3.90\\
  Ne & Neon        & 8.06   &Ge           & Germanium   & 3.62\\
  N  & Nitrogen    & 7.83   &\textbf{Se}  & Selenium    & 3.34\tablefootmark{a}\\
  Mg & Magnesium   & 7.55   &\textbf{Sc}  & Scandium    & 3.14\\    
  Si & Silicon     & 7.51   &\textbf{Kr}  & Krypton     & 3.12\\
  Fe & Iron        & 7.46   &\textbf{Ga}  & Gallium     & 3.02\\
  S  & Sulphur     & 7.12   &\textbf{Sr}  & Strontium   & 2.83\\
  Al & Aluminium   & 6.43   &\textbf{B}   & Boron       & 2.70\\ 
  Ar & Argon       & 6.38   &\textbf{Zr}  & Zirconium   & 2.59\\
  Ca & Calcium     & 6.30   &\textbf{Br}  & Bromine     & 2.54\tablefootmark{a}\\
  Na & Sodium      & 6.22   &\textbf{Rb}  & Rubidium    & 2.32\\
  Ni & Nickel      & 6.20   &\textbf{As}  & Arsenic     & 2.30\tablefootmark{a}\\
  Cr & Chromium    & 5.62   &\textbf{Ba}  & Barium      & 2.27\\ 
  Mn & Manganese   & 5.42   &\textbf{Xe}  & Xenon       & 2.22\\
  P  & Phosphorus  & 5.41   &\textbf{Y}   & Yttrium     & 2.21\\
  Cl & Chlorine    & 5.31   &\textbf{Te}  & Tellurium   & 2.18\tablefootmark{a}\\
  K  & Potassium   & 5.07   &\textbf{Sn}  & Tin         & 2.02\\
  Ti & Titanium    & 4.97   &\textbf{Pb}  & Lead        & 1.95\\
  Co & Cobalt      & 4.94   &\textbf{Li}  & Lithium     & 0.96\\ 
  \hline
  \end{tabular}
  \tablefoot{Elements listed in bold are added to \fc as part of this work.
  \tablefoottext{a}{Based on meteoritic element abundances (see also \citet{Palme2014pacs.book...15P}).}
}
\end{table}
 
One of the aims of this study is to expand the set of elements included in \fc. In previous \fc releases, we included gas-phase species and condensates for elements with solar abundances at least as high as that of germanium. For the studies by \citet{Hoeijmakers2019A&A...627A.165H} and \citet{Kitzmann2023A&A...669A.113K}, which focussed on modelling ultra-hot Jupiters, the hottest known exoplanets, we limited the gas-phase chemistry calculations to atoms and ions, omitting molecular species, but expanded the selection of elements up to uranium. This input data is available as an alternative to the default choice in the \fc repository.
 
It is important to note that there are two common ways to define the abundance of an element $j$. These two forms, denoted as $x_j$ and $\epsilon_j$, are related by the following expression:
\begin{equation}
   x_j = \log \epsilon_j + 12 \ .
\end{equation}
The quantity $x_j$ is widely used in the astronomical literature. In this notation, the abundance of hydrogen is defined as $x_\mathrm{H} = 12$, so that $\epsilon_\mathrm{H} = 1$ by construction.
To maintain consistency with astronomical notations, the user provides elemental abundances in the $x_j$ format in the corresponding input file.
 
\begin{figure}[ht]
    {
    \centering
    \includegraphics[width=0.9\linewidth,clip]{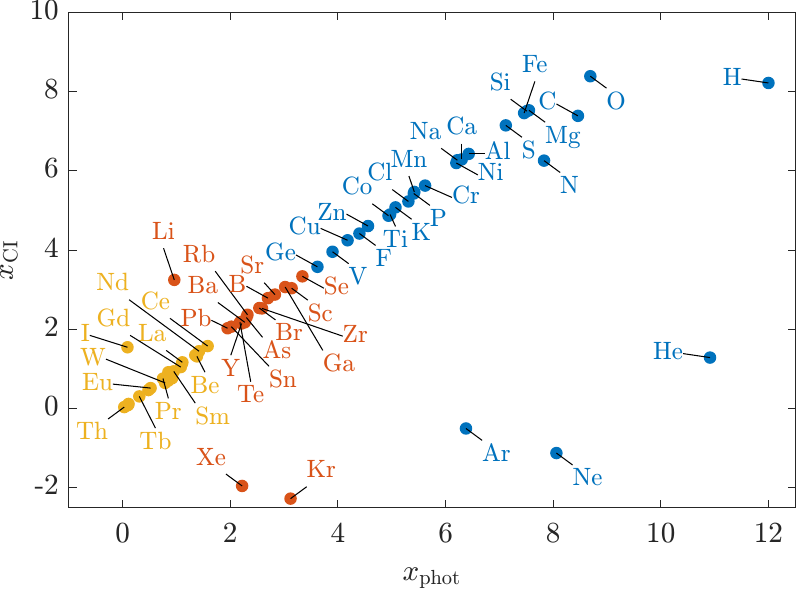}
    \caption{Element abundances of various elements in the solar photosphere ($x_\mathrm{phot}$) and CI chondrites ($x_\mathrm{CI}$) from \citet{Asplund2021A&A...653A.141A}. The elements labelled in blue are part of the original \fc release \citep{Stock2018MNRAS.479..865S, Stock2022MNRAS.517.4070S}. Elements that are added to \fc as part of this study are coloured in red, while remaining elements are depicted in yellow. For elements where no photospheric abundance is available, such as Se, Br, and Te, we use their CI abundances instead for $x_\mathrm{phot}$.}
    \label{fig:element_abundances}
    }
\end{figure}
 
Figure~\ref{fig:element_abundances} shows the abundances of various elements in the solar photosphere and in carbonaceous (CI) chondrites, based on the compilation by \citet{Asplund2021A&A...653A.141A}. Elements included in the original \fc dataset are marked in blue. The remaining elements appear to form two distinct groups, separated by a gap near $x_\mathrm{phot} \approx 2$. For this study, we include the group marked in red, which contains elements such as bromine, arsenic, and tellurium. We also added elements recently detected in the atmospheres of ultra-hot Jupiters, such as scandium, yttrium, barium, and strontium \citep[e.g.,][]{Borsato2023A&A...673A.158B, Hoeijmakers2019A&A...627A.165H}.
 
We additionally included lithium (Li) in the updated \fc element set. Although its solar photospheric abundance is lower than those of the other newly added elements, lithium is an important age indicator in young low-mass stars and is used to establish the substellar nature of brown dwarf candidates. The adopted solar lithium abundance may not be appropriate in all contexts, though. In the Sun, lithium is heavily depleted by proton-capture reactions that destroy it once convective mixing transports it to sufficiently hot layers near the base of the convection zone. When modelling cooler systems where this process does not occur, users may want to adjust the Li abundance to better reflect its primordial value, closer to that found in CI chondrites.
 
\begin{figure*}
    {
    \centering
    \includegraphics[width=0.8\linewidth,clip]{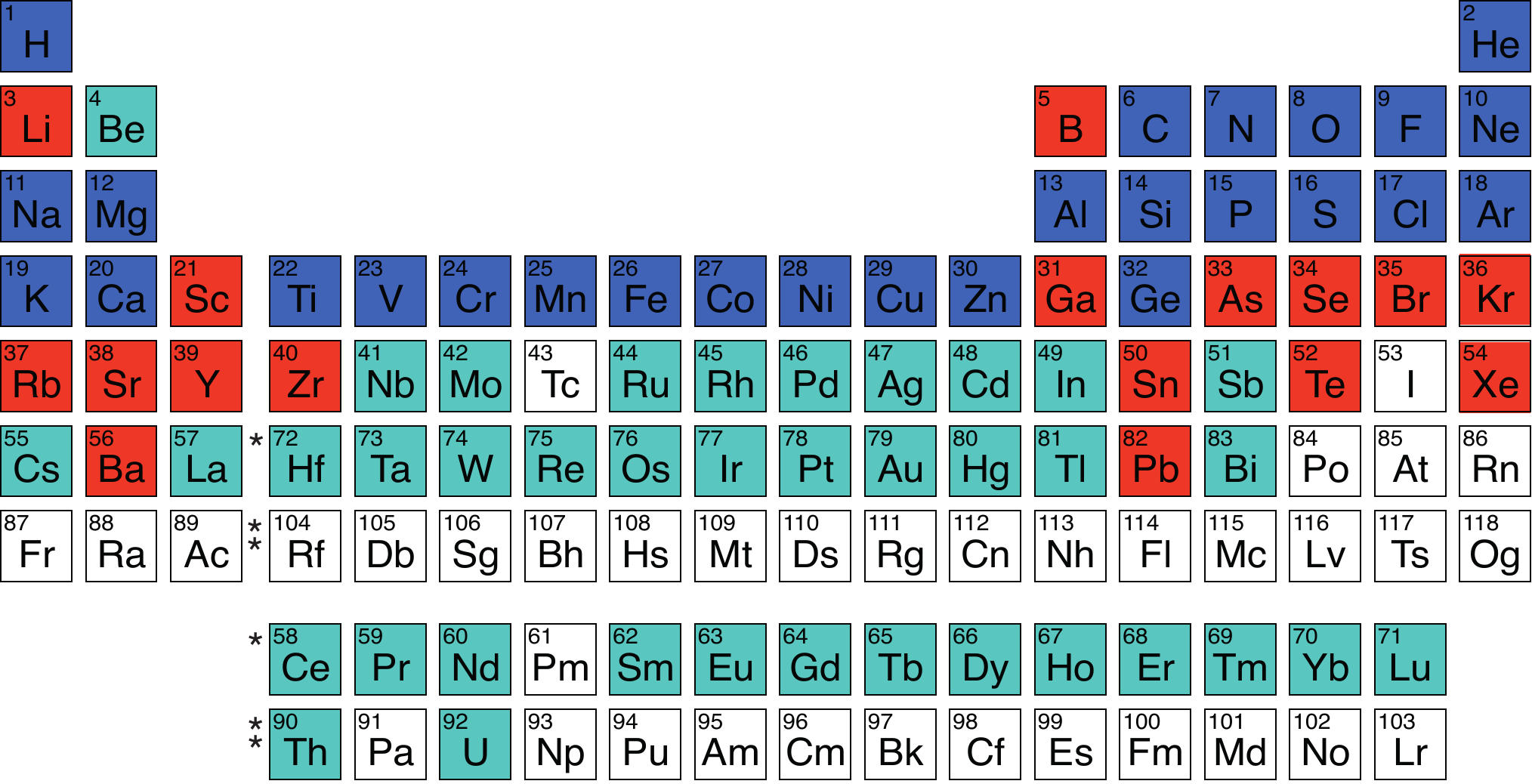}
    \caption{Overview of the elements included in \fc in the form of the periodic table. Elements coloured in blue are part of the previous versions of \fc \citep{Stock2018MNRAS.479..865S, Stock2022MNRAS.517.4070S, Kitzmann2024MNRAS.527.7263K}, while those in cyan are additionally available in the data set by \citet{Hoeijmakers2019A&A...627A.165H} that contains only atoms as well as singly and doubly ionised gas-phase species. Elements for which new molecules and condensates are added within this study are marked in red.}
    \label{fig:fastchem_elements}
    }
\end{figure*}

Figure~\ref{fig:fastchem_elements} summarises the original elements, the newly added elements, and those provided by \citet{Hoeijmakers2019A&A...627A.165H} in the form of the periodic table. In previous \fc releases, we used the elemental abundances from \citet{Asplund09}. For this update, we have adopted the revised values from \citet{Asplund2021A&A...653A.141A}, as shown in Table~\ref{table:abundances_asplund}. Most values are based on solar photospheric measurements. Exceptions include selenium (Se), bromine (Br), arsenic (As), and tellurium (Te), which are not detectable in high-resolution solar spectra. For these, we used meteoritic abundances from \citet{Asplund2021A&A...653A.141A}.
 
\subsection{Chemical elemental abundances}
 
In addition to the compilation by \citet{Asplund2021A&A...653A.141A}, another widely used source of elemental abundances is \citet{Lodders2003ApJ...591.1220L}, which has recently been updated by \citet{Lodders2025SSRv..221...23L}. Although \fc uses the \citet{Asplund2021A&A...653A.141A} abundances by default, the updated \citet{Lodders2025SSRv..221...23L} dataset is also included in the \fc repository for users who prefer this reference. For completeness, \fc also retains the older abundance compilations of \citet{Lodders2003ApJ...591.1220L} and \citet{Asplund09}.
 
For each dataset, three different options are provided. The default input files include only the elements present in previous releases of \fc and are intended for users who do not require the elements added in this study. Using a reduced set of elements generally speeds up the chemistry calculations and is therefore advantageous if the additional elements are not needed. The \emph{extended} input files include both the original elements and the new elements introduced here. In addition, for each compilation we provide \emph{full} input files containing most elements up to uranium. These can be used in conjunction with the expanded set of ions and anions introduced by \citet{Hoeijmakers2019A&A...627A.165H}. As in previous versions of \fc, users may also specify elemental abundances manually if desired.

\subsection{Gas-phase species}
 
The introduction of more chemical elements to \fc allows for the treatment of more molecules and ions (this subsection), as well as condensates (Sect.~\ref{ssec:new_condensates}). Our selection of molecules and ions in the gas phase is essentially based on the NIST-JANAF Thermochemical Tables \citep{Chase1998}, the work of \citet{Barklem2016AA...588A..96B}, and \citet{Hoeijmakers2019A&A...627A.165H}. To calculate the fit coefficients $\lbrace a_0,\,a_1,\,b_0,\,b_1,\,b_2\rbrace$ for the temperature-dependent mass action constants
\begin{equation}
    \ln \bar{K}(T)=\frac{a_0}{T}+a_1 \ln T +b_0 + b_1 T+b_2 T^2
\label{eq:mass_action_constant_fit}
\end{equation}
required by \fc \citep[see][]{Stock2018MNRAS.479..865S}, we mainly use the well-established data of \citet{Chase1998}. In some cases, when errors were reported or the data differed significantly from independent sources in the literature, we relied on the sources indicated in Table~\ref{table:gas-phase_species}, which lists all molecules and ions used in the present version of \fc.

\subsubsection{Diatomic molecules of elements more abundant than vanadium}
 
\citet{Barklem2016AA...588A..96B} provide equilibrium constants for diatomic molecules in tabulated form, from which the fit coefficients for equation~(\ref{eq:mass_action_constant_fit}) can be calculated\footnote{This is, however, not possible for their ions, due to different definitions of the equilibrium constants in this case.}. Additional molecules not included in previous versions of \fc are \ch{MnCl}, \ch{NCl}, \ch{ZnCl}, \ch{CoH}, \ch{CrS}, \ch{MnF}, \ch{NiF}, \ch{ZnF}, and \ch{ZnH}.
Furthermore, we revised the data for \ch{AlN}, \ch{CCl}, \ch{CP}, \ch{MgO}, \ch{MgS}, \ch{MnO}, \ch{TiN}, and \ch{NiO} using values from \citet{Barklem2016AA...588A..96B}, and updated the data for \ch{CaH} using values from \citet{McBride2002}.
 
\subsubsection{Molecules of elements less abundant than vanadium}
Data for molecules that include elements less abundant than vanadium were taken mainly from the NIST-JANAF Thermochemical Tables \citep{Chase1998}. Other sources in the literature we adopted include \citet{Tsuji1973A&A....23..411T}, \citet{McBride2002}, \citet{Barklem2016AA...588A..96B}, and \citet{Goo23} \citep[see also][]{Bur05}, as indicated in Table~\ref{table:gas-phase_species}.
 
\subsubsection{Ions}

For applications of \fc to the atmospheres of ultra-hot Jupiters, \citet{Hoeijmakers2019A&A...627A.165H} computed thermochemical data for ions of most elements up to uranium using the Saha equation. At the extreme temperatures encountered in these atmospheres, most elements exist in atomic or ionised form, while molecules and condensates are largely absent. 

All remaining ion data were originally taken from the JANAF tables. However, a comparison for ions such as \ch{Ni+}, \ch{Cu+}, or \ch{Co+} revealed large differences in some cases relative to our Saha equation approach. These likely stem from outdated partition functions in JANAF. The NASA Glenn Thermodynamic Database \citep{McBride2002} also shows deviations for species such as \ch{B+} or \ch{Ag+}.

Following the approach described in \citet{Hoeijmakers2019A&A...627A.165H}, we therefore refit the singly and doubly ionised cations, as well as the anions, of all elements. For the cations we used the latest partition functions from NIST \citep{NIST_ASD}, while those for the anions were taken from \citet{Andersen1999JPCRD..28.1511A}.
 
\subsection{Condensates}
\label{ssec:new_condensates}
 
Following the same approach used for gas-phase species, the thermochemical data for condensates were taken primarily from the JANAF tables \citep{Chase1998}, if available. For elements not covered by \citet{Chase1998}, we supplemented the dataset with values from the compilation by \citet{Barin1995}. In some cases, additional species from \citet{Barin1995} were also included to extend the coverage for selected elements already included in the previous releases of \fc.
 
For example, the JANAF tables list only a single zinc-bearing condensate, zinc sulfate (\ch{ZnSO4(s)}). Other zinc species, such as zinc sulfide (\ch{ZnS(s)}), are, however, potentially important in astrophysical environments, particularly as dust-forming species in the atmospheres of substellar objects. To account for this limitation, we augmented the zinc condensate data with additional species from \citet{Barin1995}.
 
Similar limitations exist for other elements, such as vanadium and germanium, for which JANAF provides only sparse condensed-phase data. In these cases, relevant condensates were likewise added from the Barin database. A complete list of all condensates included in \fc is provided in Table~\ref{table:cond_new}.
 
As described by \citet{Kitzmann2024MNRAS.527.7263K}, the mass action constants of each available phase were fitted independently using Eq.~\ref{eq:mass_action_constant_fit}. In the standard input data of \fc, multiple phases of a given condensate are combined, when present in the source data, into a single dataset, with phase transition temperatures taken from the corresponding databases. This approach is optional, and users may instead provide custom datasets in which each phase of a condensate is treated as a separate species.
 
As also noted by \citet{Kitzmann2024MNRAS.527.7263K}, the evaluation of Eq.~\ref{eq:mass_action_constant_fit} for condensates is typically restricted to the temperature range covered by the original thermochemical data. Condensate properties are often tabulated over relatively narrow temperature intervals, and extrapolating the fitted expression beyond these ranges can lead to unphysical values of $\ln \bar{K}(T)$. Nevertheless, \fc provides a dedicated configuration option that allows the analytical fit to be evaluated outside the original tabulated temperature range. This option should be used with great caution, as large extrapolations may yield physically unrealistic results.

\subsubsection{Corrections for arsenic species in the Barin database}

Due to the lack of arsenic species in the JANAF tables, we used the Barin database as basis for As-bearing condensates. The data for the arsenic species \ch{As(s)}, \ch{As(g)}, \ch{As2(g)}, \ch{As3(g)}, and \ch{As4(g)} in the Barin database were originally sourced from the compilation by \citet{Hultgren1973}. According to that dataset, solid arsenic would sublimate at 1407 K via the reaction:
\begin{equation}
  \ch{As(s) -> 1/2 As2(g)} \ .
\end{equation}
However, as noted by, for example, \citet{Dabbs1983}, some of the arsenic data reported by \citet{Hultgren1973} are likely incorrect. More recent studies \citep{PankratzL.B1984Tdfm, Zhangdoi:10.1021/je1011086} indicate that the sublimation of \ch{As(s)} occurs at a significantly lower temperature, approximately 887 K, via the reaction:
\begin{equation}
  \ch{As(s) -> 1/4 As4(g)} \ .
\end{equation}
To correct these discrepancies in the Barin database, we adopted the thermodynamic data for \ch{As(s)}, \ch{As(g)}, \ch{As2(g)}, \ch{As3(g)}, and \ch{As4(g)} compiled by \citet{Gokcen1989}. 
 
\citet{Barin1995} used \ch{As(s)} and \ch{As2(g)} as reference species for calculating the enthalpy of formation ($\Delta_\mathrm{f}H^\minuso$) and Gibbs free energy of formation ($\Delta_\mathrm{f}G^\minuso$) for all arsenic-bearing species. Using the updated data from \citet{Gokcen1989}, we redefined the reference state of arsenic to include \ch{As(g)} and \ch{As4(g)} with the correct sublimation temperature of 887 K, and updated the thermodynamic properties of all arsenic-containing species in the original Barin database accordingly.

\subsection{Available input data files}
 
As with the elemental abundances, we provide several thermochemical data compilations of varying complexity. The standard input files contain data for the original set of elements and are intended for users who do not require the newly added elements. These files nevertheless include updated thermochemical data and additional species that were not part of earlier \fc versions (see Tables~\ref{table:gas-phase_species} and~\ref{table:cond_new}).
 
In addition, we provide \emph{extended} input files that include all species associated with both the original and newly added elements. Finally, as noted above, we also supply an input file based on the compilation by \citet{Hoeijmakers2019A&A...627A.165H}, which includes thermochemical data for ions and anions of elements up to uranium.
 
The extended dataset introduced in this publication comprises a total of 800 gas-phase molecules and ions, as well as 511 condensates, spanning 44 elements. As described above, the \fc input data combine solid and liquid phases of a condensate into a single data entry. If these phases were treated as independent species, the total number of included condensates would increase to 740.
 
For user convenience, the input directory of the \fc repository now also contains a dedicated Python script, \texttt{create\_custom\_compilation.py}, which allows users to generate custom input files for a selected subset of elements. This enables, for example, the rapid creation of much smaller sets of chemical species for commonly used elements such as \ch{H}, \ch{He}, C, O, and N, or the construction of tailored datasets for non-solar elemental compositions.

\section{Chemical sequences of the elements}

\subsection{Grid calculations}

In this section, we use the updated version of \fc, together with the newly added elements and chemical species, to compute the chemical equilibrium composition, including condensation, over a wide pressure–temperature ($p$–$T$) range. Specifically, we performed \fc calculations on a grid spanning temperatures from 100 to 6000~K and pressures from $10^{-12}$ to 100~bar in accordance with \citet{Stock2018MNRAS.479..865S}. This temperature range was chosen to match the domain over which most thermochemical data remain within their tabulated limits; for example, the JANAF thermochemical tables \citep{Chase1998} are generally provided for temperatures between 100 and 6000~K. 

The calculations were carried out for two sets of elemental abundances. The first adopts solar abundances from \citet{Asplund2021A&A...653A.141A} (see Table~\ref{table:abundances_asplund}), while the second assumes a carbon-to-oxygen ratio of $\mathrm{C/O}=2$ to represent a carbon-rich environment. In the latter case, the elemental oxygen abundance was fixed to its solar value, and the carbon abundance was adjusted accordingly to achieve the enhanced C/O ratio.

Each grid comprises 250\,000 individual \fc model evaluations, resulting in 500\,000 \fc calculations in total. All \fc options were left at their default settings. The number of chemistry iterations required for convergence was generally fewer than 400 at the lowest temperatures and fewer than 20 at higher temperatures. At low temperatures, determining the set of thermodynamically stable condensates required fewer than 500 additional iterations, owing to the large number of potential condensate species (see Table~\ref{table:cond_new}) and the increased number of elements included.

We note that, especially at low temperatures, the chemistry is likely no longer fully in chemical equilibrium and becomes increasingly dominated by non-equilibrium processes, such as transport. However, in many cases even rather cold systems can be reasonably described by equilibrium chemistry, at least for the main atmospheric constituents, as shown, for example, by \citet{Doucot2026MNRAS.551g1550D}.

\subsection{Results}

\begin{figure*}
  \sidecaption
  \parbox[b]{12cm}{%
    \includegraphics[width=0.49\linewidth]{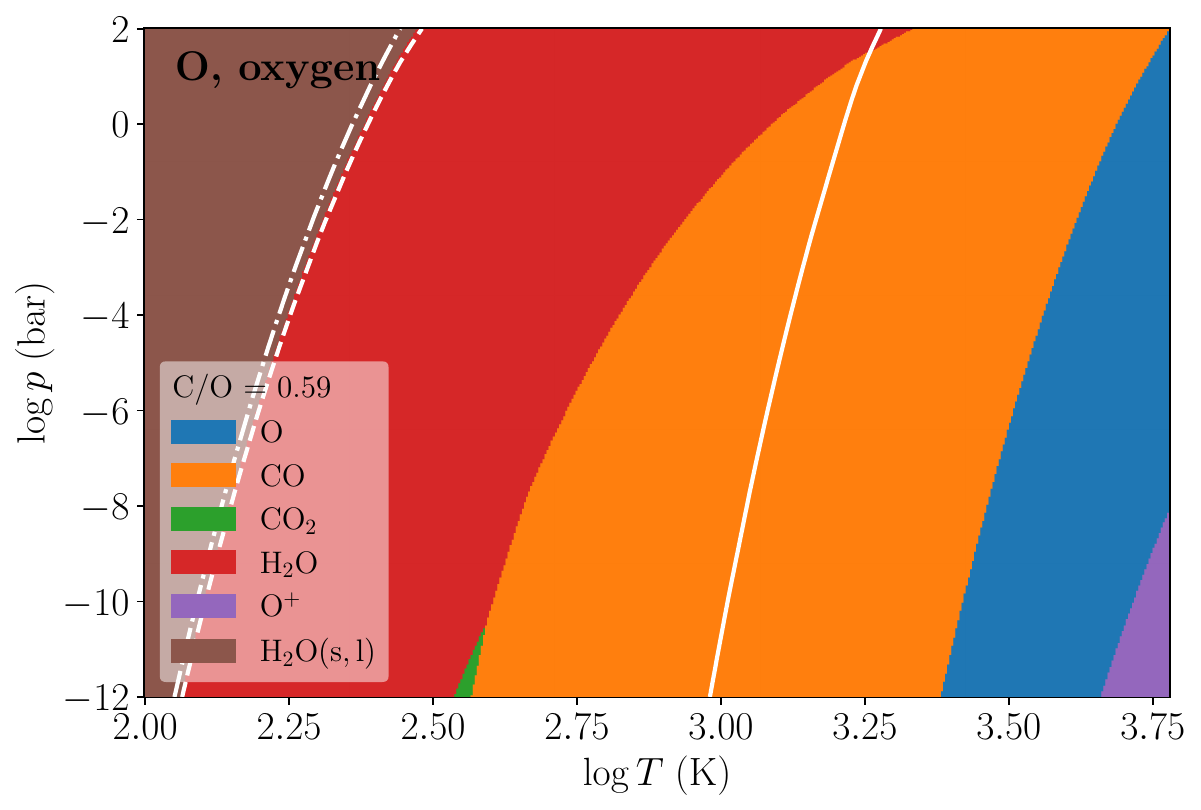}\hfill
    \includegraphics[width=0.49\linewidth]{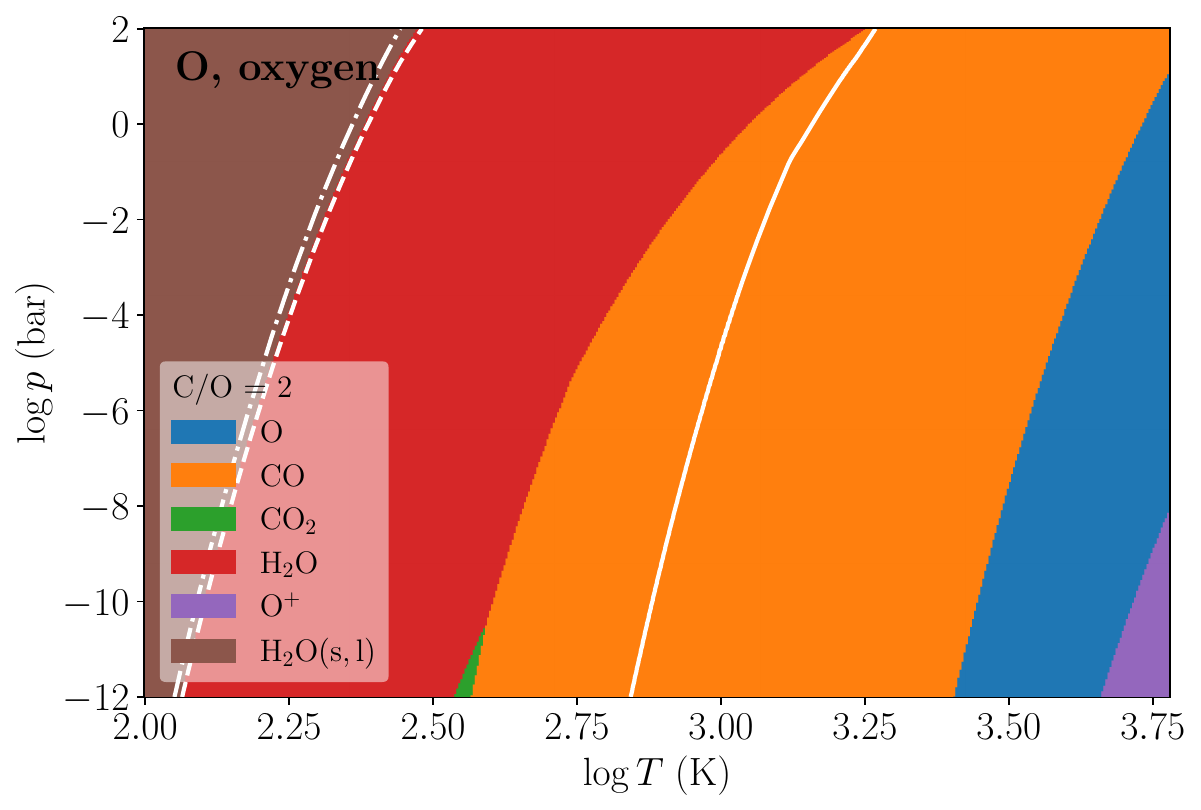}\\
    \includegraphics[width=0.49\linewidth]{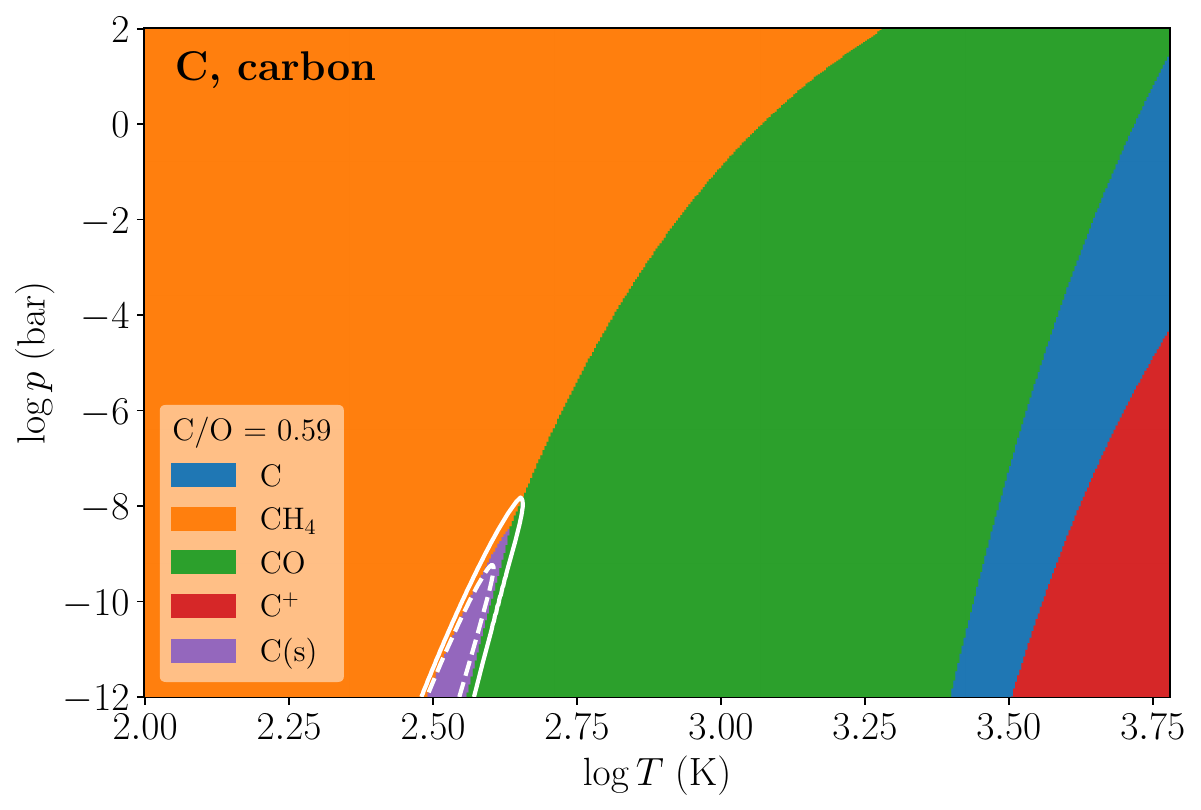}\hfill
    \includegraphics[width=0.49\linewidth]{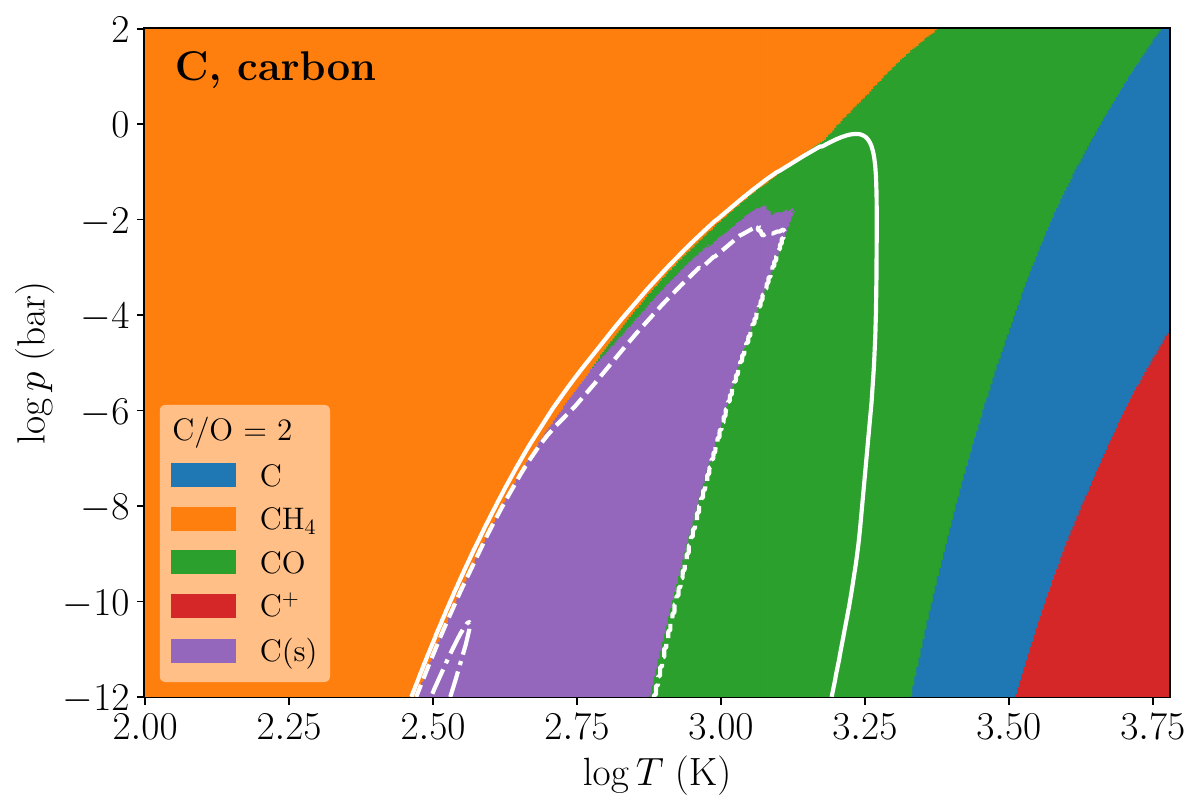}\\
    \includegraphics[width=0.49\linewidth]{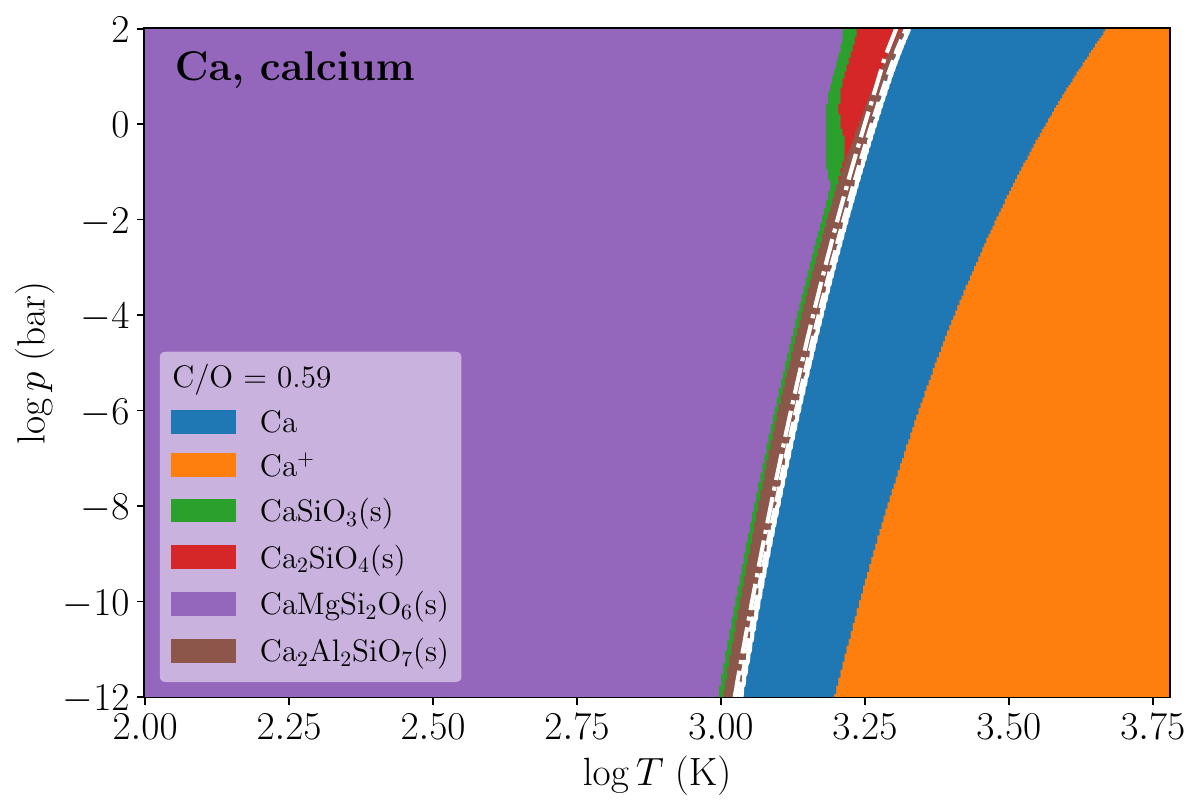}\hfill
    \includegraphics[width=0.49\linewidth]{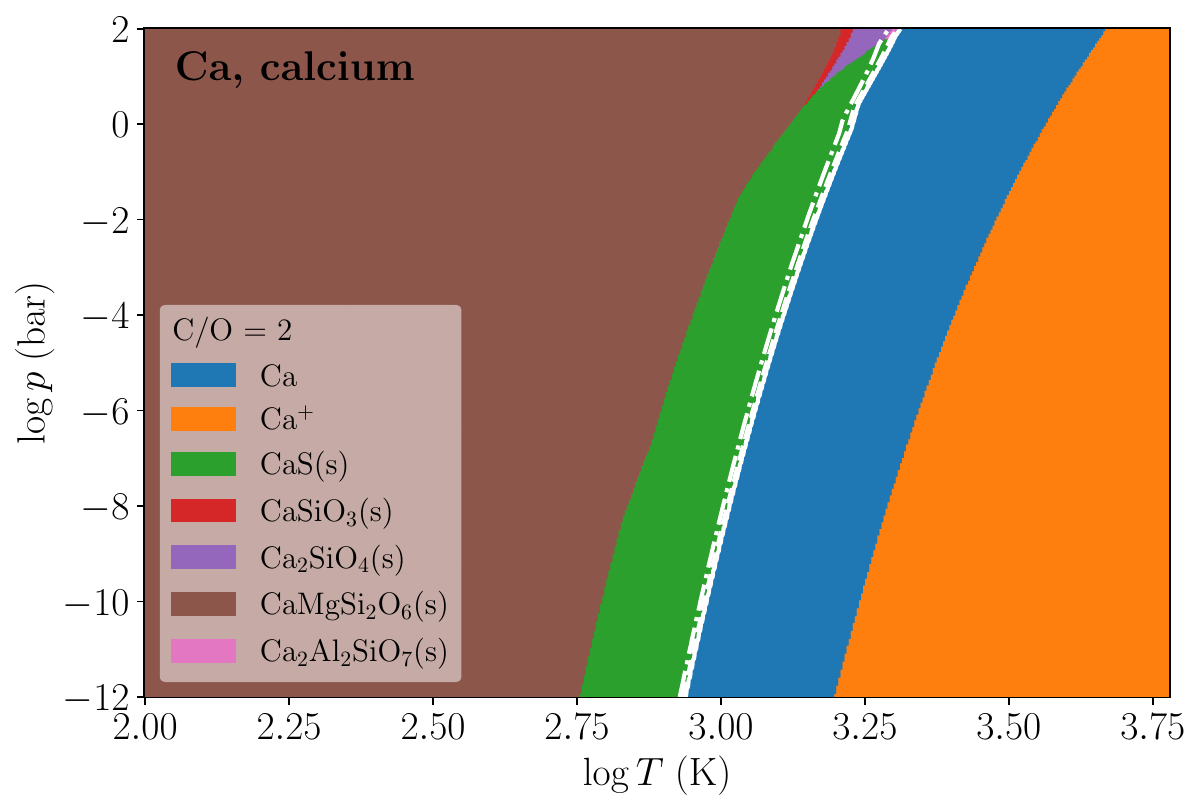}\\
    \includegraphics[width=0.49\linewidth]{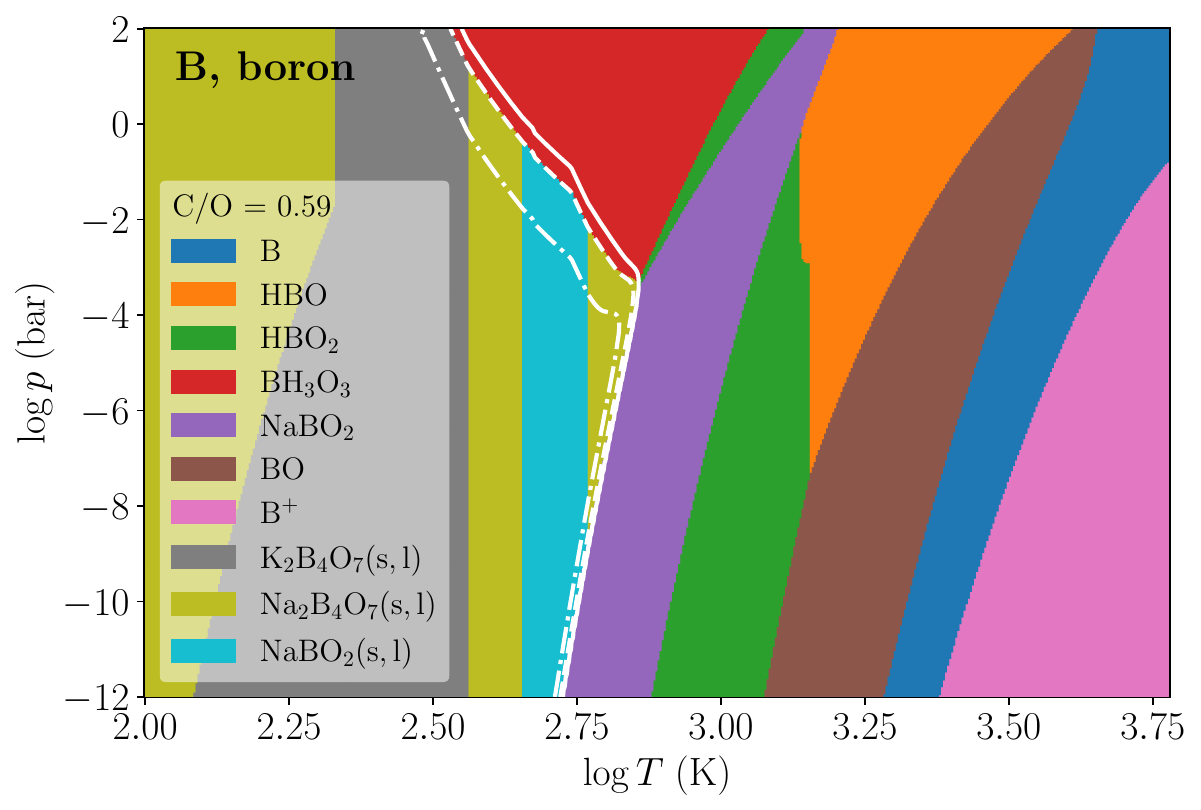}\hfill
    \includegraphics[width=0.49\linewidth]{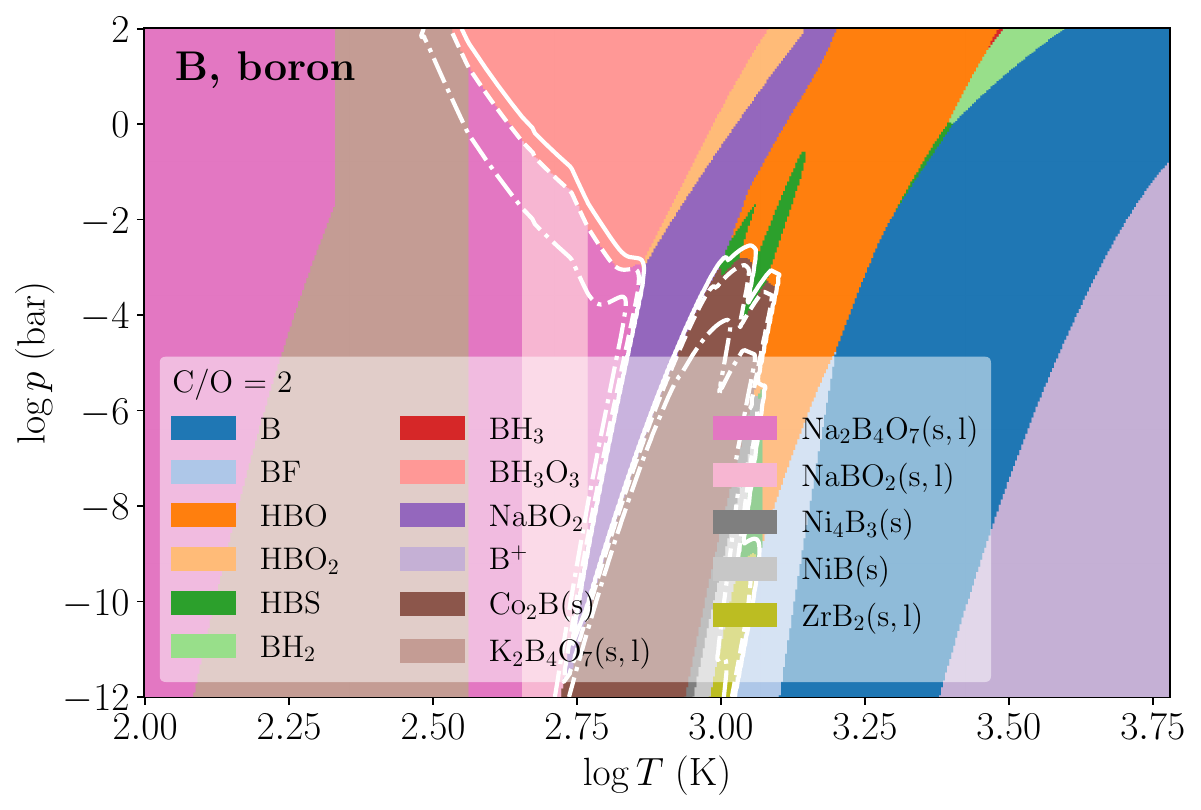}%
  }
  \caption{Results of the \fc calculations across a wide $p$-$T$ range. The panels on the left show results for solar elemental abundances, while those on the right correspond to a carbon-rich composition with a C/O ratio of 2. For each element, the plots display the dominant gas-phase and condensate species, that is, the species containing the largest fraction of the element, as a function of pressure and temperature. The white lines indicate the degree of condensation of the element: solid lines correspond to 10\% condensation, dashed lines to 50\%, and dash-dotted lines to 90\%.}
    \label{fig:chem_results}
\end{figure*}

Given the large number of gas-phase species and condensates included in \fc, it is not feasible to present the full, highly detailed output of the chemistry calculations. Instead, we show the most abundant species containing each element as a function of pressure and temperature. For the gas phase, this representation is necessarily limited, as an element is typically distributed among dozens of molecular species. For condensates, however, many elements are incorporated into only a small number of species owing to the phase rule \citep[see][]{Gib76, Gib78}; see also \citet{Kitzmann2024MNRAS.527.7263K} for further details. The results presented below therefore effectively illustrate the condensation sequences of the elements \citep[see, e.g.,][]{Grossman1972GeCoA..36..597G, Lewis1972Icar...16..241L, Lodders2003ApJ...591.1220L}.

Currently, \fc is limited to pure condensates. In more realistic situations, however, solid and liquid solutions can also play an important role, for example Fe–Ni–Co–Si alloys or silicate solid solutions such as olivine, \ch{(Mg,Fe)2SiO4}, and pyroxene, \ch{(Mg,Fe)SiO3} \citep[see, e.g.,][]{Ebel2021}. Dissolving Co in an Fe–Ni alloy at higher temperatures, for instance, would prevent the formation of pure Co(s,l) as a separate condensate at lower temperatures. \fc will, therefore, tend to predict that minor elements such as Co condense at too low temperatures and thus remain in the gas phase over too large a temperature range.

Figure~\ref{fig:chem_results} shows representative results for oxygen (O), carbon (C), calcium (Ca), and boron (B). All other elements included in \fc, excluding hydrogen and the noble gases, are presented in Appendix~\ref{sec:appendix_chemistry}. Within the selected $p$–$T$ range of our computational grid, the noble gases are present only in atomic or ionised form.

\subsubsection{Solar C/O ratio}

For solar elemental abundances, carbon and oxygen follow the expected chemical sequence \citep[see, e.g.,][]{Burrows1999ApJ...512..843B, Lodders2002Icar..155..393L, Visscher2006ApJ...648.1181V}. At high temperatures, both elements are initially present in ionised and atomic forms before combining to form carbon monoxide (CO), which is the most thermodynamically stable molecule over a wide range of pressure–temperature ($p$–$T$) conditions. As a result, CO sequesters most of the available carbon and oxygen. At lower temperatures, methane (\ch{CH4}) becomes the dominant carbon-bearing gas-phase species, while water vapour (\ch{H2O}) becomes the primary oxygen-bearing molecule. Eventually, water vapour condenses into liquid or solid phases. Although a large fraction of oxygen condenses into water, a significant amount is also incorporated into oxygen-bearing minerals such as diopside (\ch{CaMgSi2O6(s)}; see the left panel of Fig.~\ref{fig:chem_results}), as well as forsterite (\ch{Mg2SiO4}) and pyroxmangite (\ch{MnSiO3(s)}; see Appendix~\ref{sec:appendix_chemistry}). Consequently, oxygen condensation begins at relatively high temperatures and can exceed 10\% even before water condensation occurs. At solar metallicity, up to 20\% of oxygen is typically bound in high-temperature condensates \citep[see, e.g.,][]{Woitke2018A&A...614A...1W, Sharp1990ApJS...72..417S}.

In contrast, carbon condenses into graphite (\ch{C(s,l)}) only within a narrow region of the $p$–$T$ plane and otherwise remains predominantly in the gas phase. Graphite formation requires a sufficient abundance of free carbon in the gas phase. As long as carbon is largely locked in CO, the saturation ratio of graphite remains well below one. Only when the dominant oxygen carrier transitions from CO to \ch{H2O} does enough free carbon become available to allow graphite formation, and even then only over a restricted range of temperatures and pressures. At still lower temperatures, carbon may condense into solid or liquid methane (\ch{CH4(s,l)}; see, e.g., \citealt{Kitzmann2024MNRAS.527.7263K}).

Many elements exhibit a relatively simple sequence of dominant gas-phase species. As shown in Fig.~\ref{fig:chem_results}, oxygen is predominantly found in atomic form or bound in \ch{CO}, \ch{CO2}, or \ch{H2O}. Similarly, carbon is divided mainly between \ch{CO} and \ch{CH4}. Other elements, however, display much more complex behaviour. For example, the dominant boron-bearing species follow a relatively complicated sequence that depends strongly on both temperature and pressure, as illustrated in the lower panel of Fig.~\ref{fig:chem_results}.

While elements such as carbon and oxygen condense only partially over a wide $p$–$T$ range, other elements condense very rapidly once a critical temperature is reached. This behaviour is evident, for example, for calcium in Fig.~\ref{fig:chem_results}, where the temperatures corresponding to 10\% and 90\% condensation are nearly identical. Similar behaviour is observed for many other elements shown in Appendix~\ref{sec:appendix_chemistry}, including bromine (Br), copper (Cu), iron (Fe), manganese (Mn), and zinc (Zn).

Some elements follow relatively simple condensate sequences. Calcium, for instance, transitions from \ch{CaSiO3(s)} to \ch{Ca2SiO4(s)}, then to \ch{Ca2Al2SiO7(s)}, and finally to \ch{CaMgSi2O6(s)} at lower temperatures, in line with the classical solar-composition condensation sequence \citep{Grossman1972GeCoA..36..597G, Lodders2003ApJ...591.1220L}. Other elements form only a very limited number of condensates. Cobalt (Co), for example, condenses solely as elemental cobalt (\ch{Co(s,l)}), while scandium (Sc) appears only as \ch{Sc2O3(s,l)}, and zinc primarily as \ch{ZnS(s,l)}.

By contrast, several elements display much more complex condensation behaviour. Titanium, for example, forms a wide variety of condensates, including \ch{MgTi2O5(s,l)}, \ch{TiN(s,l)}, \ch{TiO(s,l)}, \ch{TiO2(s,l)}, \ch{Ti3O5(s,l)}, \ch{Ti4O7(s,l)}, and \ch{CaTiO3(s)} \citep[see Fig.~\ref{fig:chem_plots} and][]{Lodders2002ApJ...577..974L}. Other elements with similarly complex condensation sequences include vanadium (V), magnesium (Mg), and zirconium (Zr).

An interesting silicon-bearing condensate that is often neglected in equilibrium chemistry models is silicon monoxide (\ch{SiO(s)}). Most thermochemical databases lack data for this species and instead assume quartz (\ch{SiO2}) to be the more stable species. Experimental data for \ch{SiO(s)} are available from laboratory measurements by \citet{Gail2013AA...555A.119G}, which are included in the \fc data compilation as well as in \texttt{GGChem} \citep{Woitke2018A&A...614A...1W}. As shown in Fig.~\ref{fig:chem_plots}, \ch{SiO(s)} does indeed form, albeit only over a limited range of temperatures and pressures: around 1000~K and at pressures above approximately 0.1~bar. Notably, recent \textit{James Webb} Space Telescope (JWST) observations suggest that \ch{SiO(s)} may be present in the atmospheres of brown dwarfs within the temperature and pressure regime predicted by our calculations \citep{Molliere2025A&A...703A..79M}.

\subsubsection{Carbon-rich case}

In the carbon-rich case ($\mathrm{C/O}=2$), a broadly similar chemical sequence is observed, but with notable shifts in condensation behaviour (see the right panels of Figs.~\ref{fig:chem_results} and \ref{fig:chem_plots}); see, e.g., \citet{Larimer1979GeCoA..43.1455L}, \citet{Lodders1997ApJ...484L..71L}, and \citet{Lodders1999IAUS..191..279L} for analyses of the corresponding condensation sequences. Most notably, oxygen begins to condense only at lower temperatures than in the solar-composition case. This shift arises from the large chemical binding energy of CO: when the C/O ratio is larger than one, nearly all of the available oxygen is locked in CO, leaving no free oxygen to form oxygen-bearing condensates. Only at lower temperatures, where a fraction of the carbon condenses into graphite, does oxygen become available again, allowing other oxygen-bearing condensates to form.

Several elements exhibit markedly altered condensation sequences under carbon-rich conditions. As Fig.~\ref{fig:chem_results} suggests, instead of initially forming \ch{Ca2Al2SiO7(s)} as in the solar C/O case, calcium now first condenses into calcium sulfide (\ch{CaS(s)}) over a large $p$-$T$ range before finally forming the usual diopside (\ch{CaMgSi2O6(s)}) at lower temperatures. A similar example is aluminium, which in the carbon-rich case initially condenses into aluminium nitride (\ch{AlN(s)}) owing to the lack of available oxygen. Only once oxygen becomes available again does it form the usual oxygen-bearing condensates, such as corundum (\ch{Al2O3}) and spinel (\ch{MgAl2O4}). Boron also shows a strongly modified condensation sequence (see Fig.~\ref{fig:chem_results}). In the solar C/O case, boron condenses into oxygen-bearing species such as \ch{NaBO2(s,l)} and \ch{K2B4O7(s,l)}. Although these species also form in the $\mathrm{C/O}=2$ case over a similar temperature range, boron begins to condense at significantly higher temperatures into species such as \ch{NiB(s)} and \ch{ZrB(s,l)}. Sulphur likewise exhibits substantially altered behaviour. For solar C/O, sulphur condenses only into troilite (\ch{FeS}), whereas for $\mathrm{C/O}=2$ it condenses much earlier into \ch{MgS(s)}. This species cannot form in the solar case because magnesium is already sequestered in oxygen-bearing condensates such as enstatite and forsterite. In the carbon-rich case, however, the lack of freely available oxygen prevents the formation of these magnesium silicates, leaving magnesium available to form other magnesium-bearing condensates.

As expected for C/O > 1, carbon increasingly condenses into graphite. Nevertheless, even under carbon-rich conditions, solid carbon formation is confined to a relatively narrow region of the $p$–$T$ plane. This highlights a key challenge for the formation of graphite dust in environments such as the outer atmospheres of asymptotic giant branch (AGB) stars: the atmospheric $p$–$T$ profile must pass through this restricted region, the ``dust-forming window'', for graphite (\ch{C(s)}) to form and enable dust-driven winds \citep[see, e.g.,][]{GailSedlmayr2013}.

Naively, one might expect carbides to be the dominant condensates over a wide $p$–$T$ range in a $\mathrm{C/O}=2$ environment. However, although carbon-bearing condensates such as zirconium carbide (\ch{ZrC}), titanium carbide (\ch{TiC}), vanadium carbide (\ch{V2C}), and silicon carbide (\ch{SiC}) do form, they are stable only over very limited regions of the temperature–pressure plane. This is again a consequence of the strong chemical stability of gaseous CO, which sequesters most of the carbon. Furthermore, once graphite forms, most carbon is removed from the gas phase into \ch{C(s)}. Even after graphite disappears at lower temperatures and carbon returns to the gas phase, it is predominantly bound in methane (\ch{CH4}) and is therefore not available for carbide formation.

Some elements do not change their condensation sequences in the carbon-rich case. This is generally true for elements that do not form stable oxygen- or carbon-bearing condensates, such as cobalt, copper, nickel, and zinc. The chemistries of zinc and rubidium, for example, are essentially unaffected by the C/O ratio (see Fig.~\ref{fig:chem_plots}).

As in the solar C/O case, silicon also condenses into \ch{SiO(s)}. In contrast to the solar-abundance case, however, \ch{SiO(s)} now forms over a much wider temperature–pressure range, superseding silicon carbide as the dominant silicon-bearing condensate at temperatures below about 900~K. This behaviour results from the lack of free oxygen to form quartz. These results suggest that \ch{SiO(s)} should be explicitly considered in thermochemical equilibrium calculations, particularly for carbon-rich environments in which the availability of free oxygen is limited.

In general, most elements condense below a characteristic temperature and remain locked in condensed form thereafter. One exception to this general trend is carbon. In both cases, it condenses into graphite at some temperature but returns to the gas phase as methane at lower temperatures. Another example is boron in the $\mathrm{C/O}=2$ case. Here, boron shows a complex behaviour in the gas and the condensed phase as a function of temperature and pressure. It first condenses into \ch{Co2B(s)}, briefly returns to the gas phase as \ch{NaBO2}, and finally condenses again as \ch{Na2B4O7(s,l)}.

\section{Summary}

In this work, we presented a major update to the chemical equilibrium code \fc, aimed at improving its numerical robustness, expanding its chemical data, and extending its applicability to a wider range of astrophysical and planetary environments.
 
We introduced a convergence accelerator based on a multidimensional Newton method that significantly improves the stability and performance of the gas-phase chemistry solver, particularly for strong deviations from solar elemental abundances. This addresses convergence issues that previously arose when elements with comparable abundances were tightly coupled through shared molecular species. We also reformulated the gas-phase equations in logarithmic element densities, removing the dependence on quadruple-precision arithmetic and allowing \fc to be applied at temperatures as low as 1\,K on any platform that supports double precision. Additional improvements to the condensate solver, including adaptive Levenberg--Marquardt regularisation and a combined gas--condensate Newton method, further enhance robustness for strongly non-solar elemental compositions. Together, these changes also yield substantial speed gains: a representative protoplanetary-disk calculation that previously required approximately 60 minutes on a 16-core processor now completes in about 18 seconds.
 
In addition, we substantially expanded the chemical database by adding numerous new elements together with their associated gas-phase species and condensates. This extension enables \fc to model the chemistry of environments that include rare-earth, transition, and heavy metals, including elements that have recently been detected in the atmospheres of ultra-hot Jupiters. Multiple elemental abundance compilations and datasets of varying complexity are now supported, allowing users to tailor the selection of chemical data to their specific application while maintaining computational efficiency.
 
Using the updated \fc version, we computed comprehensive chemical equilibrium grids, including condensation, over a wide range of pressures and temperatures for both solar and carbon-rich ($\mathrm{C/O}=2$) elemental compositions. These calculations demonstrate the resulting condensation sequences for all included elements and highlight the diversity and complexity of chemical behaviour across different thermodynamic regimes. In particular, we showed how condensate sequences can change substantially with elemental composition, especially for carbon-rich environments where the formation of CO sequesters most of the oxygen.
 
Overall, the improvements presented here make \fc a more robust and flexible tool for equilibrium chemistry calculations. The updated code is well suited for applications ranging from atmospheric retrievals and global circulation models to large parameter-space studies, as well as the interpretation of JWST observations and high-resolution spectra from ground-based telescopes \citep[e.g.,][]{Prinoth2024AJ....168..133P}. By providing efficient and extensible equilibrium calculations, this work also lays the groundwork for more advanced modelling efforts that incorporate kinetic, transport, and photochemical processes.

\section*{Data availability}
  The equilibrium chemistry code \fc and the thermochemical data is available at \url{https://github.com/NewStrangeWorlds/FastChem}. The corrensponding Python package \textsc{pyfastchem} is available at \url{https://pypi.org/project/pyfastchem}.

\begin{acknowledgements}
 D.K. acknowledges the support from the Swiss National Science Foundation under the grant 10002706. This work has been carried out within the framework of the National Centre of Competence in Research PlanetS supported by the Swiss National Science Foundation under grants 51NF40\_182901 and 51NF40\_205606. D.K. acknowledges the financial support of the SNSF. 
\end{acknowledgements}

\bibliographystyle{aa} 
\bibliography{references}

\clearpage

\begin{appendix}

\onecolumn

\section{Overview of gas-phase and condensate species included in \fc}

\begin{table}[h!]
\caption{\label{table:gas-phase_species} List of all gas-phase species included in the present version of \fc.}
\centering
\scriptsize
\begin{tabular}{lp{0.9\textwidth}}
\hline\hline
Element & Molecules, Ions \\
\hline
\ch{H}  & \underline{\ch{H-}}$^j$, \underline{\ch{H+}}$^j$, \ch{H2}, \ch{H2-}, \ch{H2+} \\
\ch{He} &  \underline{\ch{He+}}$^j$ \\
\ch{O}  & \ch{HO}, \ch{HO-}, \ch{HO+}, \ch{HO2}, \ch{H2O}, \ch{H2O2}$^e$, \ch{H3O+},  \underline{\ch{O-}}$^j$,  \underline{\ch{O+}}$^j$, \ch{O2}, \ch{O2-}, \ch{O2+}, \ch{O3}\\
\ch{C}  &  \underline{\ch{C-}}$^j$,  \underline{\ch{C+}}$^j$, \ch{CH}, \ch{CH-}, \ch{CH+}, \ch{CHO}, \ch{CHO+}, \ch{CH2}, \ch{CH2O}, \ch{CH3}, \ch{CH4}, \ch{CH4O2}$^c$, \ch{CO}, \ch{CO2}, \ch{CO2-}, \ch{C2}, \ch{C2-}, \ch{C2H}, \ch{C2H2}, \ch{C2H2O2}$^c$, \ch{C2H2O4}$^c$, \ch{C2H4}, \ch{C2H4O}, \ch{C2H4O3}$^c$, \ch{C2H6O2}$^c$, \ch{C2O}, \ch{C3}, \ch{C3H}, \ch{C3O2}, \ch{C4}, \ch{C4H6O4}$^c$, \ch{C5} \\
\ch{Ne} & \underline{\ch{Ne+}}$^j$ \\
\ch{N}  & \underline{\ch{HCN}}$^k$, \ch{HNC}$^k$, \ch{CHNO}, \ch{CN}, \ch{CN-}, \ch{CN+}, \ch{CNO}, \ch{CNN}, \ch{NCN} \ch{C2N}, \ch{C2NO}$^c$, \ch{C2N2}, \ch{C3N2O}$^c$, \ch{C4N2}, \ch{HN}$^k$, \ch{HNO}, \textsl{cis}-\ch{HNO2}, \textsl{trans}-\ch{HNO2}, \ch{HNO3}$^e$, \ch{H2N}, \ch{H2N2}, \ch{H3N}, \ch{H4N2},  \underline{\ch{N-}}$^j$,  \underline{\ch{N+}}$^j$, \ch{NO}, \ch{NO+}, \ch{NO2}, \ch{NO2-}, \ch{NO3}, \ch{N2}, \ch{N2-}, \ch{N2+}, \ch{N2O}, \ch{N2O+}, \ch{N2O3}, \ch{N2O4}, \ch{N2O5}, \ch{N3} \\
\ch{Mg} & \ch{HMg}$^k$, \ch{HMgO}, \ch{HMgO+}, \ch{H2MgO2},  \underline{\ch{Mg+}}$^j$, \ch{MgN}, \underline{\ch{MgO}}$^i$, \ch{Mg2} \\
\ch{Si} & \ch{CSi}, \ch{CSi2}, \ch{C2Si}, \ch{C2Si2}, \ch{C4H2Si}, \ch{HSi}, \ch{HSi+}, \ch{H2Si}, \ch{H3Si}, \ch{H4Si}, \ch{NSi}, \ch{NSi2}, \ch{OSi}, \ch{O2Si},  \underline{\ch{Si-}}$^j$,  \underline{\ch{Si+}}$^j$, \ch{Si2}, \ch{Si3} \\
\ch{Fe} & \ch{C5FeO5}, \underline{\ch{Fe-}}$^j$, \underline{\ch{Fe+}}$^j$, \ch{FeH}$^k$, \ch{FeH2O2}, \ch{FeO} \\
\ch{S}  & \ch{COS}, \ch{CS}, \ch{CS2}, \ch{FeS}, \ch{HS}$^f$, \ch{HS-}, \ch{H2O4S}$^e$, \ch{H2S}, \underline{\ch{MgS}}$^i$, \ch{NS}$^f$, \ch{OS}, \ch{OS2}$^f$, \ch{O2S}, \ch{O3S}, \underline{\ch{S-}}$^j$, \underline{\ch{S+}}$^j$, \ch{SSi}, \ch{S2}, \ch{S3}, \ch{S4}, \ch{S5}, \ch{S6}, \ch{S7}, \ch{S8} \\
\ch{Al} & \underline{\ch{Al-}}$^j$, \underline{\ch{Al+}}$^j$, \ch{AlH}, \ch{OAlH}, \ch{AlOH} \ch{AlHO-}, \ch{AlHO+}, \ch{AlHO2}, \underline{\ch{AlN}}$^i$, \ch{AlO}, \ch{AlO-}, \ch{AlO+}, \ch{AlO2}, \ch{AlO2-}, \ch{AlS}, \ch{Al2}, \ch{Al2O}, \ch{Al2O+}, \ch{Al2O2}, \ch{Al2O2+}, \ch{CAl} \\
\ch{Ar} & \underline{\ch{Ar+}}$^j$ \\
\ch{Ca} & \underline{\ch{Ca-}}$^j$, \underline{\ch{Ca+}}$^j$, \underline{\ch{CaH}}$^d$, \ch{CaHO}, \ch{CaHO+}, \ch{CaH2O2}, \ch{CaO}, \ch{CaS}, \ch{Ca2} \\
\ch{Na} & \ch{CNNa}, \ch{C2N2Na2}, \ch{HNa}$^k$, \ch{HNaO}, \ch{HNaO+}, \ch{H2Na2O2}, \underline{\ch{Na-}}$^j$, \underline{\ch{Na+}}$^j$, \ch{NaO}, \ch{NaO-}, \ch{Na2}, \ch{Na2O4S} \\
\ch{Ni} & \ch{C4NiO4}, \ch{HNi}$^k$, \underline{\ch{Ni-}}$^j$, \underline{\ch{Ni+}}$^j$, \underline{\ch{NiO}}$^i$, \ch{NiS} \\
\ch{Cr} & \ch{C2Cr}, \underline{\ch{Cr-}}$^j$, \underline{\ch{Cr+}}$^j$, \ch{CrH}$^k$, \ch{CrN}, \ch{CrO}, \ch{CrO2}, \ch{CrO3}, \textbf{\ch{CrS}}$^i$ \\
\ch{Mn} & \ch{HMn}$^k$, \underline{\ch{Mn+}}$^j$, \underline{\ch{MnO}}$^i$, \ch{MnS} \\
\ch{P}  & \ch{CHP}, \ch{CP}$^i$, \underline{\ch{HP}}$^b$, \ch{H2P}, \underline{\ch{H3P}}$^b$, \ch{NP}$^b$, \ch{OP}$^k$, \ch{O2P}, \ch{O6P4}, \ch{O10P4}, \underline{\ch{P-}}$^j$, \underline{\ch{P+}}$^j$, \ch{PS}$^f$, \ch{P2}, \ch{P4}, \ch{P4S3} \\
\ch{Cl} & \ch{AlCl}, \ch{AlCl+}, \ch{AlClO}, \ch{AlCl2}, \ch{AlCl2-}, \ch{AlCl2+}, \ch{AlCl3}, \ch{Al2Cl6}, \underline{\ch{CCl}}$^i$, \ch{CClN}, \ch{CClO}, \ch{CCl2}, \ch{CCl2O}, \ch{CCl3}, \ch{CCl4}, \ch{CHCl}, \ch{CHCl3}, \ch{CH2Cl2}, \ch{CH3Cl}, \ch{CH3Cl3Si}, \ch{C2Cl2}, \ch{C2Cl4}, \ch{C2Cl6}, \ch{C2HCl}, \ch{C2H3ClO2}$^c$, \ch{CaCl}, \ch{CaCl2}, \underline{\ch{Cl-}}$^j$, \underline{\ch{Cl+}}$^j$, \ch{ClFe}, \ch{ClH}$^g$, \ch{ClHO}, \ch{ClH3Si}, \ch{ClMg}, \ch{ClMg+}, \textbf{\ch{ClMn}}$^i$, \textbf{\ch{ClN}}$^i$, \ch{ClNO}, \ch{ClNO2}, \ch{ClNa}, \ch{ClNi}, \ch{ClO}, \ch{ClO2}, \ch{ClO3}, \underline{\ch{ClP}}, \ch{ClS}, \ch{ClS+}, \ch{ClS2}, \ch{ClSi}, \ch{Cl2}, \ch{Cl2Fe}, \ch{Cl2H2Si}, \ch{Cl2Mg}, \ch{Cl2Na2}, \ch{Cl2Ni}, \ch{ClOCl}, \ch{ClClO} \ch{ClO2Cl}, \ch{ClOClO} \ch{Cl2O2S}, \ch{Cl2S}, \ch{Cl2S+}, \ch{Cl2S2}, \ch{Cl2Si}, \ch{Cl3Fe}, \ch{Cl3HSi}, \ch{Cl3OP}, \ch{Cl3P}, \ch{Cl3PS}, \ch{Cl3Si}, \ch{Cl4Fe2}, \ch{Cl4Mg2}, \ch{Cl4Si}, \ch{Cl5P}, \ch{Cl6Fe2} \\
\ch{K}  & \ch{CKN}, \ch{C2K2N2}, \ch{ClK}, \ch{Cl2K2}, \ch{HK}, \ch{HKO}, \ch{HKO+}, \ch{H2K2O2}, \underline{\ch{K-}}$^j$, \underline{\ch{K+}}$^j$, \ch{KO}, \ch{KO-}, \ch{K2}, \ch{K2O4S}\\
\ch{Ti} & \ch{C2Ti}, \ch{C4Ti}, \ch{ClOTi}, \ch{ClTi}, \ch{Cl2OTi}, \ch{Cl2Ti}, \ch{Cl3Ti}, \ch{Cl4Ti}, \ch{HTi}$^h$, \underline{\ch{NTi}}$^i$, \ch{OTi}, \ch{O2Ti}, \ch{STi}$^a$, \underline{\ch{Ti-}}$^j$, \underline{\ch{Ti+}}$^j$ \\
\ch{Co} & \ch{ClCo}, \ch{Cl2Co}, \ch{Cl3Co}, \ch{Cl4Co2}, \underline{\ch{Co-}}$^j$, \underline{\ch{Co+}}$^j$, \textbf{\ch{CoH}}$^i$ \\
\ch{Zn} & \textbf{\ch{ClZn}}$^i$, \textbf{\ch{HZn}}$^i$, \underline{\ch{Zn-}}$^j$, \underline{\ch{Zn+}}$^j$ \\
\ch{F}  & \ch{AlClF}, \ch{AlClF+}, \ch{AlClF2}, \ch{AlCl2F}, \ch{AlF}, \ch{AlF+}, \ch{AlFO}, \ch{AlF2}, \ch{AlF2-}, \ch{AlF2+}, \ch{AlF2O}, \ch{AlF2O-}, \ch{AlF3}, \ch{AlF4-}, \ch{AlF4Na}, \ch{Al2F6}, \ch{CClFO}, \ch{CClF3}, \ch{CCl2F2}, \ch{CCl3F}, \ch{CF}, \ch{CF+}, \ch{CFN}, \ch{CFO}, \ch{CF2}, \ch{CF2+}, \ch{CF2O}, \ch{CF3}, \ch{CF3+}, \ch{CF4}, \ch{CF4O}, \ch{CF8S}, \ch{CHClF2}, \ch{CHCl2F}, \ch{CHF}, \ch{CHFO}, \ch{CHF3}, \ch{CH2ClF}, \ch{CH2F2}, \ch{CH3F}, \ch{CH3F3Si}, \ch{C2F2}, \ch{C2F3N}, \ch{C2F4}, \ch{C2F6}, \ch{C2HF}, \ch{CaF}, \ch{CaF2}, \ch{ClF}, \ch{ClFMg}, \ch{ClFO2S}, \ch{ClFO3}, \ch{ClF2OP}, \ch{ClF3}, \ch{ClF3Si}, \ch{ClF5}, \ch{ClF5S}, \ch{Cl2FOP}, \ch{Cl3FSi}, \ch{CoF2}, \underline{\ch{F-}}$^j$, \underline{\ch{F+}}$^j$, \ch{F0S2}, \ch{FFe}, \ch{FH}$^g$, \ch{FHO}, \ch{FHO3S}, \ch{FH3Si}, \ch{FK}, \ch{FMg}, \ch{FMg+}, \textbf{\ch{FMn}}$^i$, \ch{FN}, \ch{FNO}, \ch{FNO2}, \ch{FNO3}, \ch{FNa}, \textbf{\ch{FNi}}$^i$, \ch{FO}, \ch{FOTi}, \ch{OFO}, \ch{FOO} \ch{FP}, \ch{FP-}, \ch{FP+}, \ch{FPS}, \ch{FS}, \ch{FS-}, \ch{FS+}, \ch{FSi}, \ch{FTi}, \textbf{\ch{FZn}}$^i$, \ch{F2}, \ch{F2Fe}, \ch{F2H2}, \ch{F2H2Si}, \ch{F2K-}, \ch{F2K2}, \ch{F2Mg}, \ch{F2Mg+}, \ch{F2N}, \textsl{cis}-\ch{F2N2}, \textsl{trans}-\ch{F2N2} \ch{F2Na-}, \ch{F2Na2}, \ch{F2O}, \ch{F2OS}, \ch{F2OSi}, \ch{F2OTi}, \ch{F2O2}, \ch{F2O2S}, \ch{F2P}, \ch{F2P-}, \ch{F2P+}, \ch{F2S}, \ch{F2S-}, \ch{F2S+}, \ch{FS2F}, \ch{SSF2} \ch{F2Si}, \ch{F2Ti}, \ch{F3Fe}, \ch{F3HSi}, \ch{F3H3}, \ch{F3N}, \ch{F3NO}, \ch{F3OP}, \ch{F3P}, \ch{F3PS}, \ch{F3S}, \ch{F3S-}, \ch{F3S+}, \ch{F3Si}, \ch{F3Ti}, \ch{F4H4}, \ch{F4Mg2}, \ch{F4N2}, \ch{F4S}, \ch{F4S-}, \ch{F4S+}, \ch{F4Si}, \ch{F4Ti}, \ch{F5H5}, \ch{F5P}, \ch{F5S}, \ch{F5S-}, \ch{F5S+}, \ch{F6H6}, \ch{F6S}, \ch{F6S-}, \ch{F7H7} \\
\ch{Cu} & \ch{ClCu}, \ch{Cl3Cu3}, \underline{\ch{Cu-}}$^j$, \underline{\ch{Cu+}}$^j$, \ch{CuF}, \ch{CuF2}, \ch{CuH}$^k$, \ch{CuO}, \ch{CuS}$^a$, \ch{Cu2} \\
\ch{V}  & \ch{C2V}, \ch{C4V}, \ch{NV}, \ch{OV}, \ch{O2V}, \underline{\ch{V-}}$^j$, \underline{\ch{V+}}$^j$ \\
\ch{Ge} & \textbf{\ch{ClGe}}$^i$, \textbf{\ch{FGe}}$^i$, \textbf{\ch{Ge-}}$^j$, \textbf{\ch{Ge+}}$^j$, \textbf{\ch{GeH}}$^i$, \textbf{\ch{GeO}}$^i$, \textbf{\ch{GeS}}$^i$ \\
\textbf{\ch{Se}} & \textbf{\ch{AlSe}}$^i$, \textbf{\ch{CSe}}$^i$, \textbf{\ch{ClSe}}$^i$, \textbf{\ch{FSe}}$^i$, \textbf{\ch{GeSe}}$^i$, \textbf{\ch{HSe}}$^i$, \textbf{\ch{NSe}}$^i$, \textbf{\ch{OSe}}$^i$, \textbf{\ch{SSe}}$^i$, \textbf{\ch{Se-}}$^j$, \textbf{\ch{Se+}}$^j$, \textbf{\ch{Se2}}$^i$ \\
\textbf{\ch{Sc}} & \textbf{\ch{ClSc}}$^i$, \textbf{\ch{FSc}}$^i$, \textbf{\ch{OSc}}$^i$, \textbf{\ch{O2Sc}}$^a$, \textbf{\ch{SSc}}$^i$, \textbf{\ch{Sc-}}$^j$, \textbf{\ch{Sc+}}$^j$ \\
\textbf{\ch{Kr}} & \textbf{\ch{Kr+}}$^j$ \\
\textbf{\ch{Ga}} & \textbf{\ch{ClGa}}$^i$, \textbf{\ch{FGa}}$^i$, \textbf{\ch{Ga-}}$^j$, \textbf{\ch{Ga+}}$^j$, \textbf{\ch{GaH}}$^i$, \textbf{\ch{GaO}}$^i$ \\
\textbf{\ch{Sr}} & \textbf{\ch{ClSr}}, \textbf{\ch{Cl2Sr}}, \textbf{\ch{FSr}}, \textbf{\ch{FSr+}}, \textbf{\ch{F2Sr}}, \textbf{\ch{HOSr}}, \textbf{\ch{HOSr+}}, \textbf{\ch{HSr}}$^i$, \textbf{\ch{H2O2Sr}}, \textbf{\ch{OSr}}$^i$, \textbf{\ch{SSr}}$^d$, \textbf{\ch{Sr-}}$^j$, \textbf{\ch{Sr+}}$^j$ \\
\textbf{\ch{B}}  & \textbf{\ch{AlBO2}}, \textbf{\ch{B-}}$^j$, \textbf{\ch{B+}}$^j$, \textbf{\ch{BCl}}$^i$, \textbf{\ch{BCl+}}$^k$, \textbf{\ch{BClF}}$^d$, \textbf{\ch{BClF2}}, \textbf{\ch{BClO}}, \textbf{\ch{BCl2}}$^d$, \textbf{\ch{BCl2-}}, \textbf{\ch{BCl2+}}, \textbf{\ch{BCl2F}}, \textbf{\ch{BCl2H}}, \textbf{\ch{BCl3}}, \textbf{\ch{BF}}, \textbf{\ch{BFO}}, \textbf{\ch{BF2}}, \textbf{\ch{BF2-}}, \textbf{\ch{BF2+}}, \textbf{\ch{BF2H}}, \textbf{\ch{BF2HO}}, \textbf{\ch{BF2O}}, \textbf{\ch{BF3}}, \textbf{\ch{BF4K}}, \textbf{\ch{BH}}$^i$, \textbf{\ch{BHO}}, \textbf{\ch{BHO-}}, \textbf{\ch{BHO+}}, \textbf{\ch{BHO2}}, \textbf{\ch{BHS}}, \textbf{\ch{BHS+}}, \textbf{\ch{BH2}}$^d$, \textbf{\ch{BH2O2}}, \textbf{\ch{BH3}}, \textbf{\ch{BH3O3}}, \textbf{\ch{BKO2}}, \textbf{\ch{BN}}$^d$, \textbf{\ch{BNaO2}}, \textbf{\ch{BO}}, \textbf{\ch{BO2}}$^k$, \textbf{\ch{BO2-}}$^k$, \textbf{\ch{BS}}$^d$, \textbf{\ch{B2}}, \textbf{\ch{B2Cl4}}, \textbf{\ch{B2F4}}, \textbf{\ch{B2F4O}}, \textbf{\ch{B2H4O4}}, \textbf{\ch{B2H6}}, \textbf{\ch{B2O}}$^d$, \textbf{\ch{B2O2}}, \textbf{\ch{B2O3}}, \textbf{\ch{B3Cl3O3}}, \textbf{\ch{B3FH2O3}}, \textbf{\ch{B3F2HO3}}, \textbf{\ch{B3F3O3}}, \textbf{\ch{B3H3O3}}, \textbf{\ch{B3H3O6}}, \textbf{\ch{B3H6N3}}, \textbf{\ch{B5H9}}, \textbf{\ch{B10H4}}, \textbf{\ch{CB}} \\
\textbf{\ch{Zr}} & \textbf{\ch{C2Zr}}$^a$, \textbf{\ch{C4Zr}}$^a$, \textbf{\ch{ClZr}}, \textbf{\ch{Cl2Zr}}, \textbf{\ch{Cl3Zr}}, \textbf{\ch{Cl4Zr}}, \textbf{\ch{FZr}}, \textbf{\ch{F2Zr}}, \textbf{\ch{F3Zr}}, \textbf{\ch{F4Zr}}, \textbf{\ch{HZr}}, \textbf{\ch{NZr}}$^i$, \textbf{\ch{OZr}}$^i$, \textbf{\ch{O2Zr}}, \textbf{\ch{SZr}}$^a$, \textbf{\ch{Zr-}}$^j$, \textbf{\ch{Zr+}}$^j$ \\
\textbf{\ch{Br}} & \textbf{\ch{AlBr}}, \textbf{\ch{AlBr3}}, \textbf{\ch{Al2Br6}}, \textbf{\ch{BBr}}, \textbf{\ch{BBrCl}}, \textbf{\ch{BBrCl2}}, \textbf{\ch{BBrF}}, \textbf{\ch{BBrF2}}, \textbf{\ch{BBrO}}, \textbf{\ch{BBr2}}$^k$, \textbf{\ch{BBr2Cl}}, \textbf{\ch{BBr2F}}, \textbf{\ch{BBr2H}}, \textbf{\ch{BBr3}}, \textbf{\ch{Br-}}$^j$, \textbf{\ch{Br+}}$^j$, \textbf{\ch{BrCa}}, \textbf{\ch{BrCl}}, \textbf{\ch{BrF}}, \textbf{\ch{BrF3}}, \textbf{\ch{BrF5}}, \textbf{\ch{BrF5S}}, \textbf{\ch{BrH}}, \textbf{\ch{BrH3Si}}, \textbf{\ch{BrK}}, \textbf{\ch{BrMg}}, \textbf{\ch{BrN}}, \textbf{\ch{BrNO}}, \textbf{\ch{BrNa}}, \textbf{\ch{BrO}}, \textbf{\ch{OBrO}}, \textbf{\ch{BrOO}} \textbf{\ch{BrO3}}, \textbf{\ch{BrP}}, \textbf{\ch{BrSi}}, \textbf{\ch{BrSr}}, \textbf{\ch{BrTi}}, \textbf{\ch{BrZr}}, \textbf{\ch{Br2}}, \textbf{\ch{Br2Ca}}, \textbf{\ch{Br2Fe}}, \textbf{\ch{Br2H2Si}}, \textbf{\ch{Br2K2}}, \textbf{\ch{Br2Mg}}, \textbf{\ch{Br2Mg+}}, \textbf{\ch{Br2Na2}}, \textbf{\ch{BrOBr}}, \textbf{\ch{BrBrO}} \textbf{\ch{Br2Si}}, \textbf{\ch{Br2Sr}}, \textbf{\ch{Br2Ti}}, \textbf{\ch{Br2Zr}}, \textbf{\ch{Br3HSi}}, \textbf{\ch{Br3OP}}, \textbf{\ch{Br3P}}, \textbf{\ch{Br3PS}}, \textbf{\ch{Br3Si}}, \textbf{\ch{Br3Ti}}, \textbf{\ch{Br3Zr}}, \textbf{\ch{Br4Fe2}}, \textbf{\ch{Br4Mg2}}, \textbf{\ch{Br4Si}}, \textbf{\ch{Br4Ti}}, \textbf{\ch{Br4Zr}}, \textbf{\ch{CBr}}, \textbf{\ch{CBrF3}}, \textbf{\ch{CBrN}}, \textbf{\ch{CBr4}} \\
\textbf{\ch{Rb}} & \textbf{\ch{BrRb}}$^a$, \textbf{\ch{ClRb}}$^i$, \textbf{\ch{FRb}}$^i$, \textbf{\ch{HRb}}$^i$, \textbf{\ch{ORb}}$^i$, \textbf{\ch{Rb-}}$^j$, \textbf{\ch{Rb+}}$^j$, \textbf{\ch{Rb2}} \\
\textbf{\ch{As}} & \textbf{\ch{As-}}$^j$, \textbf{\ch{As+}}$^j$, \textbf{\ch{AsCl}}$^i$, \textbf{\ch{AsF}}$^i$, \textbf{\ch{AsH}}$^i$, \textbf{\ch{AsN}}$^i$, \textbf{\ch{AsO}}$^i$, \textbf{\ch{AsP}}$^i$, \textbf{\ch{AsS}}$^i$, \textbf{\ch{As2}}$^i$ \\
\textbf{\ch{Ba}} & \textbf{\ch{Ba-}}$^j$, \textbf{\ch{Ba+}}$^j$, \textbf{\ch{BaBr}}, \textbf{\ch{BaBr2}}, \textbf{\ch{BaCl}}, \textbf{\ch{BaCl2}}, \textbf{\ch{BaF}}, \textbf{\ch{BaF+}}, \textbf{\ch{BaF2}}, \textbf{\ch{BaH}}$^i$, \textbf{\ch{BaHO}}, \textbf{\ch{BaHO+}}, \textbf{\ch{BaH2O2}}, \textbf{\ch{BaO}}, \textbf{\ch{BaS}} \\
\textbf{\ch{Xe}} & \textbf{\ch{Xe+}}$^j$ \\
\textbf{\ch{Y}}  & \textbf{\ch{C2Y}}$^a$, \textbf{\ch{C4Y}}$^a$, \textbf{\ch{ClY}}$^i$, \textbf{\ch{FY}}$^i$, \textbf{\ch{OY}}$^i$, \textbf{\ch{O2Y}}$^a$, \textbf{\ch{SY}}$^i$, \textbf{\ch{Y-}}$^j$, \textbf{\ch{Y+}}$^j$ \\
\textbf{\ch{Te}} & \textbf{\ch{GeTe}}$^i$, \textbf{\ch{HTe}}$^i$, \textbf{\ch{OTe}}$^i$, \textbf{\ch{STe}}$^i$, \textbf{\ch{SiTe}}$^i$, \textbf{\ch{Te-}}$^j$, \textbf{\ch{Te+}}$^j$, \textbf{\ch{Te2}}$^i$ \\
\textbf{\ch{Sn}} & \textbf{\ch{ClSn}}, \textbf{\ch{FSn}}, \textbf{\ch{HSn}}, \textbf{\ch{OSn}}, \textbf{\ch{SSn}}, \textbf{\ch{Sn-}}$^j$, \textbf{\ch{Sn+}}$^j$ \\
\textbf{\ch{Pb}} & \textbf{\ch{BrPb}}, \textbf{\ch{Br2Pb}}, \textbf{\ch{Br4Pb}}$^k$, \textbf{\ch{ClPb}}$^k$, \textbf{\ch{ClPb+}}, \textbf{\ch{Cl2Pb}}, \textbf{\ch{Cl2Pb+}}, \textbf{\ch{Cl4Pb}}$^k$, \textbf{\ch{FPb}}, \textbf{\ch{F2Pb}}, \textbf{\ch{F4Pb}}$^k$, \textbf{\ch{HPb}}, \textbf{\ch{OPb}}, \textbf{\ch{Pb-}}$^j$, \textbf{\ch{Pb+}}$^j$, \textbf{\ch{PbS}}, \textbf{\ch{Pb2}} \\
\textbf{\ch{Li}} & \textbf{\ch{AlF4Li}}, \textbf{\ch{BLiO2}}, \textbf{\ch{BrLi}}, \textbf{\ch{Br2Li2}}, \textbf{\ch{ClFLi2}}, \textbf{\ch{ClLi}}, \textbf{\ch{ClLiO}}, \textbf{\ch{Cl2Li2}}, \textbf{\ch{Cl3Li3}}, \textbf{\ch{FLi}}, \textbf{\ch{FLiO}}, \textbf{\ch{F2Li-}}, \textbf{\ch{F2Li2}}, \textbf{\ch{F3Li3}}, \textbf{\ch{HLi}}, \textbf{\ch{HLiO}}, \textbf{\ch{HLiO+}}, \textbf{\ch{H2Li2O2}}, \textbf{\ch{Li-}}$^j$, \textbf{\ch{Li+}}$^j$, \textbf{\ch{LiN}}, \textbf{\ch{LiNO}}, \textbf{\ch{LiNa}}$^i$, \textbf{\ch{LiNaO}}, \textbf{\ch{LiO}}$^i$, \textbf{\ch{LiO-}}, \textbf{\ch{Li2}}, \textbf{\ch{Li2O}}, \textbf{\ch{Li2O2}}, \textbf{\ch{Li2O4S}} \\
\hline
\end{tabular}
\tablefoot{Newly added species are highlighted in bold fond. Species with revised thermochemical data since the release of \texttt{FastChem Cond} are underlined. For convenience Hill-notation is used for species names except in case of isomers. Unless stated otherwise, thermochemical data from \citet{Chase1998} were used to fit $\ln\bar{K}_i$.}
\tablebib{
  (a)~\citet{Tsuji1973A&A....23..411T};
  (b)~\citet{Lodders1999JPCRD..28.1705L};
  (c)~\citet{Dorofeeva2001JPCRD..30..475D};
  (d)~\citet{McBride2002};
  (e)~\citet{Dorofeeva2003JPCRD..32..879D};
  (f)~\citet{Lodders2004JPCRD..33..357L};
  (g)~\citet{Shenyavskaya2004};
  (h)~\citet{Burrows2005apj};
  (i)~\citet{Barklem2016AA...588A..96B};
  (j)~This work;
  (k)~\citet{Goo23}.
}
\end{table}

\begin{table}[t]
\caption{List of all condensed-phase species included in the present version of \fc.}
\label{table:cond_new}
\centering
\scriptsize
\begin{tabular}{lp{0.9\textwidth}}
\hline\hline
Element & Condensates \\
\hline
    O & \ch{H2O(s,l)}$^e$\\
    C & \ch{C(s)}, \ch{CH4(s,l)}$^a$, \ch{CO2(s,l)}$^b$, \ch{CO(l)}$^c$\\
    N & \ch{N2H4(l)}, \ch{N2O4(s,l)}, \ch{N2(s,l)}$^f$, \ch{NH3(s,l)}$^g$\\
    Mg & \ch{MgCO3(s)}, \ch{MgC2(s)}, \ch{Mg2C3(s)}, \ch{MgH2(s)}, \ch{Mg(OH)2(s)}, \ch{Mg(s,l)}, \ch{MgO(s,l)}, \ch{Mg3N2(s)}\\
    Si & \ch{SiC(s)}, \ch{MgSiO3(s,l)}, \ch{Mg2SiO4(s,l)}, \ch{Mg2Si(s,l)}, \ch{Si3N4(s)}, \ch{SiO2(s,l)}$^i$, \ch{Si(s,l)}, \ch{SiO(s)}$^h$\\
    Fe & \ch{Fe(CO)5(l)}, \ch{Fe(s,l)}, \ch{Fe(OH)2(s)}, \ch{Fe(OH)3(s)}, \ch{FeO(s,l)}, \ch{Fe2O3(s)}, \ch{Fe3O4(s)}, \ch{Fe2SiO4(s)}$^d$\\
    S & \ch{FeSO4(s)}, \ch{FeS(s,l)}, \ch{FeS2(s)}, \ch{Fe2(SO4)3(s)}, \ch{O2S(OH)2(s,l)}, \textbf{\ch{H2S(s,l)}}$^j$, \ch{H2SO4.H2O(s,l)}, \ch{H2SO4.2H2O(s,l)}, \ch{H2SO4.3H2O(s,l)}, \ch{H2SO4.4H2O(s,l)}, \ch{MgSO4(s,l)}, \ch{MgS(s)}, \textbf{\ch{NH4SH(s)}}$^k$, \ch{S(s,l)}, \ch{SiS2(s,l)}\\
    Al & \ch{Al(s)}, \ch{AlN(s)}, \ch{MgAl2O4(s,l)}, \ch{Al2O3(s,l)}, \ch{Al2SiO5(s)}, \ch{Al2S3(s)}, \ch{Al6Si2O13(s)}, \ch{Al4C3(s)}\\
    Ca & \ch{Ca(s,l)}, \ch{Ca(OH)2(s)}, \ch{CaO(s,l)}, \ch{CaS(s)}, \ch{CaSiO3(s)}$^d$, \ch{Ca2SiO4(s)}$^d$, \ch{CaMgSi2O6(s)}$^d$, \ch{Ca2Al2SiO7(s)}$^d$, \ch{CaAl2Si2O8(s)}$^d$\\
    Na & \ch{NaAlO2(s)}, \ch{NaCN(s,l)}, \ch{NaCO3(s,l)}, \ch{NaH(s)}, \ch{NaOH(s,l)}, \ch{Na(s,l)}, \ch{NaO2(s)}, \ch{Na2O(s,l)}, \ch{Na2O2(s)}, \ch{Na2SiO3(s,l)}, \ch{Na2SO4(s,l)}, \ch{Na2Si2O5(s,l)}, \ch{Na2S(s,l)}, \ch{Na2S2(s,l)}, \ch{NaAlSi3O8(s)}$^d$\\
    Ni & \ch{Ni(CO)4(l)}, \ch{Ni(s,l)}, \ch{NiS(s,l)}, \ch{NiS2(s,l)}, \ch{Ni3S2(s,l)}, \ch{Ni3S4(s)}\\
    Cr & \ch{Cr3C2(s)}, \ch{Cr7C3(s)}, \ch{Cr23C6(s)}, \ch{Cr(s,l)}, \ch{CrN(s)}, \ch{Cr2N(s)}, \ch{Cr2O3(s,l)}\\
    Cl & \ch{AlClO(s)}, \ch{AlCl3(s,l)}, \ch{NaAlCl4(s)}, \ch{Na3AlCl6(s)}, \ch{CaCl2(s,l)}, \ch{NH4Cl(s)}, \ch{NH4ClO4(s)}, \ch{NaCl(s,l)}, \ch{NaClO4(s)}, \ch{FeCl2(s,l)}, \ch{MgCl2(s,l)}, \ch{NiCl2(s,l)}, \ch{SCl2(l)}, \ch{ClSSCl(l)}, \ch{FeCl3(s,l)}\\
    Mn & \ch{Mn(s,l)}, \ch{MnS(s)}$^d$, \ch{MnSiO3(s)}$^d$\\
    P & \ch{H3PO4(s,l)}, \ch{Mg3P2O8(s,l)}, \textbf{\ch{NH4H2PO4(s)}}$^m$, \textbf{\ch{PH3(s,l)}}$^l$, \ch{P3N5(s)}, \ch{(P2O5)2(s)}, \ch{P(s,l)}, \ch{P4S3(s,l)}\\
    K & \ch{KAlCl4(s)}, \ch{K3AlCl6(s)}, \ch{K3Al2Cl9(s)}, \ch{KCN(s,l)}, \ch{K2CO3(s,l)}, \ch{KCl(s,l)}, \ch{KClO4(s)}, \ch{KH(s)}, \ch{KOH(s,l)}, \ch{K(s,l)}, \ch{KO2(s)}, \ch{K2O(s)}, \ch{K2O2(s)}, \ch{K2SiO3(s,l)}, \ch{K2SO4(s,l)}, \ch{K2S(s,l)}, \ch{KAlSi3O8(s)}$^d$\\
    Co & \ch{CoCl2(s,l)}, \ch{Co(s,l)}, \ch{CoO(s)}, \ch{CoSO4(s)}, \ch{Co3O4(s)}\\
    Ti & \ch{TiC(s,l)}, \ch{TiCl2(s)}, \ch{TiCl3(s)}, \ch{TiCl4(s,l)}, \ch{TiH2(s)}, \ch{MgTiO3(s,l)}, \ch{MgTi2O5(s,l)}, \ch{Mg2TiO4(s,l)}, \ch{TiN(s,l)}, \ch{TiO(s,l)}, \ch{TiO2(s,l)}, \ch{Ti2O3(s,l)}, \ch{Ti3O5(s,l)}, \ch{Ti4O7(s,l)}, \ch{Ti(s,l)}, \ch{CaTiO3(s)}$^d$\\
    F & \ch{AlF3(s,l)}, \ch{K3AlF6(s)}, \ch{Na3AlF6(s,l)}, \ch{Na5Al3F14(s,l)}, \ch{CaF2(s,l)}, \ch{CoF2(s,l)}, \ch{CoF3(s)}, \ch{KF(s,l)}, \ch{NaF(s,l)}, \ch{FeF2(s,l)}, \ch{K(HF2)(s,l)}, \ch{MgF2(s,l)}, \ch{FeF3(s)}, \ch{TiF3(s)}, \ch{TiF4(s)}, \textbf{\ch{ZnF2(s,l)}}$^i$\\
    Zn & \textbf{\ch{CaZn(s)}}$^i$, \textbf{\ch{CaZn2(s)}}$^i$, \textbf{\ch{Fe2ZnO4(s)}}$^i$, \textbf{\ch{Zn2SiO4(s,l)}}$^i$, \textbf{\ch{Zn2TiO4(s)}}$^i$, \textbf{\ch{Zn3(PO4)2(s)}}$^i$, \textbf{\ch{Zn3N2(s)}}$^i$, \textbf{\ch{Zn3P2(s,l)}}$^i$, \textbf{\ch{ZnCO3(s)}}$^i$, \textbf{\ch{ZnCl2(s,l)}}$^i$, \textbf{\ch{ZnO(s,l)}}$^i$, \textbf{\ch{ZnP2(s,l)}}$^i$, \textbf{\ch{ZnS(s,l)}}$^i$, \textbf{\ch{ZnSiO3(s)}}$^i$, \ch{ZnSO4(s)}, \ch{Zn(s,l)}$^i$\\
    Cu & \ch{CuCN(s)}, \ch{CuCl(s,l)}, \ch{CuCl2(s)}, \ch{Cu(s,l)}, \ch{CuF(s)}, \ch{CuF2(s,l)}, \ch{Cu(OH)2(s)}, \ch{CuO(s)}, \ch{CuSO4(s)}, \ch{Cu2O(s,l)}\\
    V & \ch{VN(s)}, \ch{VO(s,l)}, \ch{V2O3(s,l)}, \ch{V2O4(s,l)}, \ch{V2O5(s,l)}, \ch{V(s,l)}, \textbf{\ch{Ca(VO3)2(s)}}$^i$, \textbf{\ch{Ca2V2O7(s)}}$^i$, \textbf{\ch{Ca3(VO4)2(s)}}$^i$, \textbf{\ch{Fe(VO3)2(s)}}$^i$, \textbf{\ch{FeV2O4(s)}}$^i$, \textbf{\ch{V2C(s)}}$^i$\\
    Ge & \textbf{\ch{Ge(s,l)}}$^i$, \textbf{\ch{GeMg2(s)}}$^i$, \textbf{\ch{GeNi2(s)}}$^i$, \textbf{\ch{GeO2(s,l)}}$^i$, \textbf{\ch{GeP(s)}}$^i$, \textbf{\ch{GeS(s,l)}}$^i$, \textbf{\ch{GeS2(s,l)}}$^i$\\
    \textbf{Se} & \textbf{\ch{Al2Se3(s)}}$^i$, \textbf{\ch{CaSe(s)}}$^i$, \textbf{\ch{CoSeO3(s,l)}}$^i$, \textbf{\ch{Cu2Se(s)}}$^i$, \textbf{\ch{CuSe(s)}}$^i$, \textbf{\ch{CuSeO3(s)}}$^i$, \textbf{\ch{GeSe(s,l)}}$^i$, \textbf{\ch{GeSe2(s)}}$^i$, \textbf{\ch{MgSe(s)}}$^i$, \textbf{\ch{MgSeO3(s)}}$^i$, \textbf{\ch{MnSe(s)}}$^i$,\textbf{\ch{NiSe2(s)}}$^i$, \textbf{\ch{NiSeO3(s)}}$^i$, \textbf{\ch{Se(s,l)}}$^i$, \textbf{\ch{Se2Cl2(l)}}$^i$, \textbf{\ch{SeCl4(s)}}$^i$, \textbf{\ch{SeO2(s)}}$^i$, \textbf{\ch{ZnSe(s)}}$^i$, \textbf{\ch{ZnSeO3(s,l)}}$^i$\\
    \textbf{Sc} & \textbf{\ch{Sc(s,l)}}$^i$, \textbf{\ch{Sc2O3(s,l)}}$^i$, \textbf{\ch{ScCl3(s,l)}}$^i$, \textbf{\ch{ScF3(s,l)}}$^i$, \textbf{\ch{ScN(s)}}$^i$\\
    \textbf{Ga} & \textbf{\ch{Ga(s,l)}}, \textbf{\ch{Ga2O3(s,l)}}$^i$, \textbf{\ch{Ga2S3(s)}}$^i$, \textbf{\ch{Ga2Se3(s)}}$^i$, \textbf{\ch{GaCl3(s,l)}}$^i$, \textbf{\ch{GaF3(s)}}$^i$, \textbf{\ch{GaN(s)}}$^i$, \textbf{\ch{GaP(s)}}$^i$, \textbf{\ch{GaS(s)}}$^i$, \textbf{\ch{GaSe(s)}}$^i$\\
    \textbf{Sr} & \textbf{\ch{Sr(OH)2(s,l)}}, \textbf{\ch{Sr(s,l)}}, \textbf{\ch{Sr2SiO4(s)}}$^i$, \textbf{\ch{Sr2TiO4(s)}}$^i$, \textbf{\ch{Sr3N2(s)}}$^i$, \textbf{\ch{SrAl2O4(s)}}$^i$, \textbf{\ch{SrC2(s)}}$^i$, \textbf{\ch{SrCO3(s)}}$^i$, \textbf{\ch{SrCl2(s,l)}}, \textbf{\ch{SrF2(s,l)}}, \textbf{\ch{SrH2(s)}}$^i$, \textbf{\ch{SrO(s,l)}}, \textbf{\ch{SrS(s)}}, \textbf{\ch{SrSO4(s,l)}}$^i$, \textbf{\ch{SrSiO3(s)}}$^i$, \textbf{\ch{SrTiO3(s)}}$^i$\\
    \textbf{B} & \textbf{\ch{(B(OH)2)2(s)}}, \textbf{\ch{B(s,l)}}, \textbf{\ch{B2O3(s,l)}}, \textbf{\ch{B3H3O3(s)}}, \textbf{\ch{B3O3F3(s)}}, \textbf{\ch{B4C(s,l)}}, \textbf{\ch{B5H9(l)}}, \textbf{\ch{BN(s)}}, \textbf{\ch{Ca2B2O5(s,l)}}$^i$, \textbf{\ch{Ca3B2O6(s,l)}}$^i$, \textbf{\ch{CaB2O4(s,l)}}$^i$, \textbf{\ch{CaB4O7(s,l)}}$^i$, \textbf{\ch{Co2B(s)}}$^i$, \textbf{\ch{CoB(s)}}$^i$, \textbf{\ch{CrB(s)}}$^i$, \textbf{\ch{CrB2(s)}}$^i$, \textbf{\ch{Fe2B(s)}}$^i$, \textbf{\ch{FeB(s)}}$^i$, \textbf{\ch{H3BO3(s)}}, \textbf{\ch{HBO2(s)}}, \textbf{\ch{K2B4O7(s,l)}}, \textbf{\ch{KBF4(s,l)}}, \textbf{\ch{KBH4(s)}}, \textbf{\ch{KBO2(s,l)}}, \textbf{\ch{MgB2(s)}}, \textbf{\ch{MgB4(s)}}, \textbf{\ch{MnB(s)}}$^i$, \textbf{\ch{MnB2(s)}}$^i$, \textbf{\ch{Na2B4O7(s,l)}}, \textbf{\ch{NaBH4(s)}}, \textbf{\ch{NaBO2(s,l)}}, \textbf{\ch{Ni4B3(s)}}$^i$, \textbf{\ch{NiB(s)}}$^i$, \textbf{\ch{TiB(s)}}, \textbf{\ch{TiB2(s,l)}}\\
    \textbf{Zr} & \textbf{\ch{CaZrO3(s)}}$^i$, \textbf{\ch{SrZrO3(s)}}$^i$, \textbf{\ch{Zr(s,l)}}, \textbf{\ch{ZrB2(s,l)}}, \textbf{\ch{ZrC(s,l)}}, \textbf{\ch{ZrCl2(s,l)}}, \textbf{\ch{ZrCl3(s)}}, \textbf{\ch{ZrCl4(s)}}, \textbf{\ch{ZrF2(s,l)}}, \textbf{\ch{ZrF3(s)}}, \textbf{\ch{ZrF4(s)}}, \textbf{\ch{ZrN(s,l)}}, \textbf{\ch{ZrO2(s,l)}}, \textbf{\ch{ZrS2(s)}}$^i$, \textbf{\ch{ZrSiO4(s)}}\\
    \textbf{Br} & \textbf{\ch{AlBr3(s,l)}}, \textbf{\ch{BBr3(l)}}, \textbf{\ch{Br2(s,l)}}, \textbf{\ch{CaBr2(s,l)}}, \textbf{\ch{CoBr2(s)}}$^i$, \textbf{\ch{CrBr2(s)}}, \textbf{\ch{CrBr3(s)}}$^i$, \textbf{\ch{CuBr(s,l)}}$^i$, \textbf{\ch{CuBr2(s)}}$^i$, \textbf{\ch{FeBr2(s,l)}}, \textbf{\ch{GaBr3(s,l)}}$^i$, \textbf{\ch{KBr(s,l)}}, \textbf{\ch{MgBr2(s,l)}}, \textbf{\ch{MnBr2(s,l)}}$^i$, \textbf{\ch{NH4Br(s)}}, \textbf{\ch{NaBr(s,l)}}, \textbf{\ch{NiBr2(s)}}$^i$, \textbf{\ch{ScBr3(s)}}$^i$, \textbf{\ch{SiBr4(l)}}, \textbf{\ch{SrBr2(s,l)}}, \textbf{\ch{TiBr2(s)}}, \textbf{\ch{TiBr3(s)}}, \textbf{\ch{TiBr4(s,l)}}, \textbf{\ch{ZnBr2(s,l)}}$^i$, \textbf{\ch{ZrBr2(s,l)}}, \textbf{\ch{ZrBr3(s)}}, \textbf{\ch{ZrBr4(s)}}\\
    \textbf{Rb} & \textbf{\ch{Rb(s,l)}}, \textbf{\ch{Rb2CO3(s,l)}}$^i$, \textbf{\ch{Rb2O(s,l)}}$^i$, \textbf{\ch{Rb2SO4(s,l)}}$^i$, \textbf{\ch{Rb2Si2O5(s,l)}}$^i$, \textbf{\ch{Rb2Si4O9(s,l)}}$^i$, \textbf{\ch{Rb2SiO3(s,l)}}$^i$, \textbf{\ch{RbBr(s,l)}}$^i$, \textbf{\ch{RbCl(s,l)}}$^i$, \textbf{\ch{RbF(s,l)}}$^i$, \textbf{\ch{RbO2(s,l)}}$^i$\\
    \textbf{As} & \textbf{\ch{AlAs(s)}}$^i$, \textbf{\ch{AlAsO4(s)}}$^i$, \textbf{\ch{As2O3(s,l)}}$^i$, \textbf{\ch{As2O5(s)}}$^i$, \textbf{\ch{As2S2(s,l)}}$^i$, \textbf{\ch{As2S3(s,l)}}$^i$, \textbf{\ch{As2Se3(s,l)}}$^i$, \textbf{\ch{As4S4(s,l)}}$^i$, \textbf{\ch{As(s)}}$^i$, \textbf{\ch{Ca3(AsO4)2(s)}}$^i$, \textbf{\ch{Co3(AsO4)2(s)}}$^i$, \textbf{\ch{Cr3(AsO4)2(s)}}$^i$, \textbf{\ch{CrAsO4(s)}}$^i$, \textbf{\ch{Cu3(AsO4)2(s)}}$^i$, \textbf{\ch{Cu3As(s)}}$^i$, \textbf{\ch{Cu3AsO4(s)}}$^i$, \textbf{\ch{Fe3(AsO4)2(s)}}$^i$, \textbf{\ch{FeAsO4(s)}}$^i$, \textbf{\ch{GaAs(s,l)}}$^i$,  \textbf{\ch{GaAsO4(s)}}$^i$, \textbf{\ch{K3AsO4(s)}}$^i$, \textbf{\ch{Mg3(AsO4)2(s)}}$^i$, \textbf{\ch{Mn3(AsO4)2(s)}}$^i$, \textbf{\ch{MnAs(s)}}$^i$, \textbf{\ch{Na3As(s)}}$^i$, \textbf{\ch{Na3AsO4(s)}}$^i$, \textbf{\ch{Ni3(AsO4)2(s)}}$^i$, \textbf{\ch{Ni5As2(s)}}$^i$, \textbf{\ch{NiAs(s)}}$^i$, \textbf{\ch{Rb3AsO4(s)}}$^i$, \textbf{\ch{ScAsO4(s)}}$^i$, \textbf{\ch{Sr3(AsO4)2(s)}}$^i$, \textbf{\ch{Ti3(AsO4)2(s)}}$^i$, \textbf{\ch{Zn3(AsO4)2(s)}}$^i$, \textbf{\ch{Zn3As2(s,l)}}$^i$\\
    \textbf{Ba} & \textbf{\ch{Ba(OH)2(s,l)}}, \textbf{\ch{Ba(s,l)}}, \textbf{\ch{Ba2SiO4(s)}}$^i$, \textbf{\ch{Ba2TiO4(s)}}$^i$, \textbf{\ch{Ba3(AsO4)2(s)}}$^i$, \textbf{\ch{Ba3Al2O6(s)}}$^i$, \textbf{\ch{BaAl2O4(s)}}$^i$, \textbf{\ch{BaBr2(s,l)}}, \textbf{\ch{BaC2(s)}}$^i$, \textbf{\ch{BaCO3(s)}}$^i$, \textbf{\ch{BaCl2(s,l)}}, \textbf{\ch{BaF2(s,l)}}, \textbf{\ch{BaH2(s,l)}}$^i$, \textbf{\ch{BaO(s,l)}}, \textbf{\ch{BaS(s)}}, \textbf{\ch{BaSO4(s)}}$^i$, \textbf{\ch{BaSiO3(s)}}$^i$, \textbf{\ch{BaTiO3(s)}}$^i$, \textbf{\ch{BaZrO3(s)}}$^i$\\
    \textbf{Y} & \textbf{\ch{Y(s,l)}}$^i$, \textbf{\ch{Y2O3(s)}}$^i$, \textbf{\ch{Y2Zr2O7(s)}}$^i$, \textbf{\ch{YAsO4(s)}}$^i$, \textbf{\ch{YCl3(s,l)}}$^i$, \textbf{\ch{YF3(s,l)}}$^i$, \textbf{\ch{YN(s)}}$^i$\\
    \textbf{Te} & \textbf{\ch{Al2Te3(s)}}$^i$, \textbf{\ch{As2Te3(s,l)}}, \textbf{\ch{BaTe(s)}}$^i$, \textbf{\ch{CaTe(s)}}$^i$, \textbf{\ch{Cu2Te(s)}}$^i$, \textbf{\ch{CuTe(s)}}$^i$, \textbf{\ch{FeTe2(s)}}$^i$, \textbf{\ch{Ga2Te3(s)}}$^i$, \textbf{\ch{GaTe(s)}}$^i$, \textbf{\ch{GeTe(s)}}$^i$, \textbf{\ch{MgTe(s)}}$^i$, \textbf{\ch{MnTe(s)}}$^i$, \textbf{\ch{MnTe2(s)}}$^i$, \textbf{\ch{Na2Te(s,l)}}$^i$, \textbf{\ch{NaTe(s)}}$^i$, \textbf{\ch{NaTe3(s,l)}}$^i$, \textbf{\ch{Te(s,l)}}$^i$, \textbf{\ch{TeBr4(s)}}$^i$, \textbf{\ch{TeCl4(s,l)}}$^i$, \textbf{\ch{TeO2(s,l)}}$^i$, \textbf{\ch{ZnTe(s)}}$^i$\\
    \textbf{Sn} & \textbf{\ch{Ba2Sn(s)}}$^i$, \textbf{\ch{Ca2Sn(s)}}$^i$, \textbf{\ch{CaSn(s)}}$^i$, \textbf{\ch{CoSn(s)}}$^i$, \textbf{\ch{MnSn2(s)}}$^i$, \textbf{\ch{Ni3Sn(s)}}$^i$, \textbf{\ch{Ni3Sn2(s)}}$^i$, \textbf{\ch{Sn(SO4)2(s)}}$^i$, \textbf{\ch{Sn(s,l)}}$^i$, \textbf{\ch{Sn2S3(s)}}$^i$, \textbf{\ch{Sn3(AsO4)2(s)}}$^i$, \textbf{\ch{Sn3S4(s)}}$^i$, \textbf{\ch{SnBr2(s,l)}}$^i$, \textbf{\ch{SnBr4(s,l)}}$^i$, \textbf{\ch{SnCl2(s,l)}}$^i$, \textbf{\ch{SnCl4(l)}}$^i$, \textbf{\ch{SnF2(s,l)}}$^i$, \textbf{\ch{SnO(s)}}$^i$, \textbf{\ch{SnO2(s)}}$^i$, \textbf{\ch{SnS(s,l)}}$^i$, \textbf{\ch{SnS2(s)}}$^i$, \textbf{\ch{SnSO4(s)}}$^i$, \textbf{\ch{SnSe(s)}}$^i$, \textbf{\ch{SnSe2(s)}}$^i$, \textbf{\ch{SnTe(s,l)}}$^i$\\
    \textbf{Pb} & \textbf{\ch{Ca2Pb(s)}}$^i$, \textbf{\ch{CaPb(s)}}$^i$, \textbf{\ch{Mg2Pb(s)}}$^i$, \textbf{\ch{Pb(s,l)}}, \textbf{\ch{Pb2SiO4(s)}}, \textbf{\ch{Pb3(AsO4)2(s)}}$^i$, \textbf{\ch{Pb3O4(s)}}, \textbf{\ch{PbB2O4(s)}}, \textbf{\ch{PbB4O7(s)}}, \textbf{\ch{PbBr2(s,l)}}, \textbf{\ch{PbCO3(s)}}$^i$, \textbf{\ch{PbCl2(s,l)}}, \textbf{\ch{PbF2(s,l)}}, \textbf{\ch{PbO(s,l)}}, \textbf{\ch{PbO2(s)}}, \textbf{\ch{PbS(s,l)}}, \textbf{\ch{PbSO4(s,l)}}$^i$, \textbf{\ch{PbSe(s,l)}}$^i$, \textbf{\ch{PbSeO3(s)}}$^i$, \textbf{\ch{PbSeO4(s)}}$^i$, \textbf{\ch{PbSiO3(s)}}, \textbf{\ch{PbTe(s,l)}}$^i$, \textbf{\ch{PbTiO3(s)}}$^i$\\
    \textbf{Li} & \textbf{\ch{AlLi(s)}}$^i$, \textbf{\ch{Li(s,l)}}, \textbf{\ch{Li2B4O7(s,l)}}, \textbf{\ch{Li2C2(s)}}, \textbf{\ch{Li2CO3(s,l)}}, \textbf{\ch{Li2O(s,l)}}, \textbf{\ch{Li2O2(s)}}, \textbf{\ch{Li2SO4(s,l)}}, \textbf{\ch{Li2Se(s)}}$^i$, \textbf{\ch{Li2Si2O5(s,l)}}, \textbf{\ch{Li2SiO3(s,l)}}, \textbf{\ch{Li2Te(s)}}$^i$, \textbf{\ch{Li2TiO3(s,l)}}, \textbf{\ch{Li3AlF6(s,l)}}, \textbf{\ch{Li3AsO4(s)}}$^i$, \textbf{\ch{Li3N(s)}}, \textbf{\ch{LiAlH4(s)}}, \textbf{\ch{LiAlO2(s,l)}}, \textbf{\ch{LiBH4(s)}}, \textbf{\ch{LiBO2(s,l)}}, \textbf{\ch{LiBr(s,l)}}, \textbf{\ch{LiCl(s,l)}}, \textbf{\ch{LiClO4(s,l)}}, \textbf{\ch{LiF(s,l)}}, \textbf{\ch{LiH(s,l)}}, \textbf{\ch{LiOH(s,l)}}\\
\hline
\end{tabular}
\tablefoot{Newly added species are highlighted in bold fond. Unless stated otherwise, thermochemical data from \citet{Chase1998} were used to fit $\ln\bar{K}_i$.}
\tablebib{
  (a)~\citet{Moses1992Icar...99..318M}, \citet{Prydz1972}, NIST Chemistry Webbook;
  (b)~\citet{yaws1999chemical}, NIST Chemistry Webbook;
  (c)~\citet{Goodwin1985JPCRD..14..849G};
  (d)~\citet{Sharp1990ApJS...72..417S};
  (e)~\citet{Murphy2005}, \citet{Wagner2008};
  (f)~\citet{LandoltBornstein2001:sm_lbs_978-3-540-45367-3_3};
  (g)~\citet{Lide2009crc}, \citet{Haar1978JPCRD...7..635H};
  (h)~\citet{Gail2013AA...555A.119G};
  (i)~\citet{Barin1995};
  (j)~\citet{Stulldoi:10.1021/ie50448a022};
  (k)~\citet{Lewis1969Icar...10..365L};
  (l)~\citet{Stephenson1937JChPh...5..149S};
  (m)~\citet{Rossini1952SelectedValues}, \citet{stephenson1944heat}.
}
\end{table}

\onecolumn

\section{Chemistry plots}
\label{sec:appendix_chemistry}

\begin{figure*}[h!]
  \sidecaption
  \parbox[b]{12cm}{%
    \includegraphics[width=0.49\linewidth]{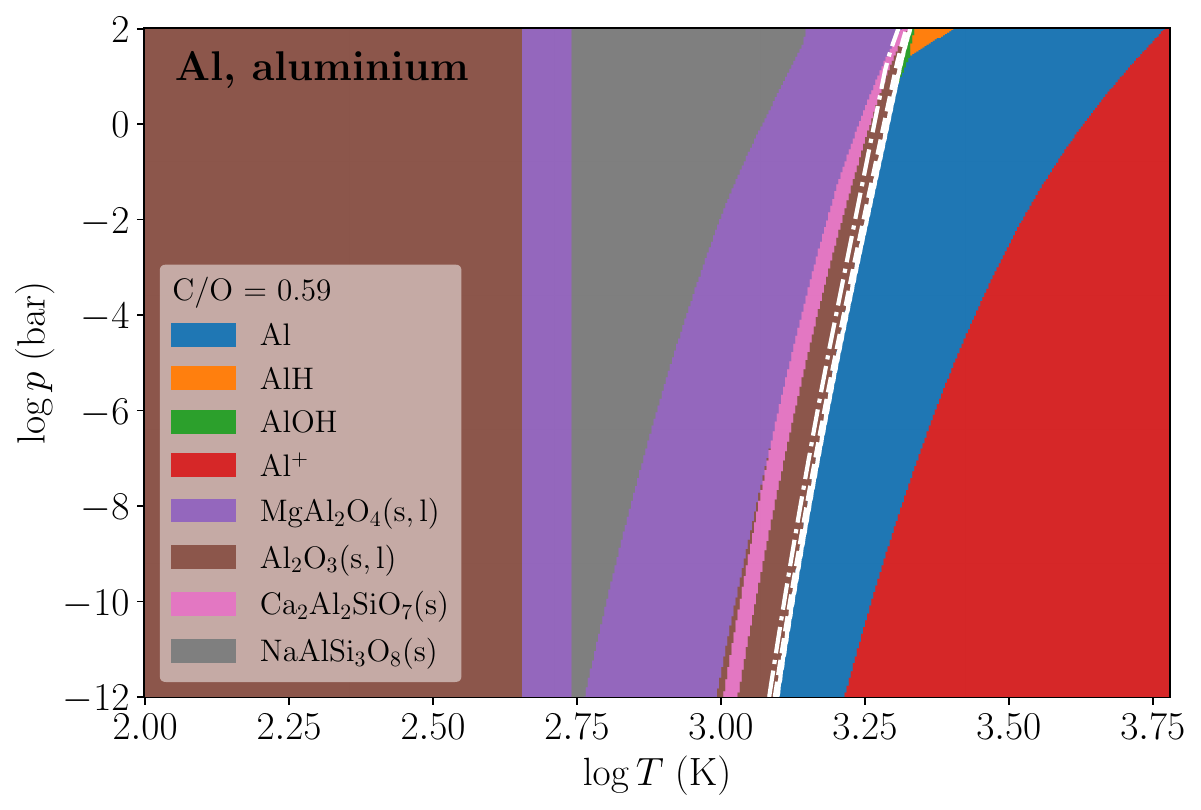}\hfill
    \includegraphics[width=0.49\linewidth]{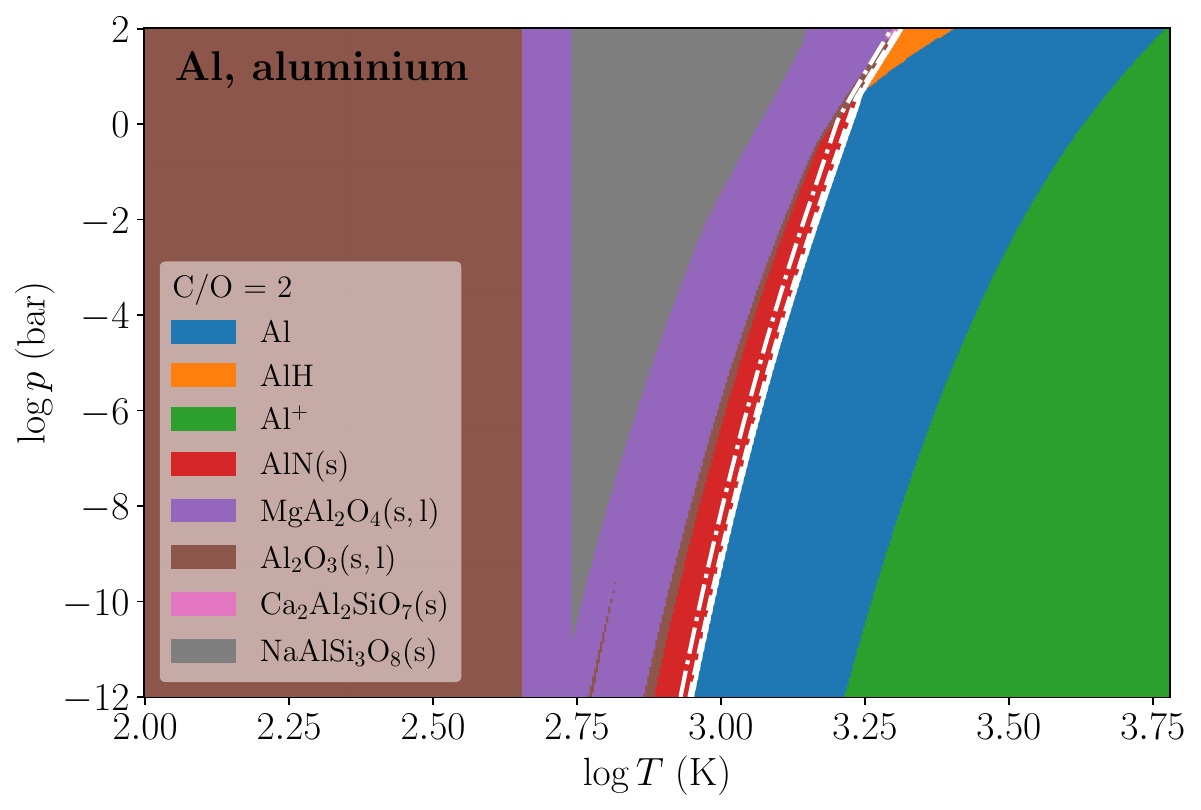}\\
    \includegraphics[width=0.49\linewidth]{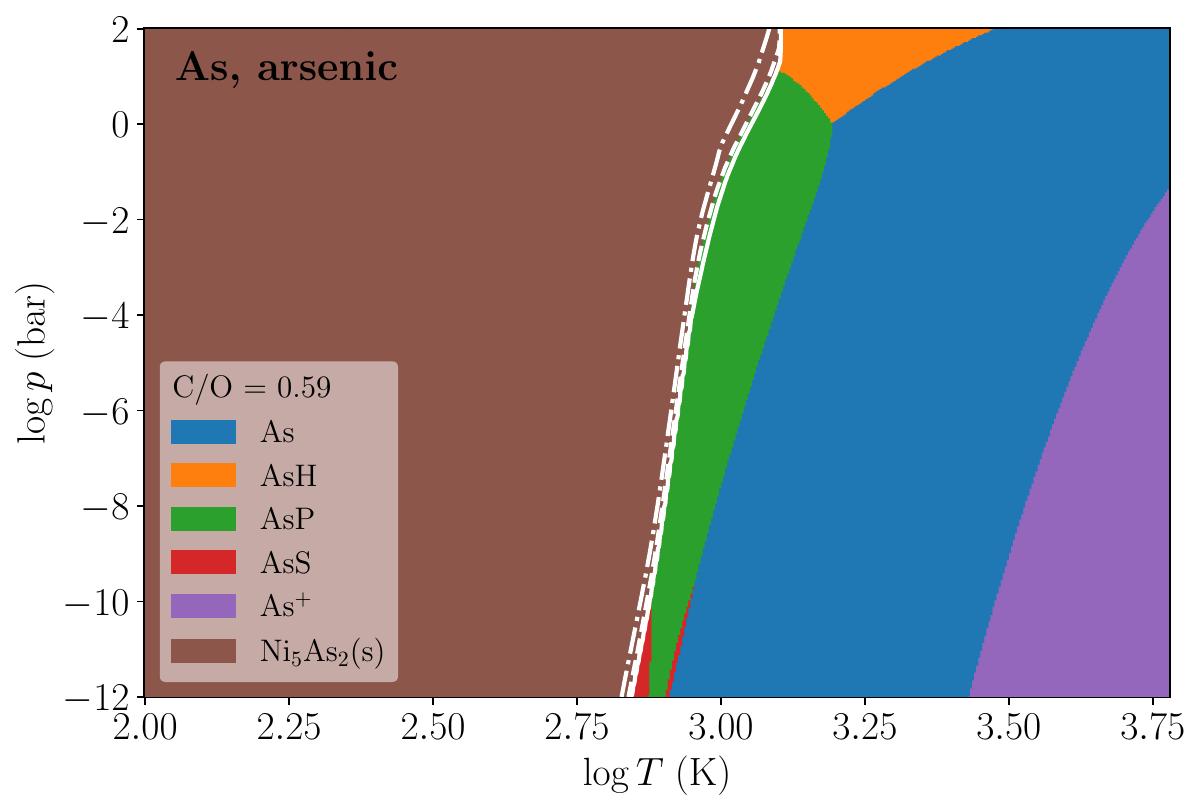}\hfill
    \includegraphics[width=0.49\linewidth]{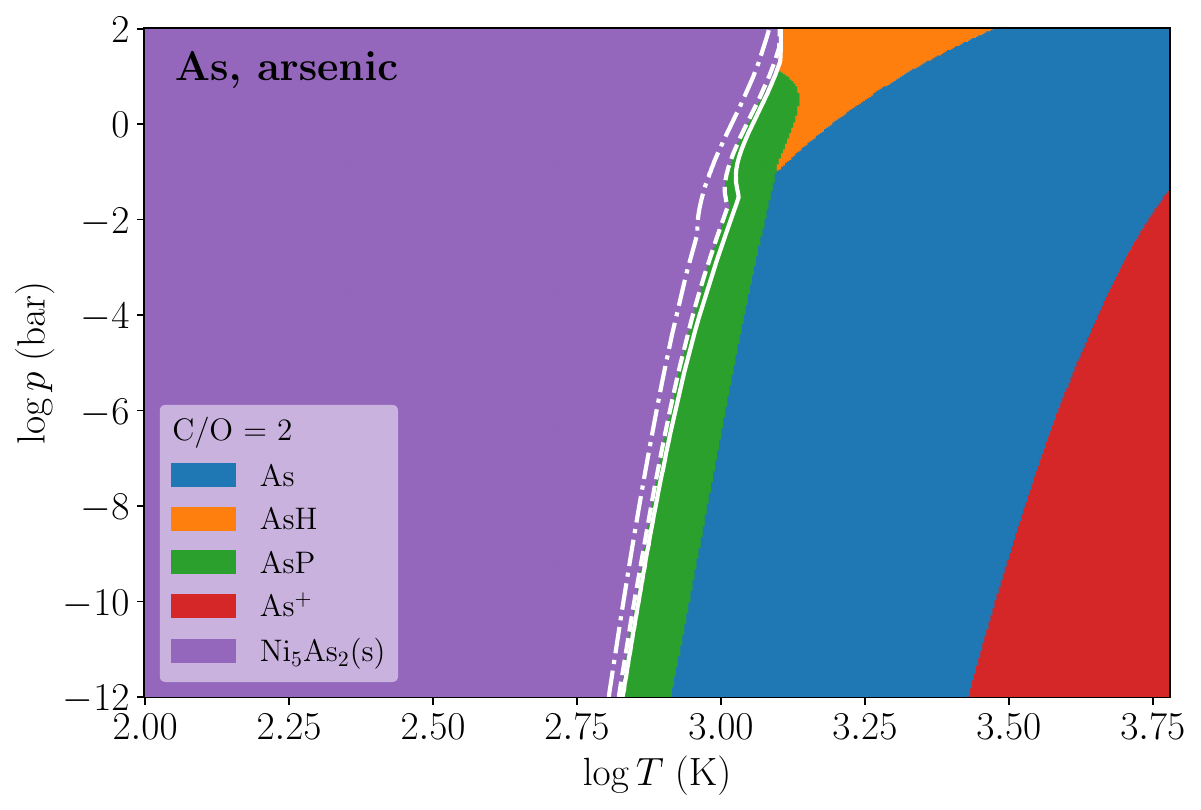}\\
    \includegraphics[width=0.49\linewidth]{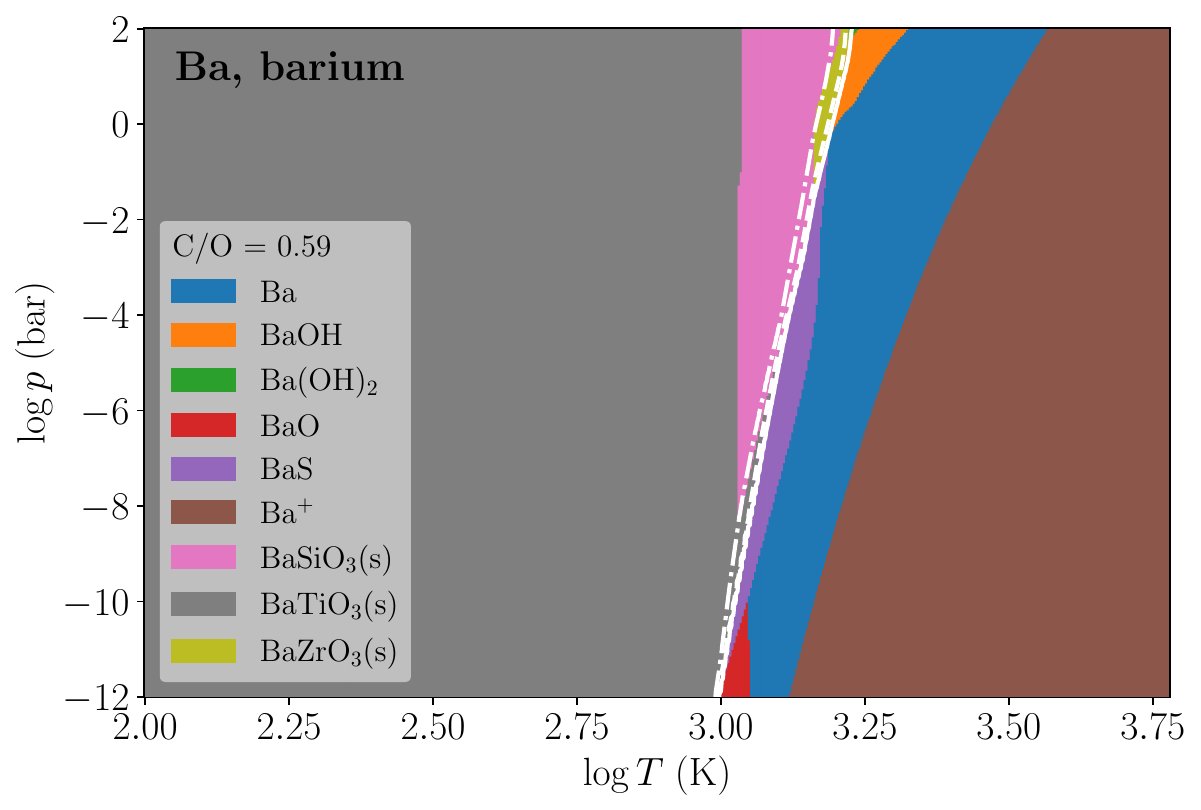}\hfill
    \includegraphics[width=0.49\linewidth]{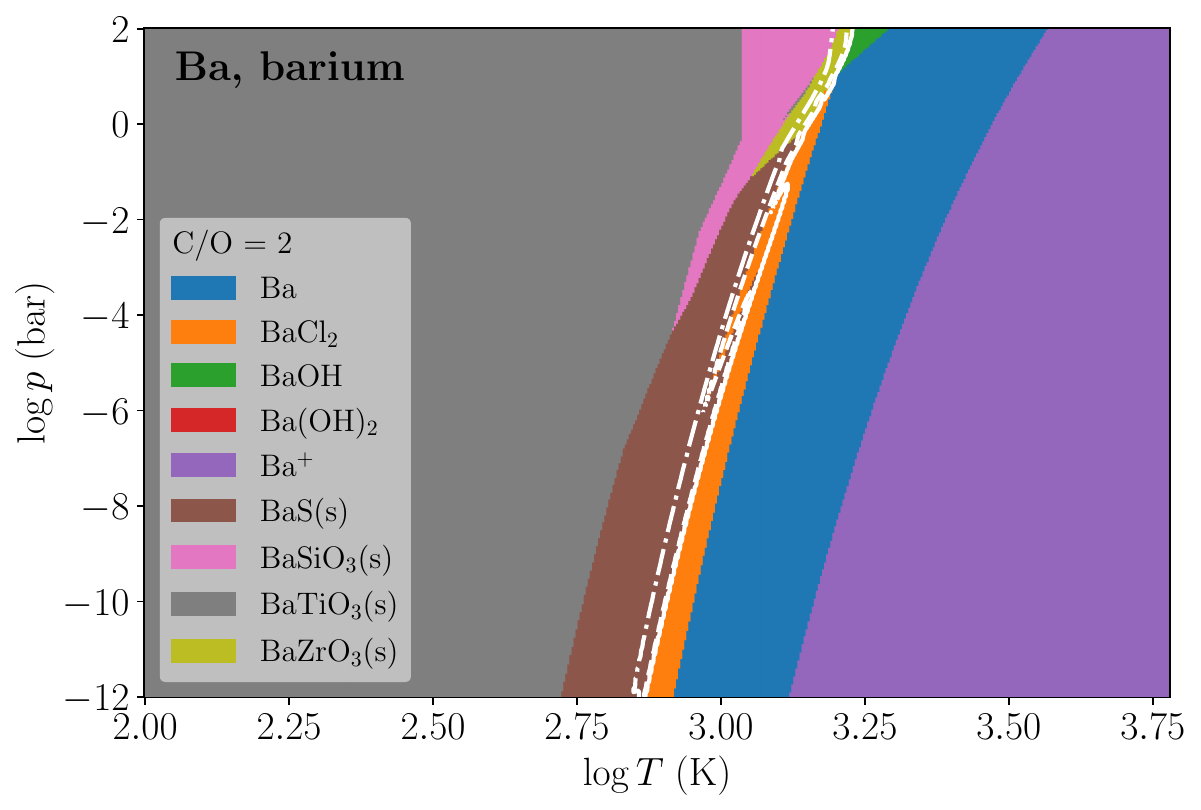}\\
    \includegraphics[width=0.49\linewidth]{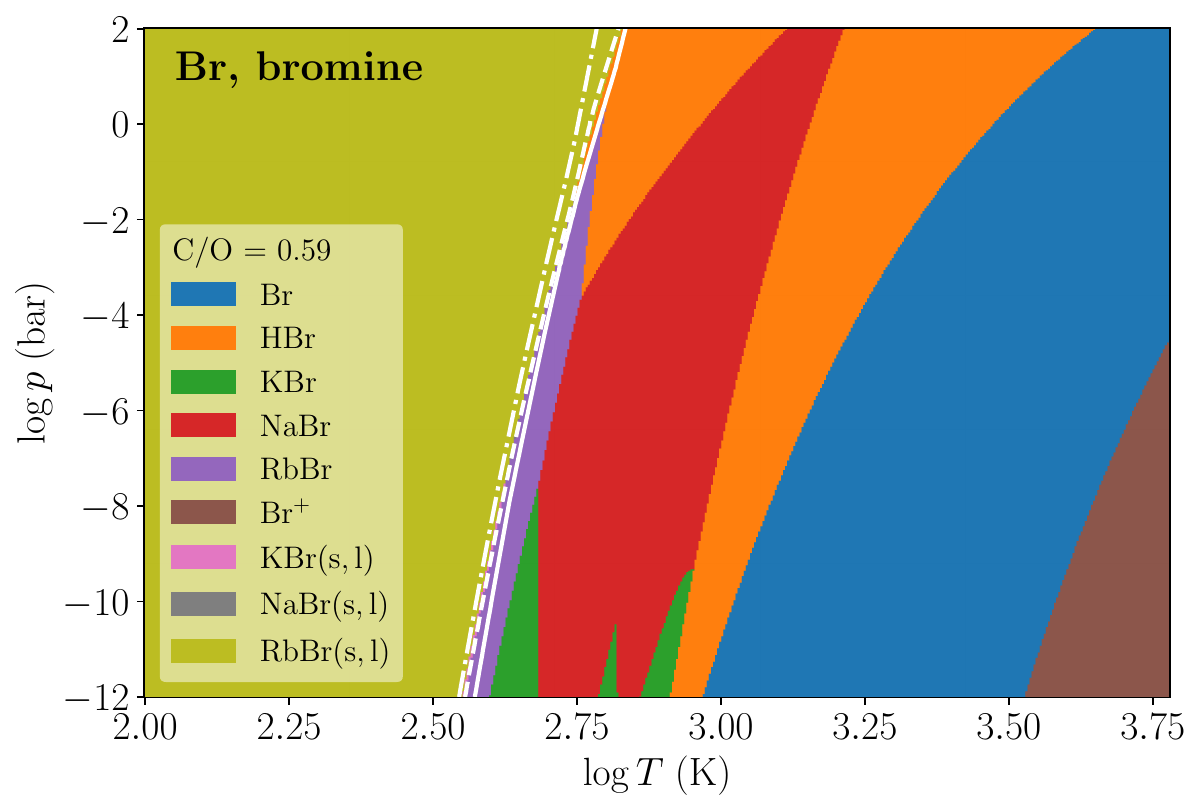}\hfill
    \includegraphics[width=0.49\linewidth]{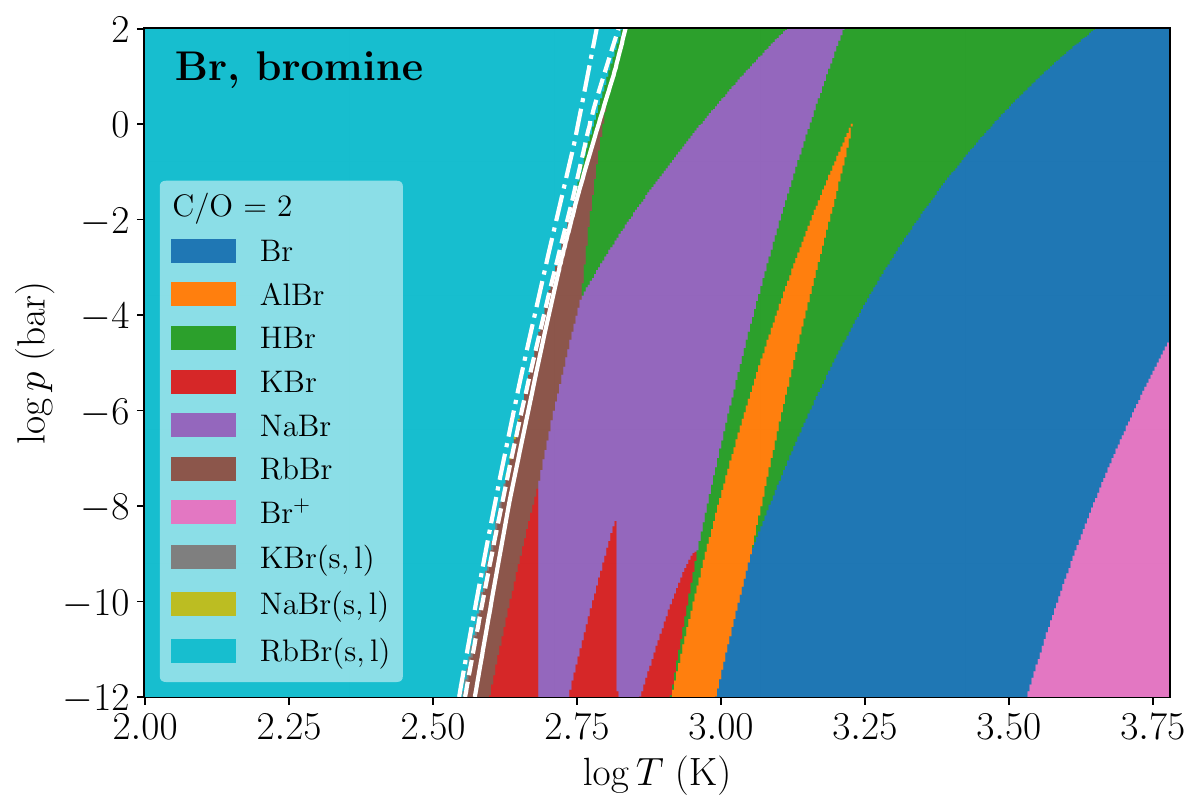}\\
    \includegraphics[width=0.49\linewidth]{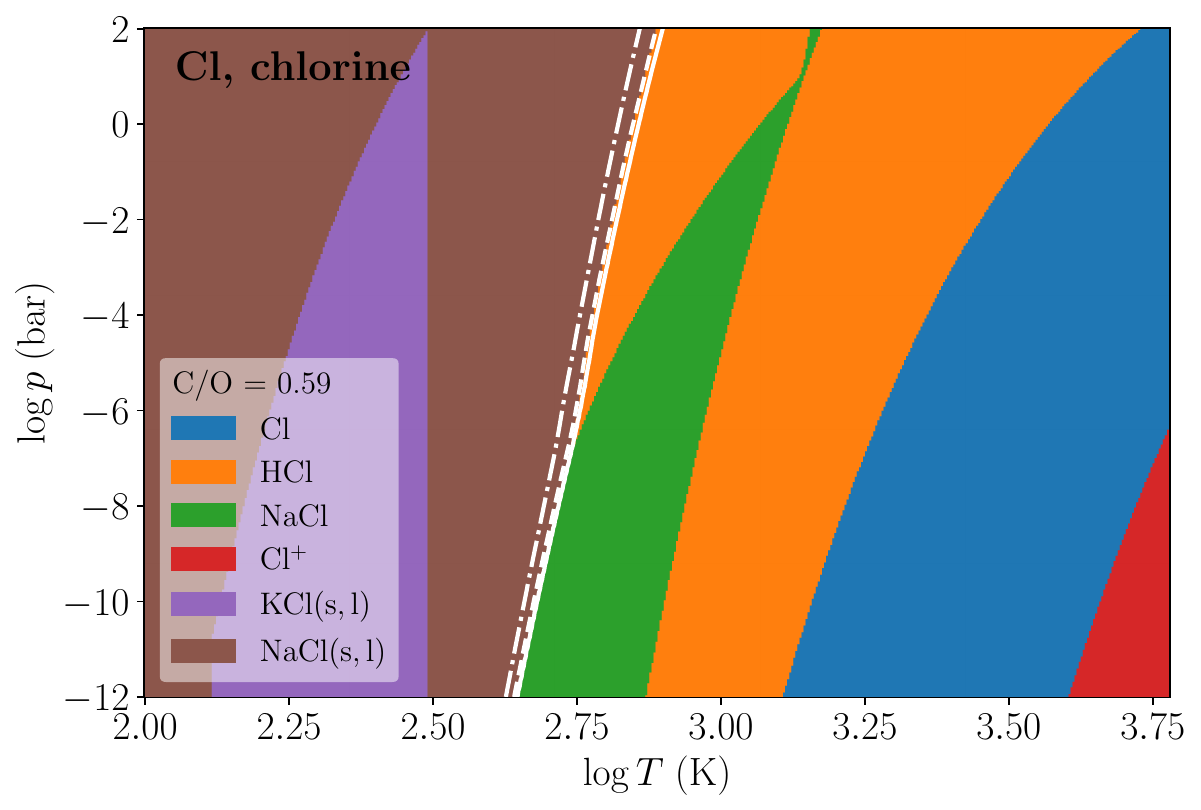}\hfill
    \includegraphics[width=0.49\linewidth]{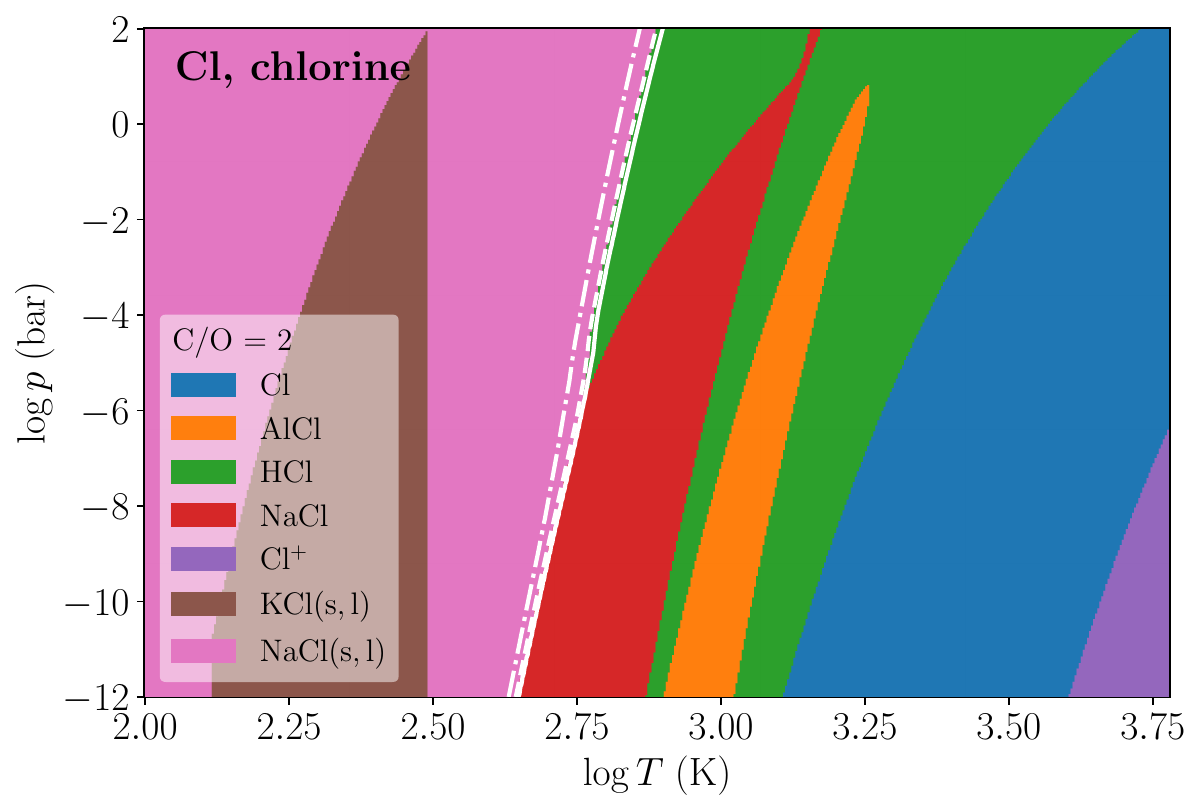}%
  }
  \caption{Same as Fig. \ref{fig:chem_results} for all remaining elements from Table \ref{table:abundances_asplund}. 
  The elements are listed in alphabetic order.}
  \label{fig:chem_plots}
\end{figure*}

\begin{figure}[h!]
  \ContinuedFloat
  \sidecaption
  \parbox[b]{12cm}{%
    \includegraphics[width=0.49\linewidth]{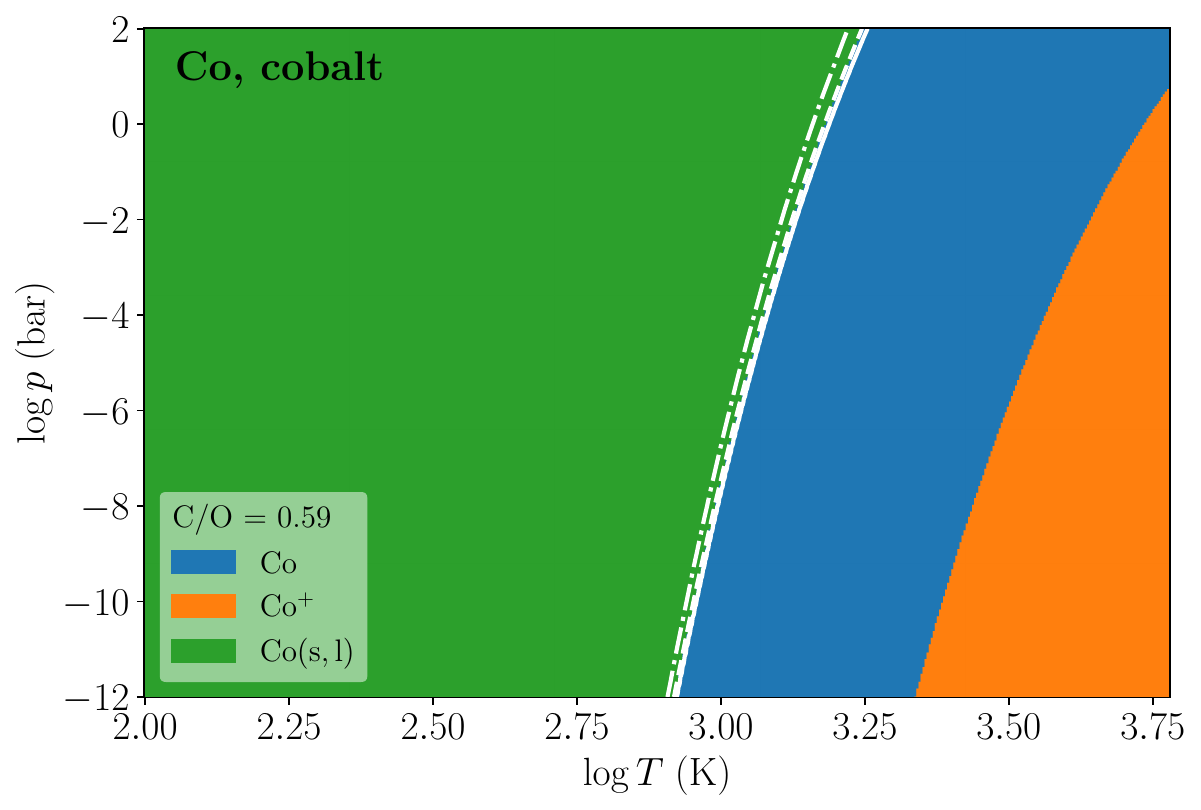}\hfill
    \includegraphics[width=0.49\linewidth]{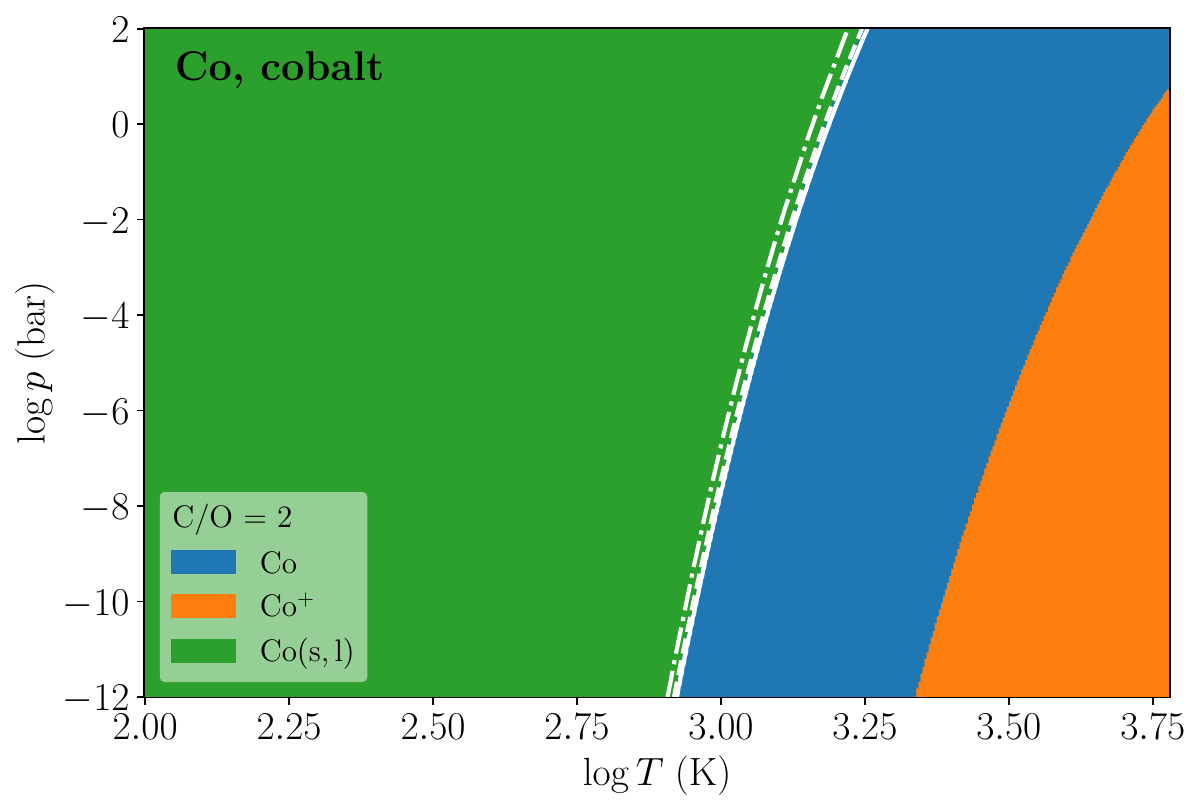}\\
    \includegraphics[width=0.49\linewidth]{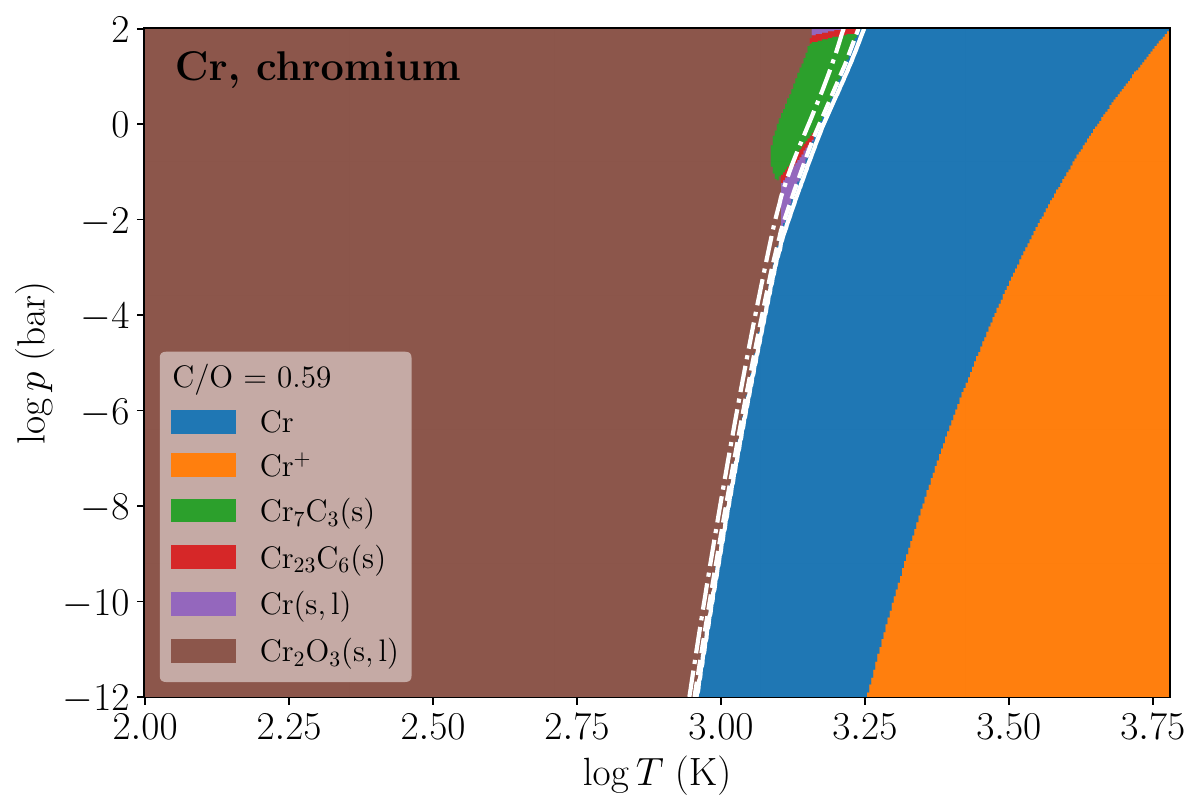}\hfill
    \includegraphics[width=0.49\linewidth]{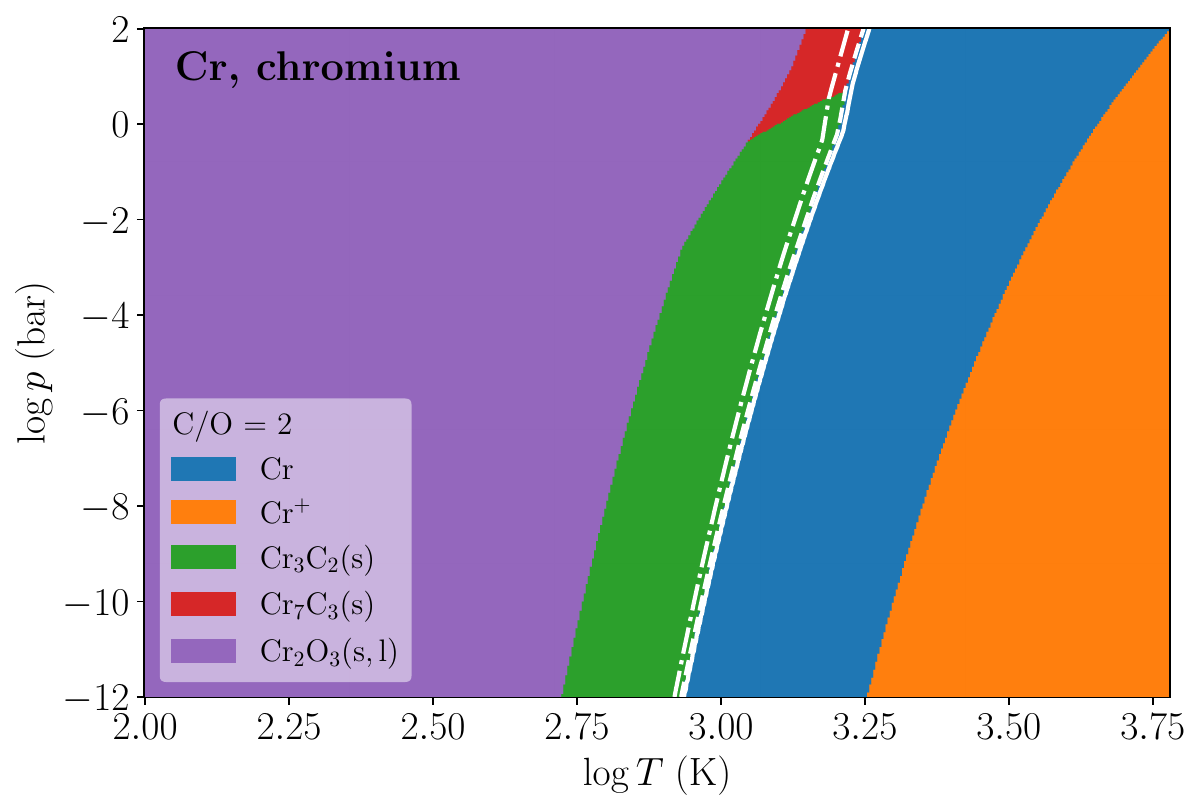}\\
    \includegraphics[width=0.49\linewidth]{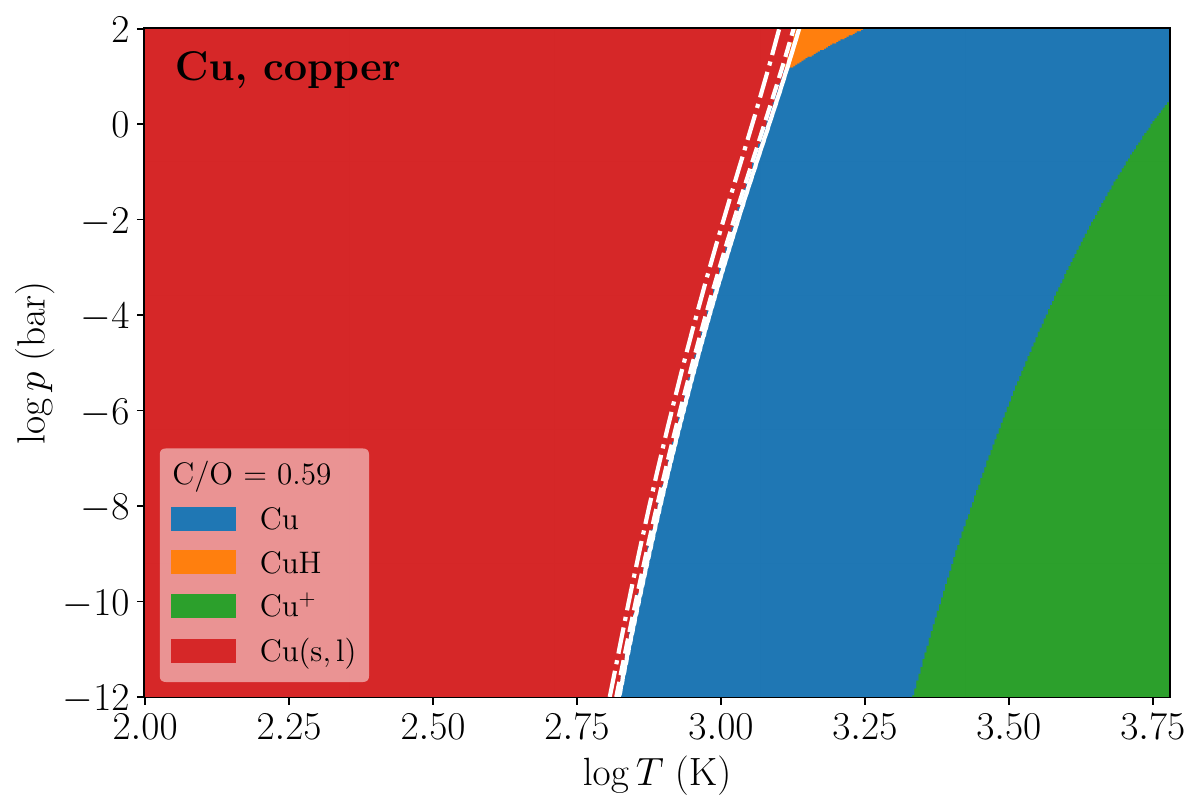}\hfill
    \includegraphics[width=0.49\linewidth]{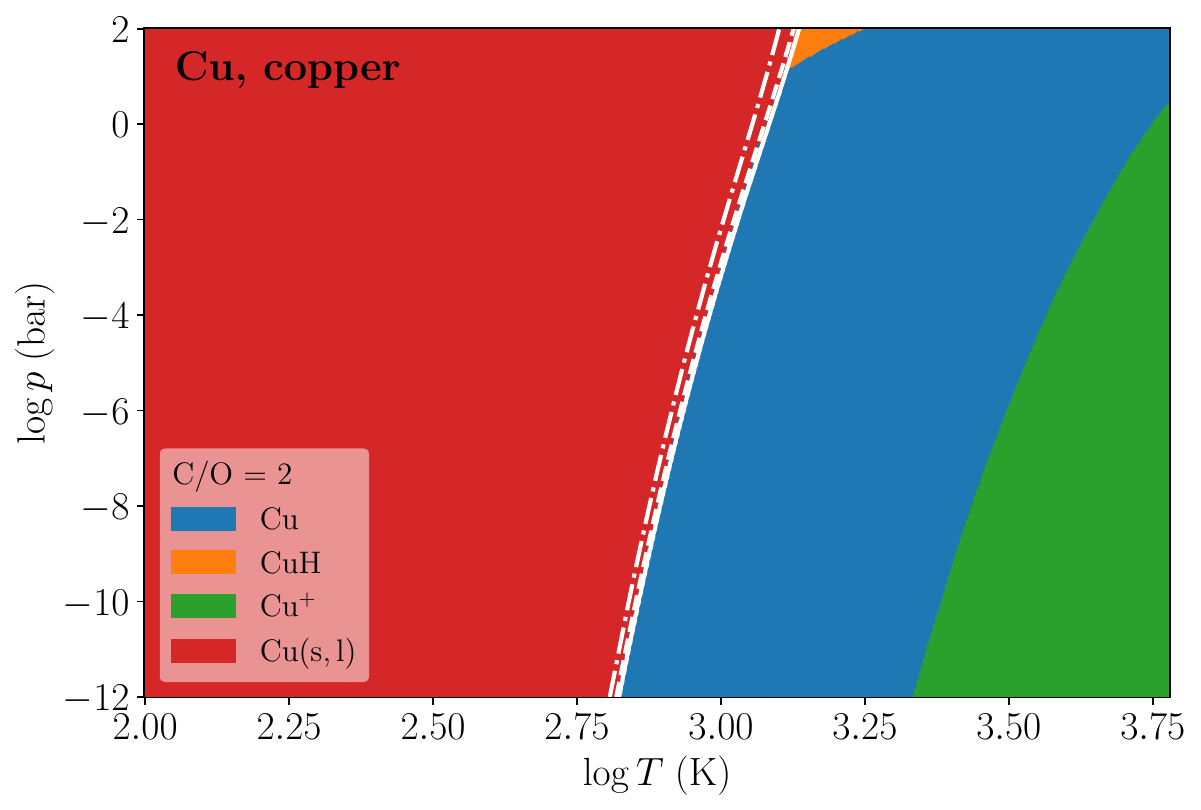}\\
    \includegraphics[width=0.49\linewidth]{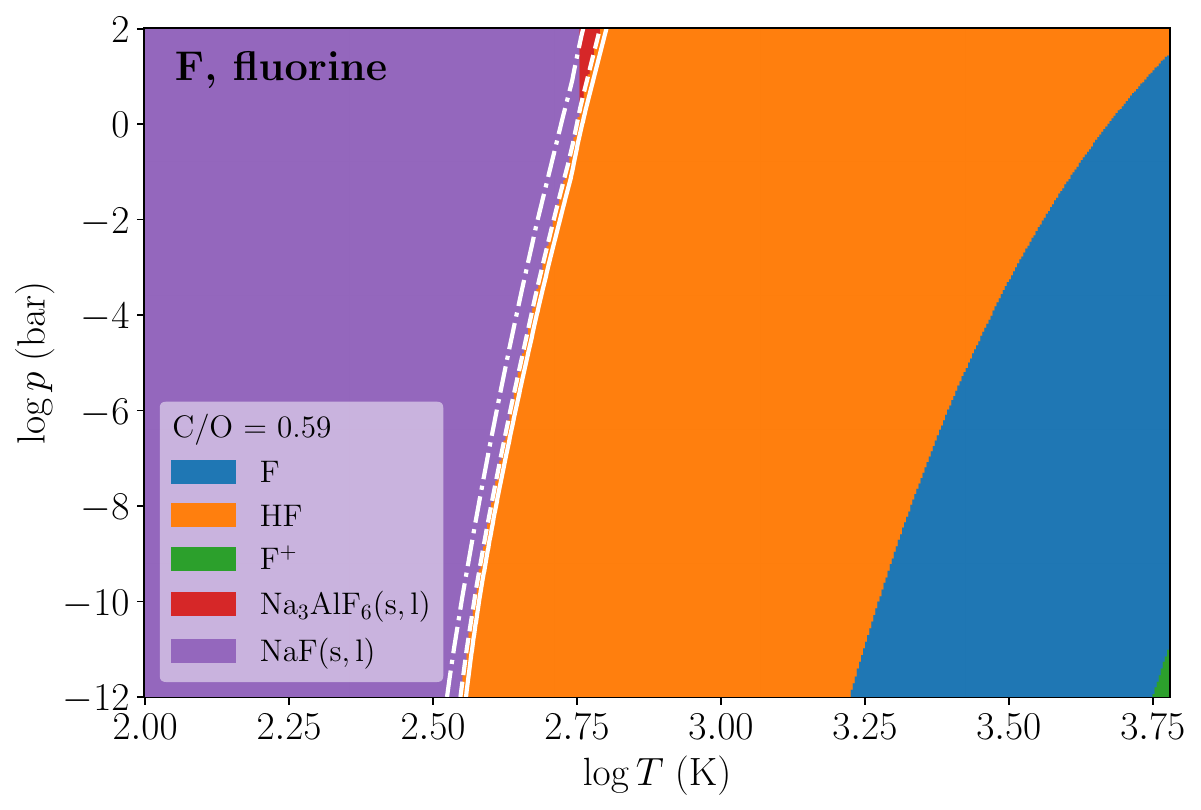}\hfill
    \includegraphics[width=0.49\linewidth]{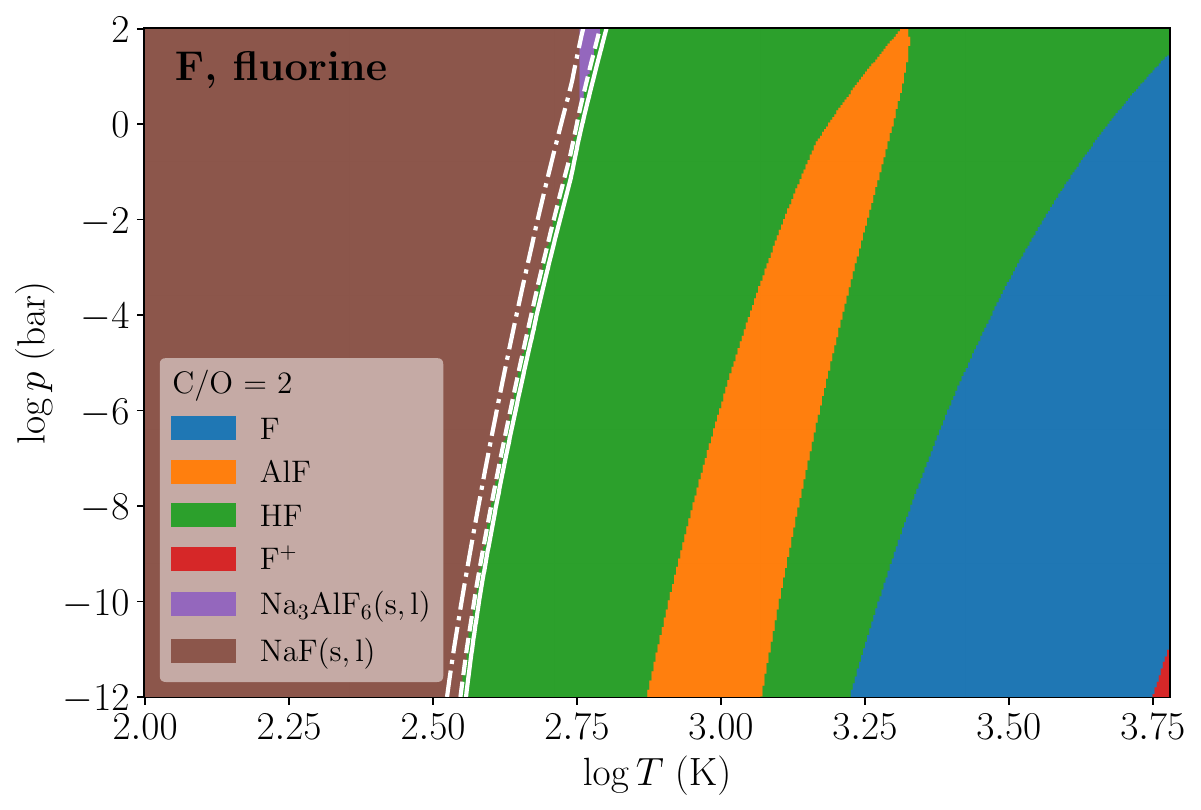}\\
    \includegraphics[width=0.49\linewidth]{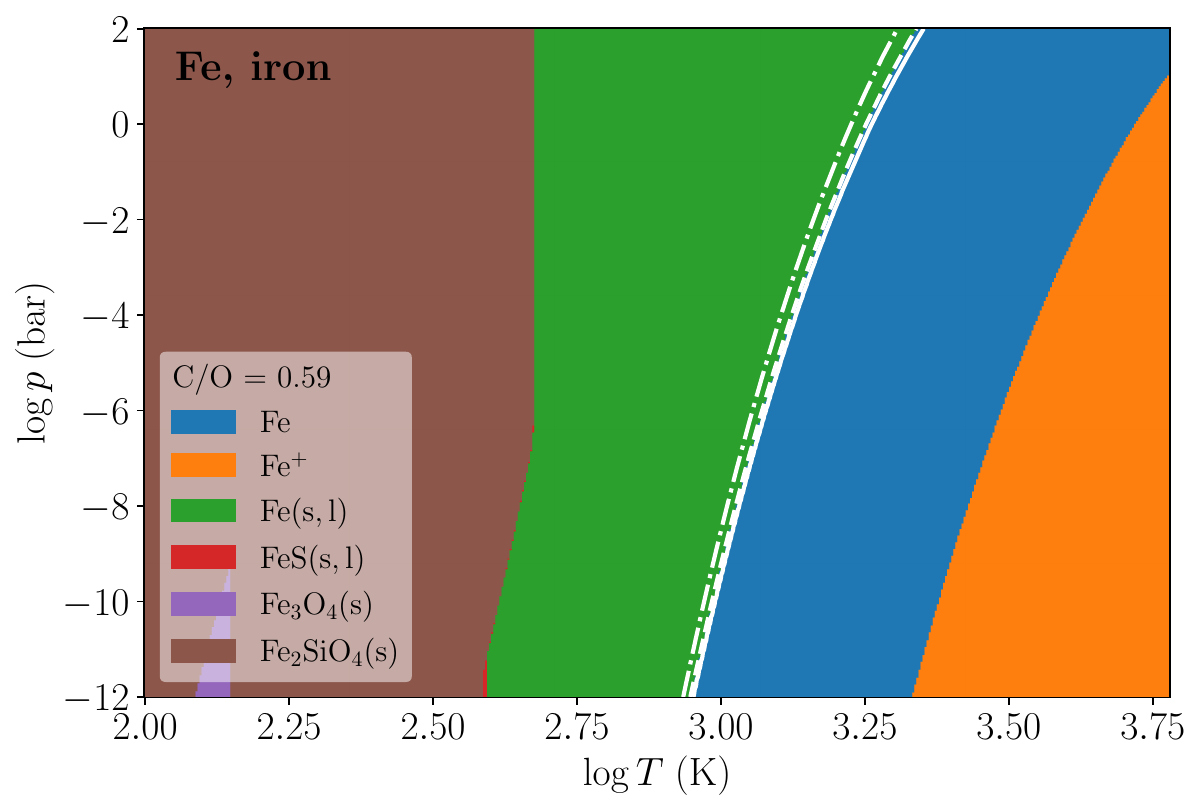}\hfill
    \includegraphics[width=0.49\linewidth]{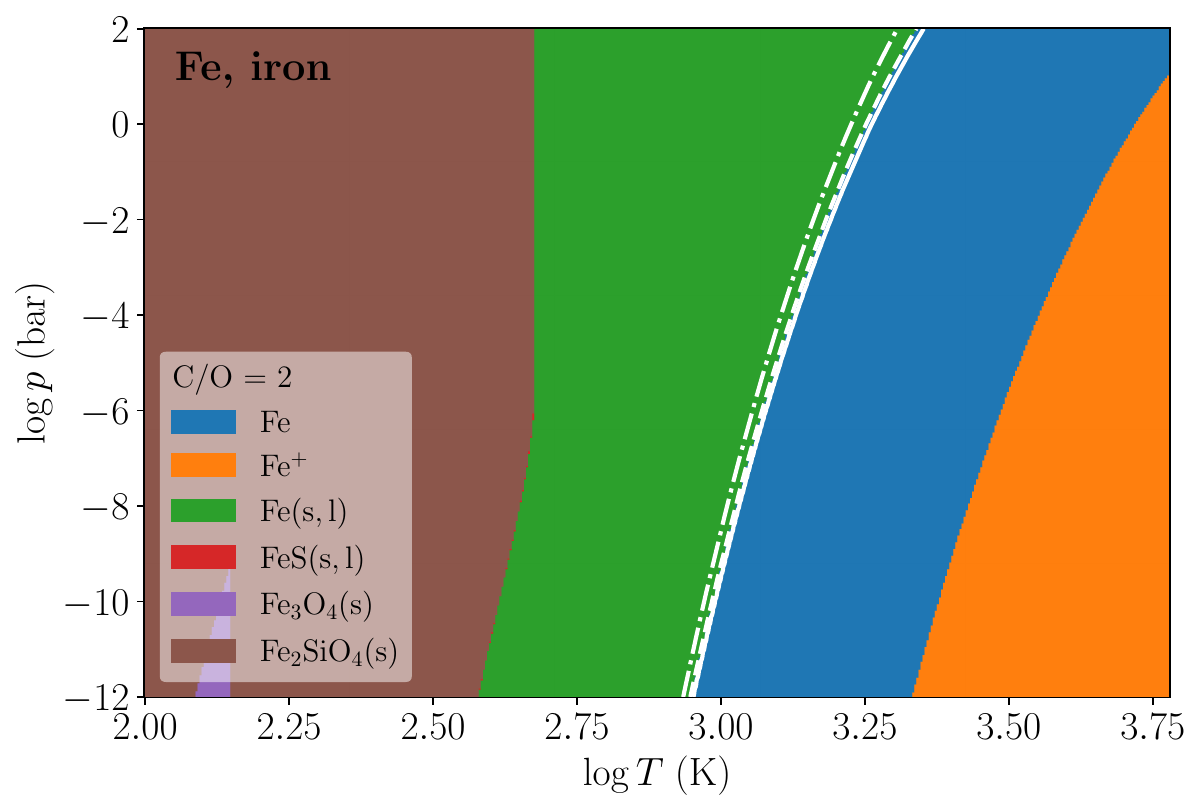}\\
    \includegraphics[width=0.49\linewidth]{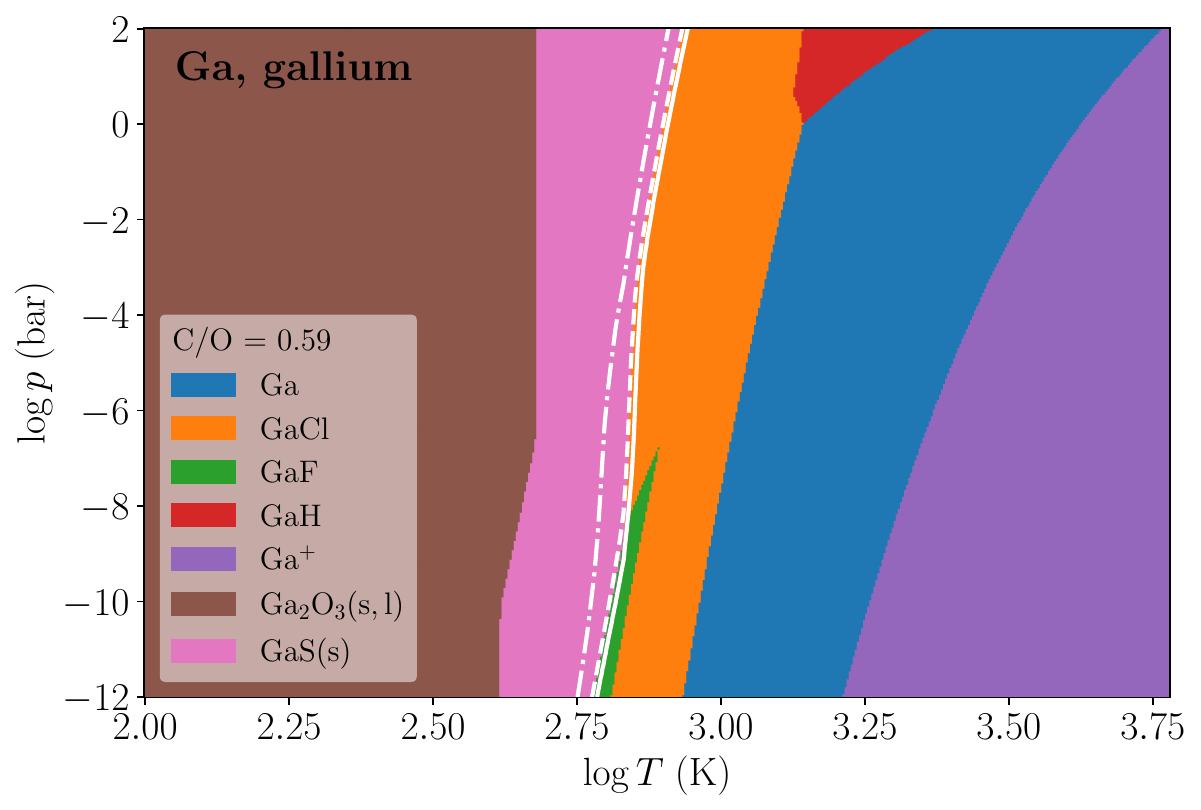}\hfill
    \includegraphics[width=0.49\linewidth]{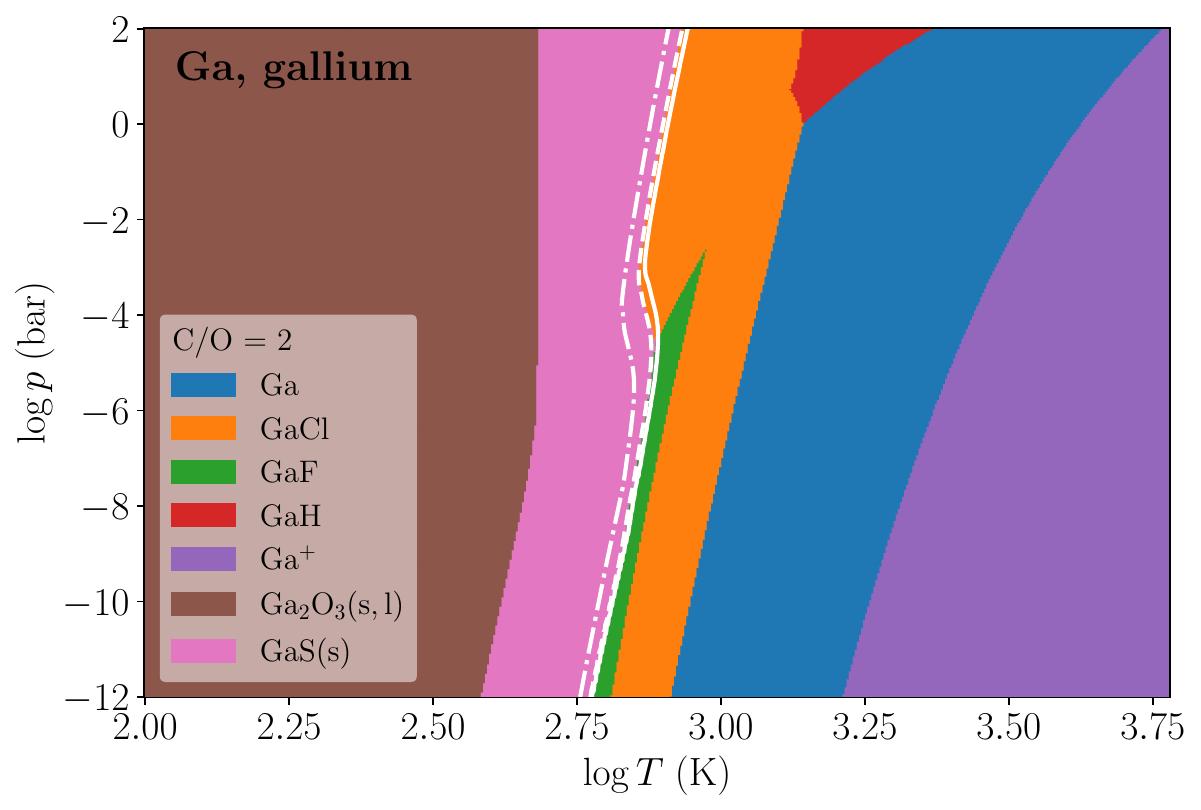}%
  }
  \caption{continued.}
\end{figure}

\begin{figure}[h!]
  \ContinuedFloat
  \sidecaption
  \parbox[b]{12cm}{%
    \includegraphics[width=0.49\linewidth]{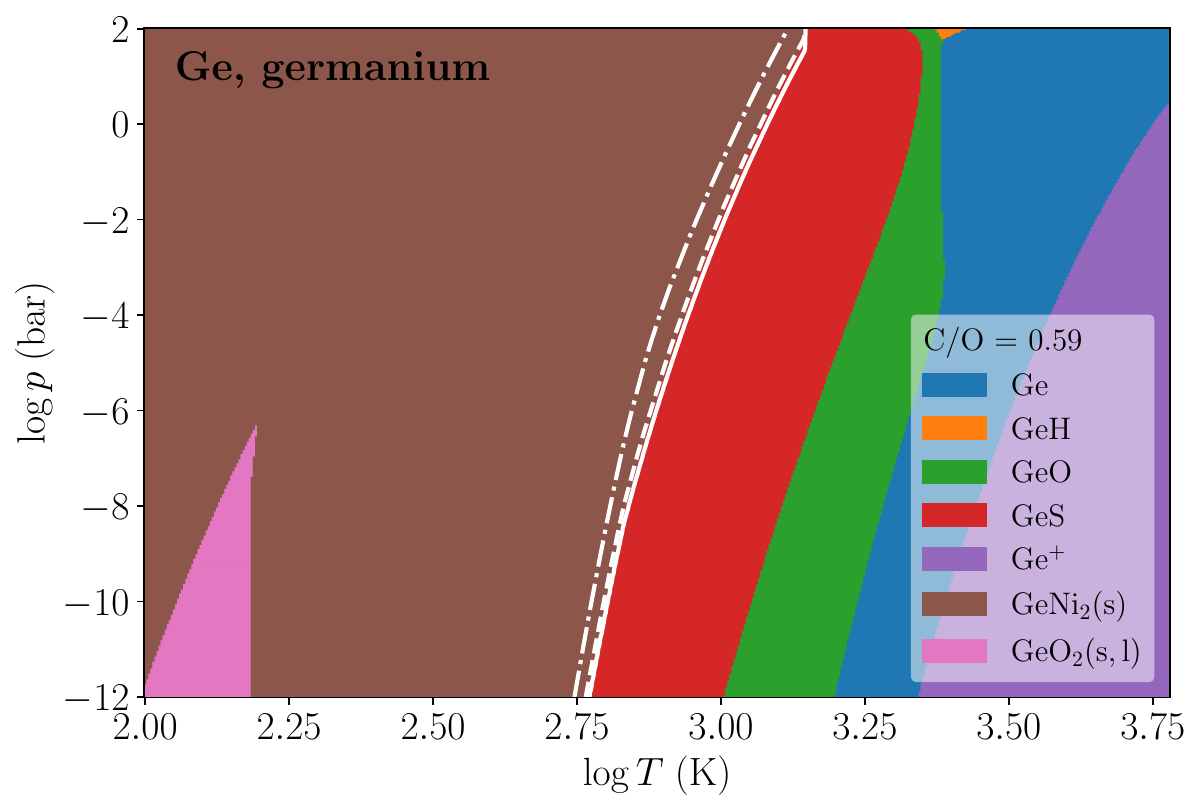}\hfill
    \includegraphics[width=0.49\linewidth]{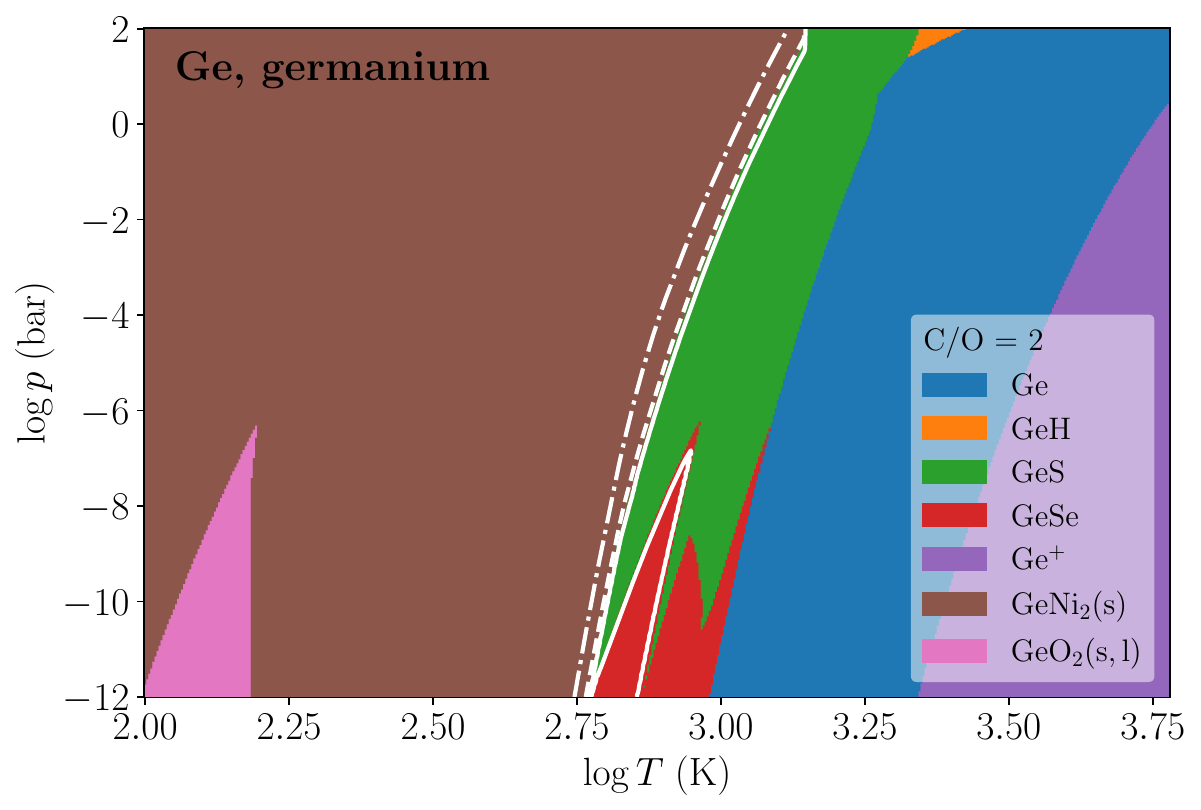}\\
    \includegraphics[width=0.49\linewidth]{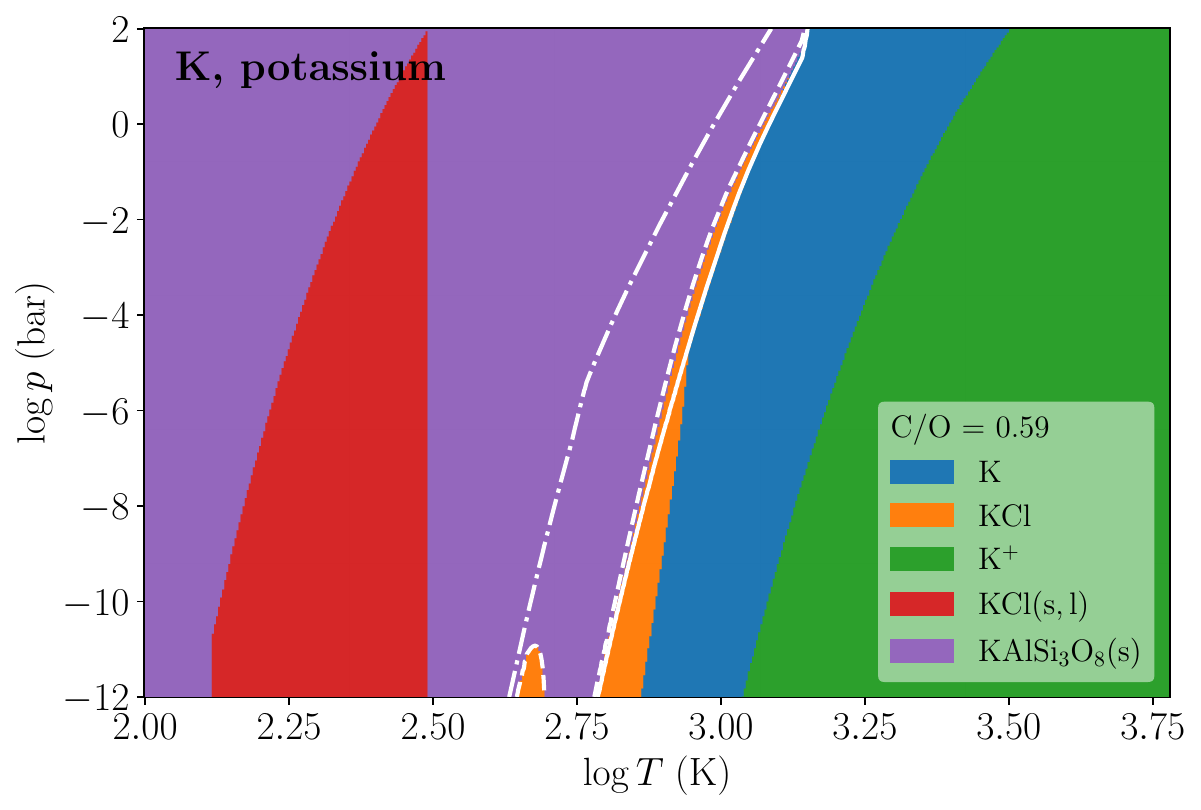}\hfill
    \includegraphics[width=0.49\linewidth]{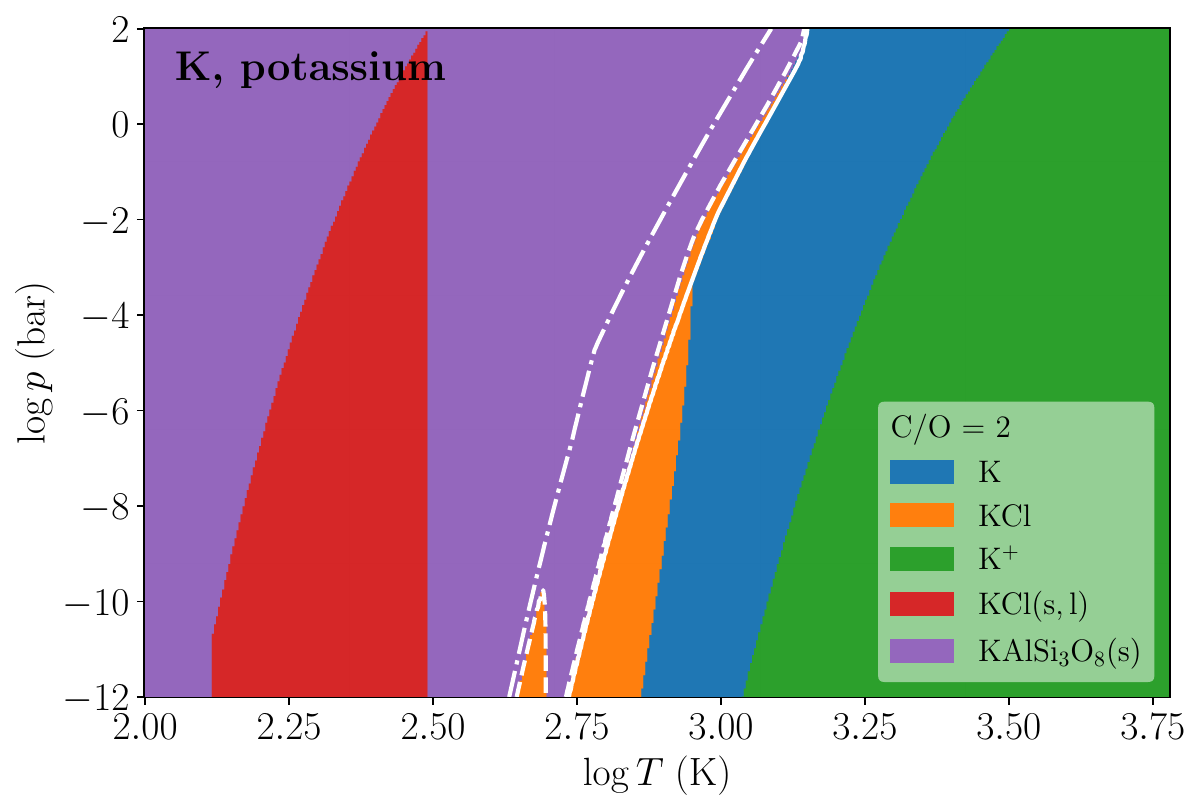}\\
    \includegraphics[width=0.49\linewidth]{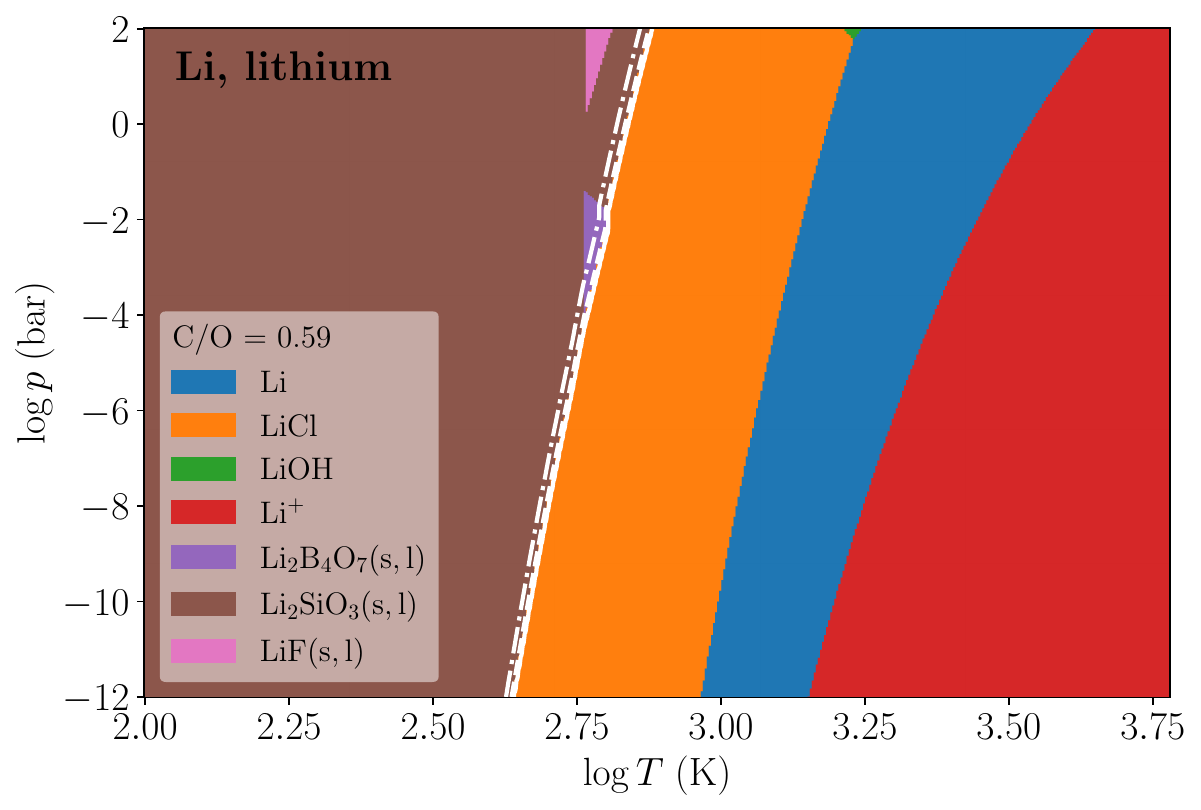}\hfill
    \includegraphics[width=0.49\linewidth]{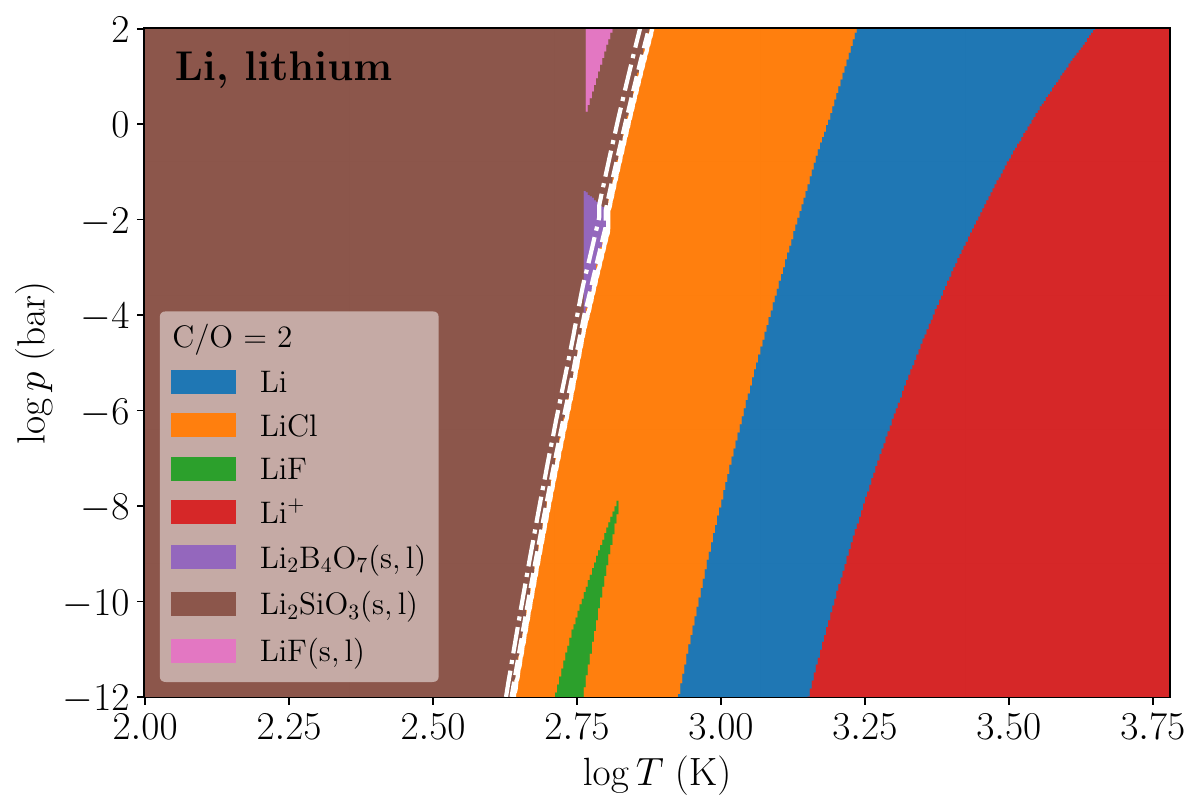}\\
    \includegraphics[width=0.49\linewidth]{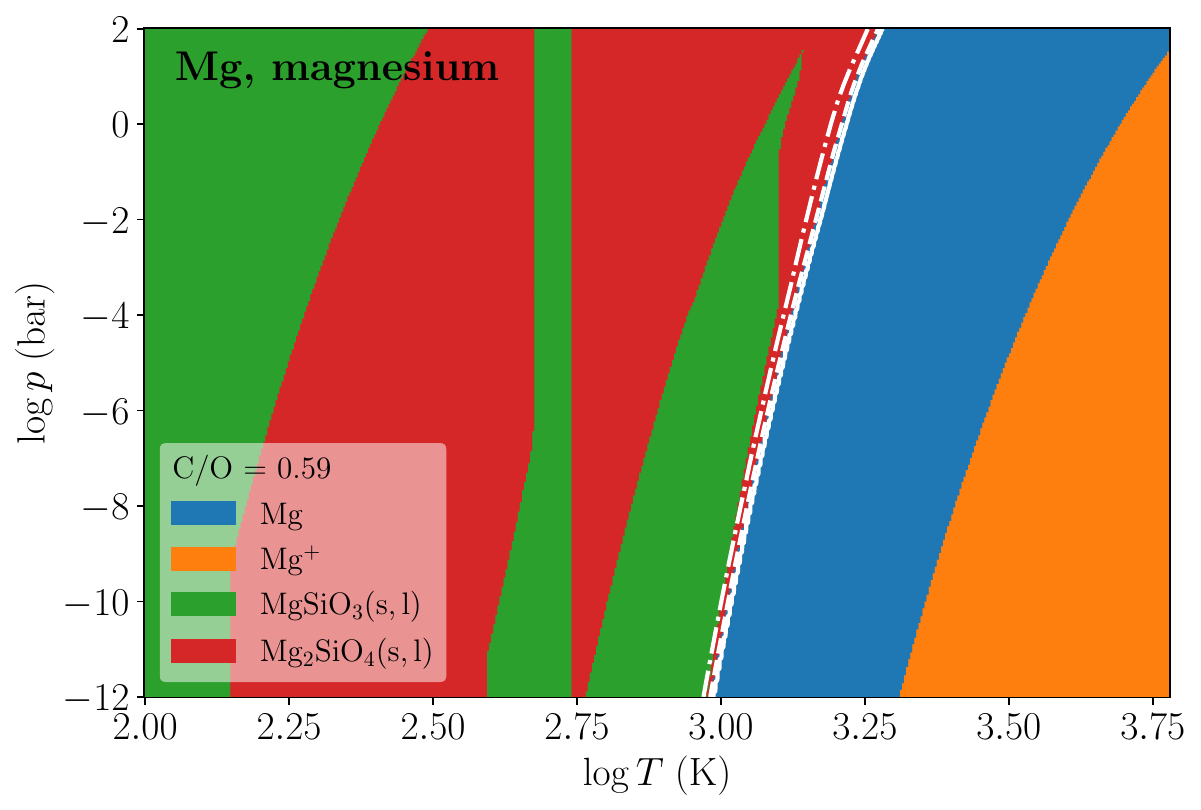}\hfill
    \includegraphics[width=0.49\linewidth]{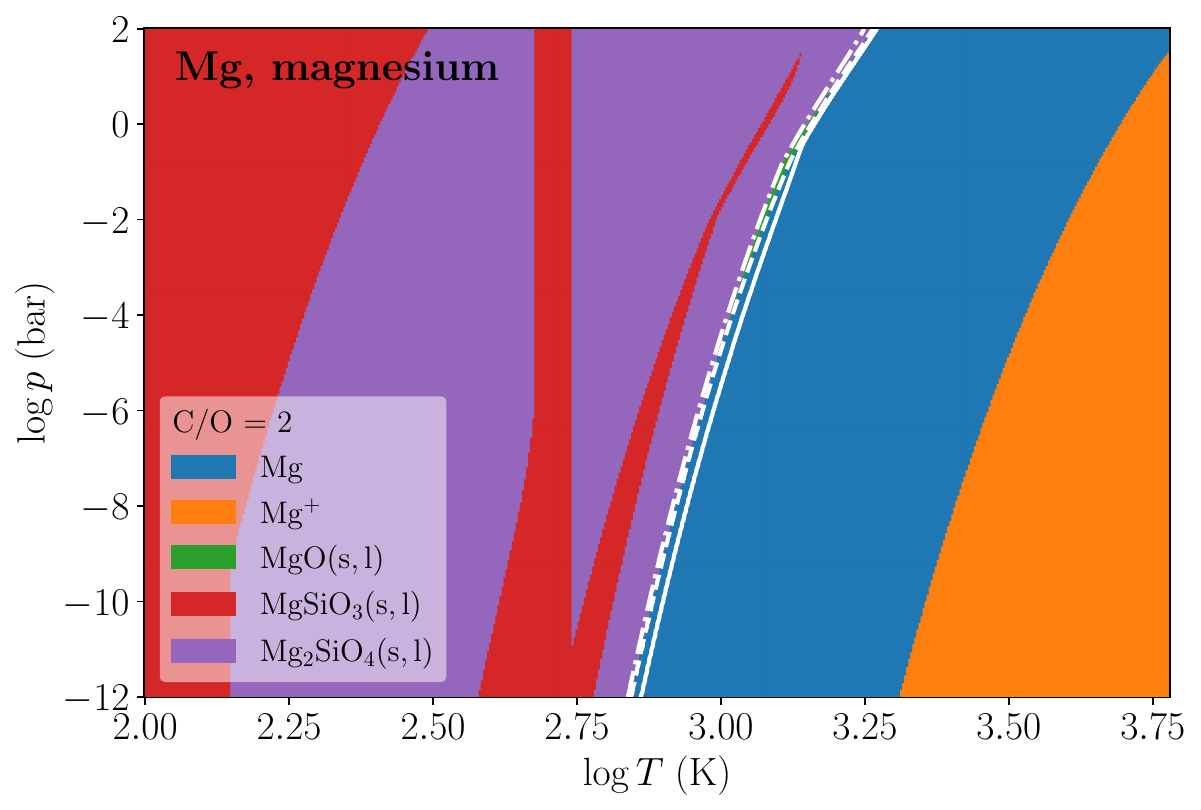}\\
    \includegraphics[width=0.49\linewidth]{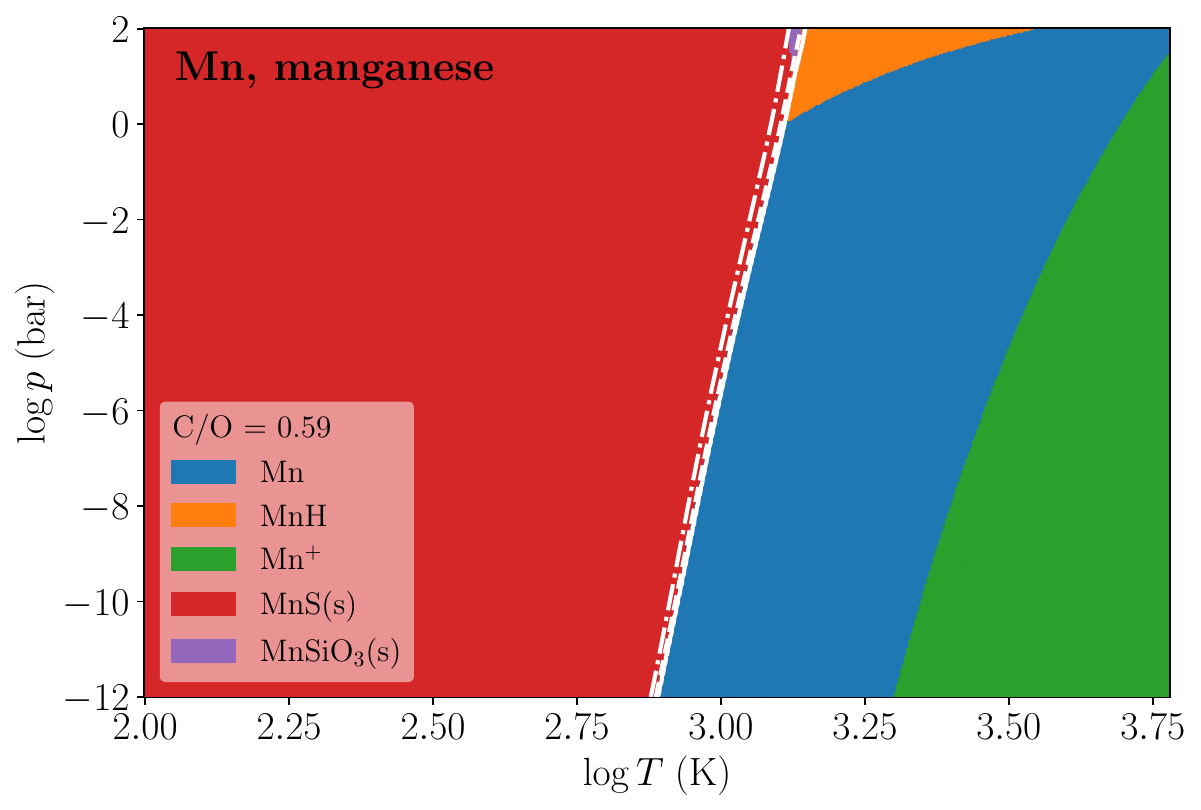}\hfill
    \includegraphics[width=0.49\linewidth]{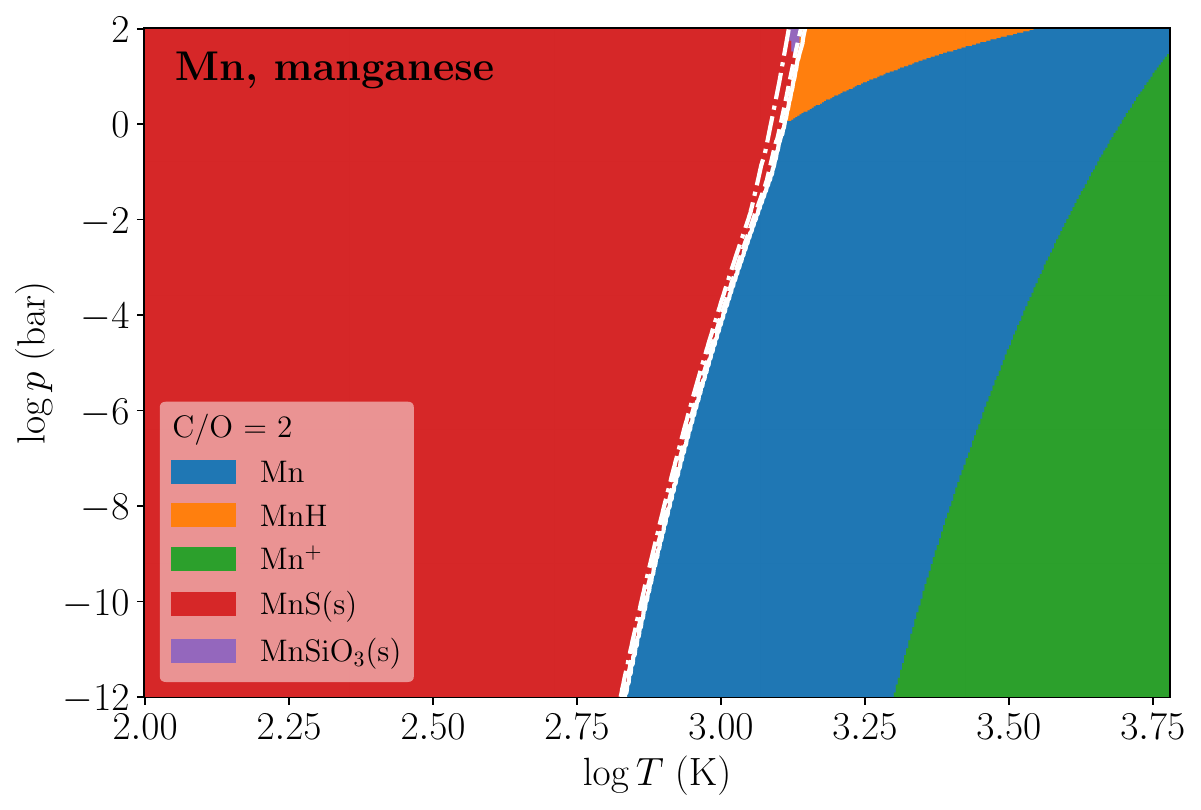}\\
    \includegraphics[width=0.49\linewidth]{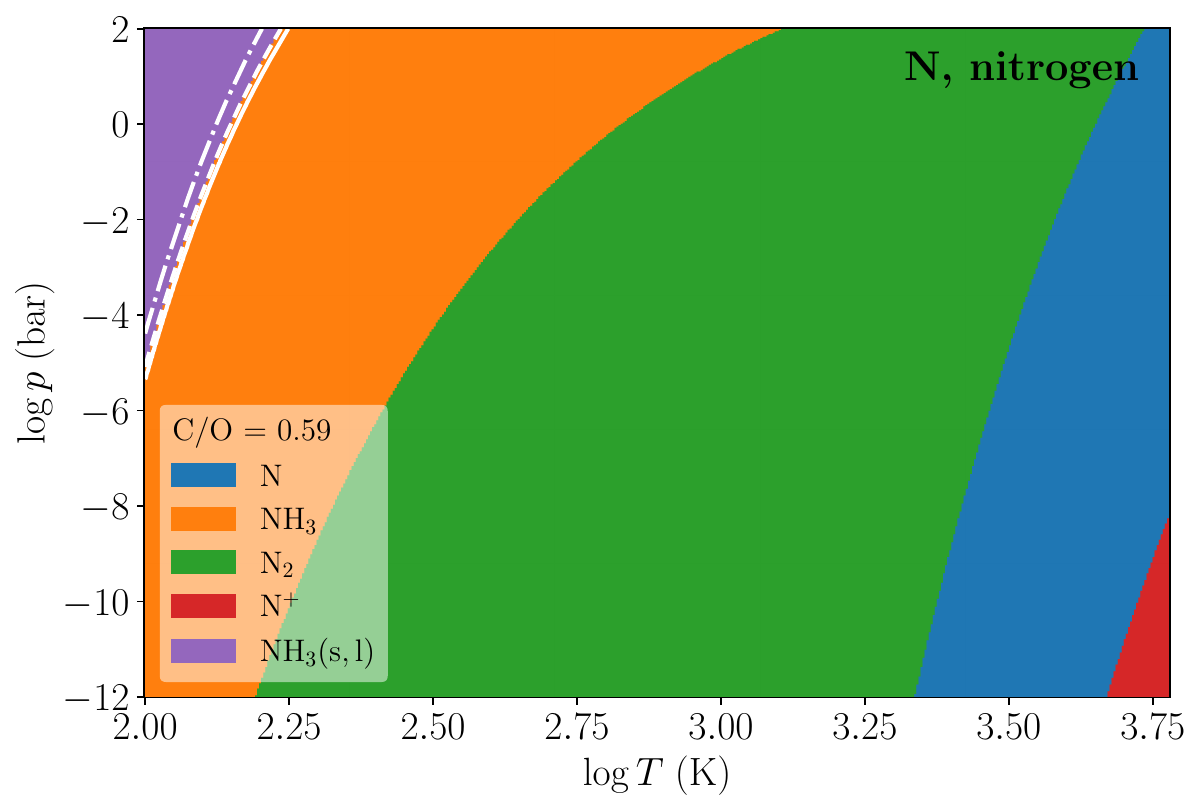}\hfill
    \includegraphics[width=0.49\linewidth]{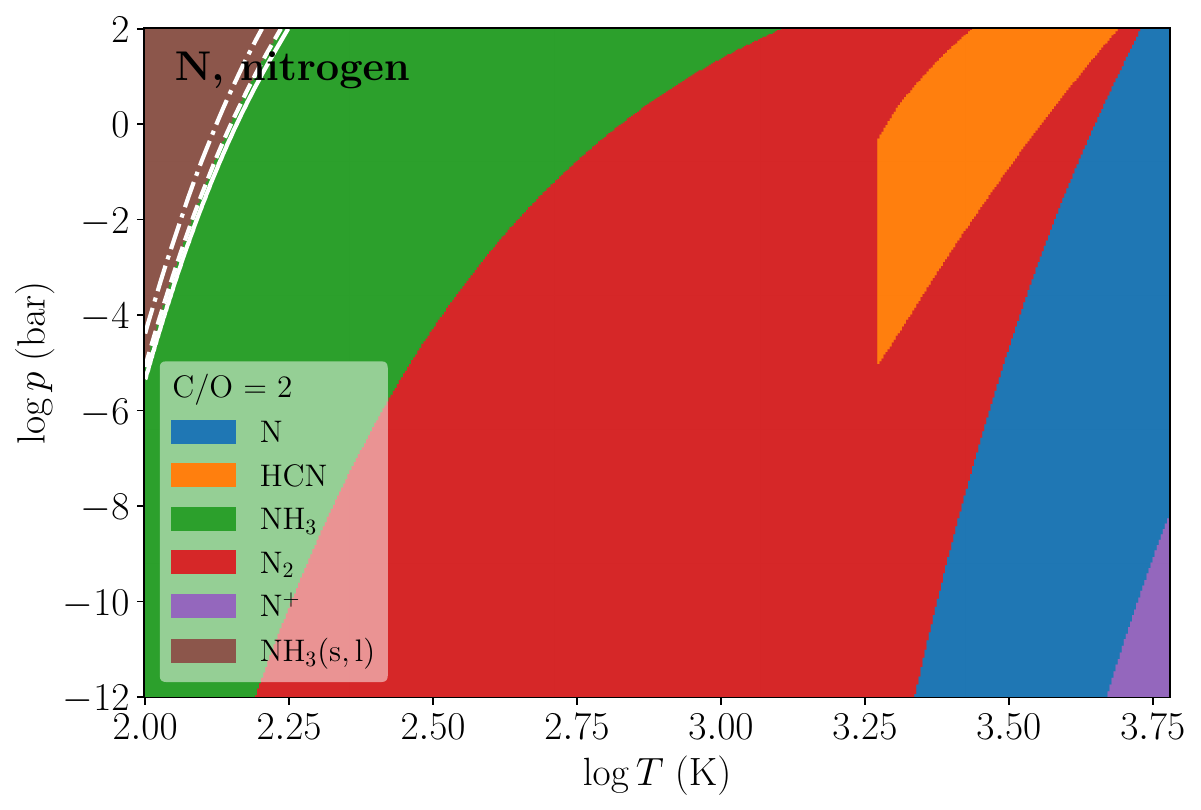}%
  }
  \caption{continued.}
\end{figure}

\begin{figure}[h!]
  \ContinuedFloat
  \sidecaption
  \parbox[b]{12cm}{%
    \includegraphics[width=0.49\linewidth]{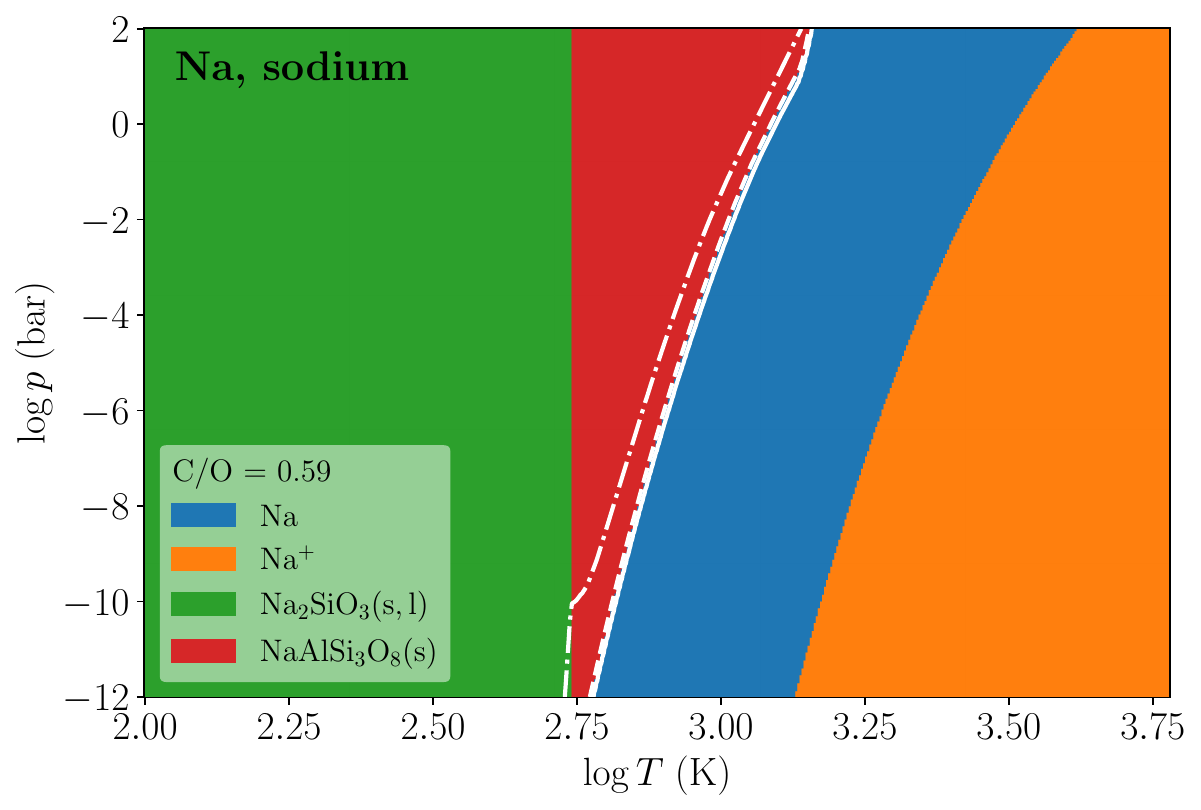}\hfill
    \includegraphics[width=0.49\linewidth]{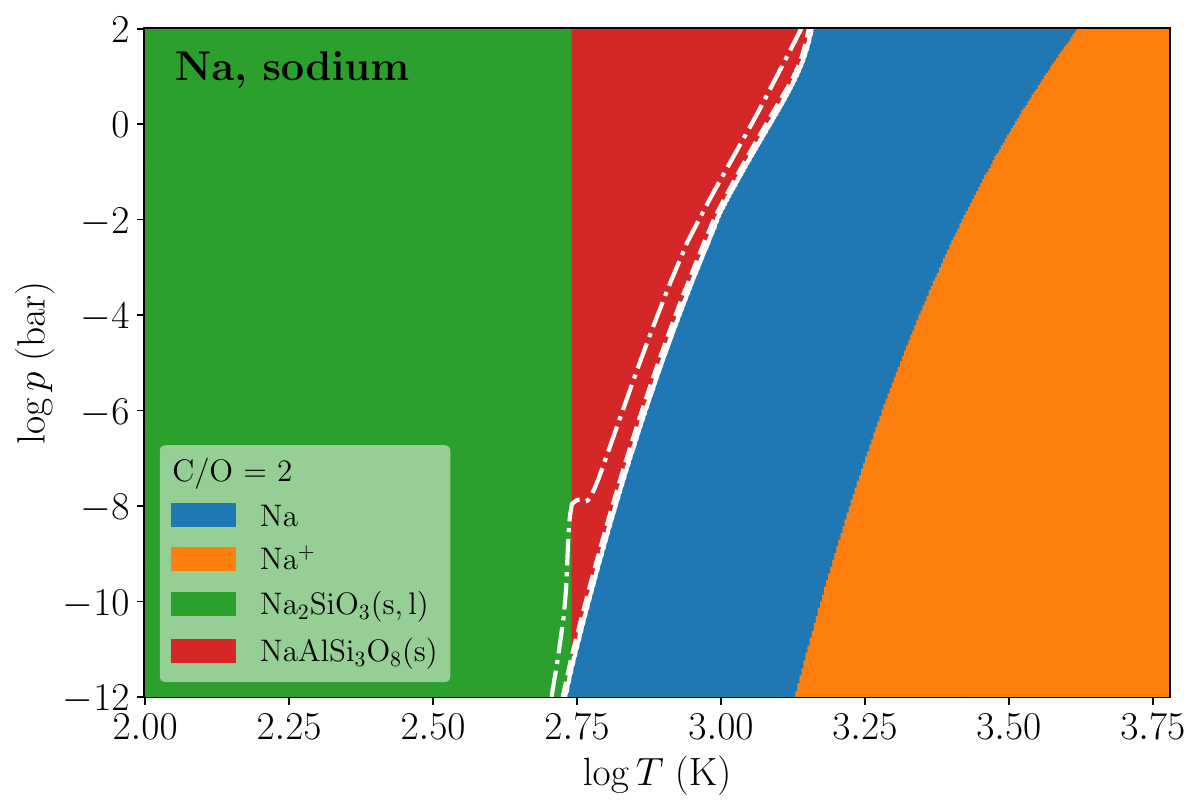}\\
    \includegraphics[width=0.49\linewidth]{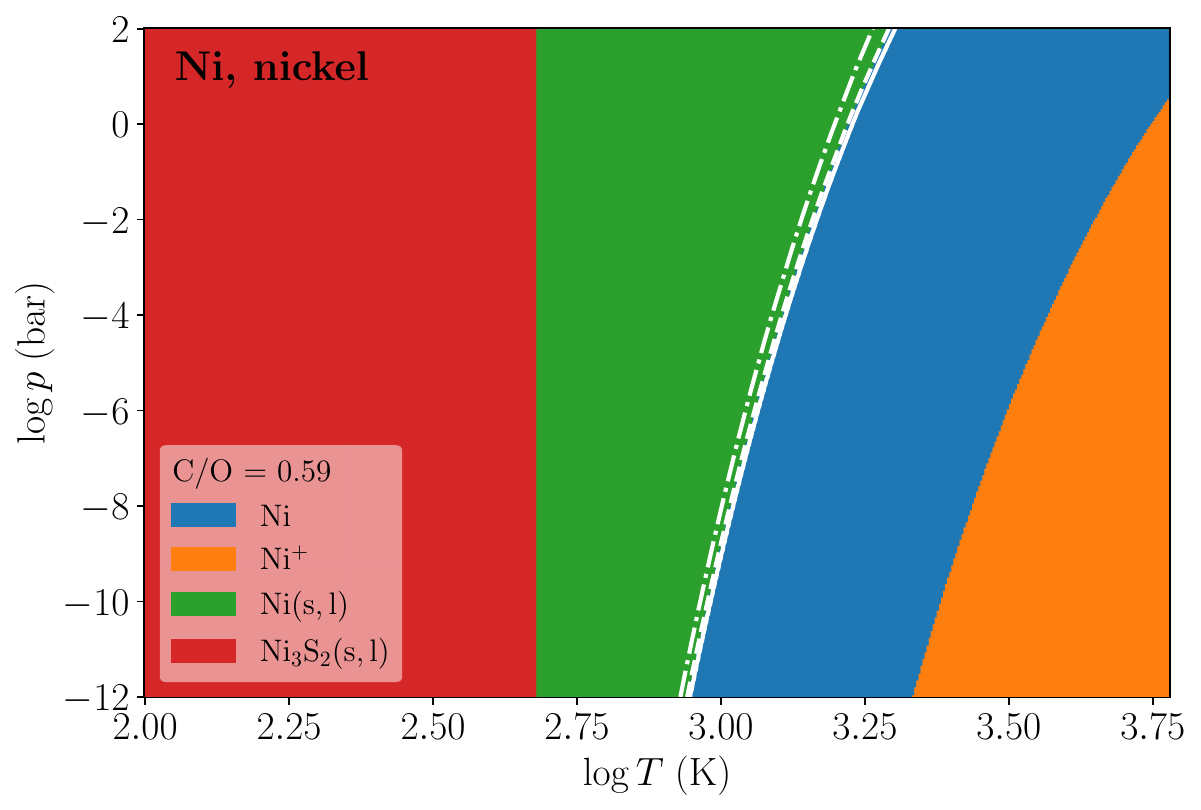}\hfill
    \includegraphics[width=0.49\linewidth]{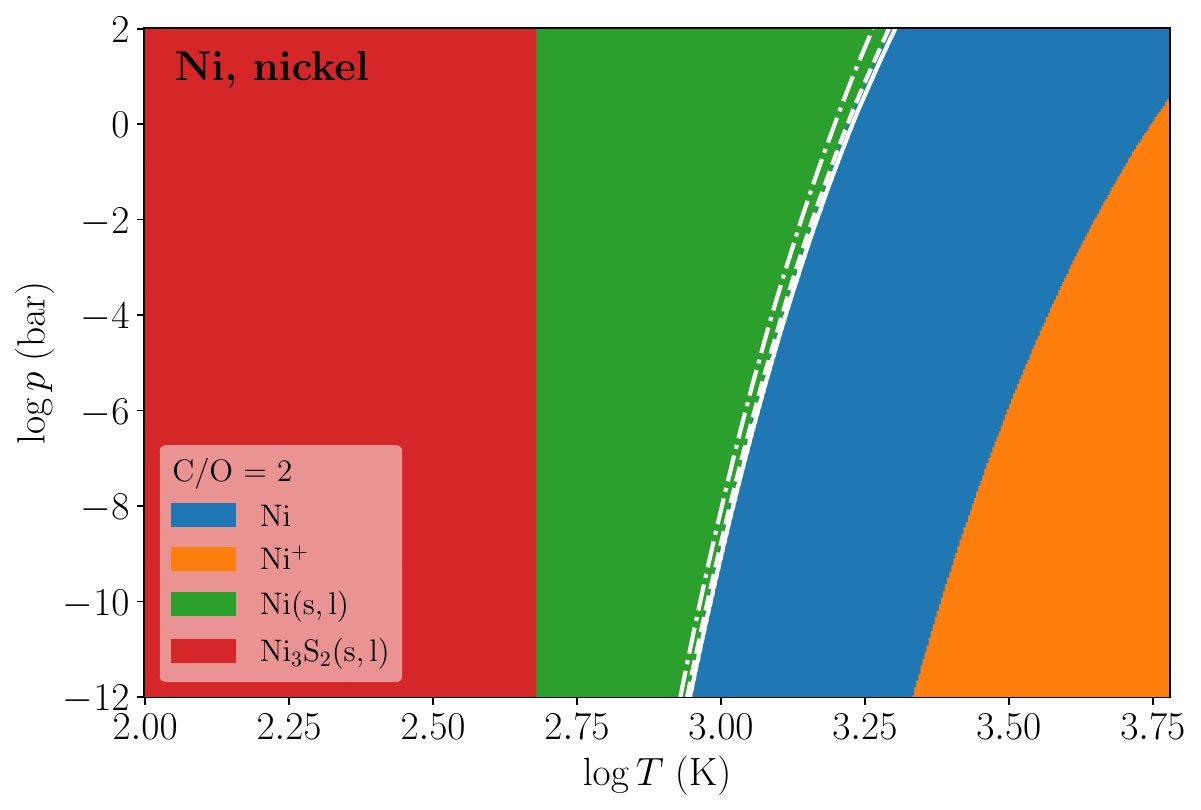}\\
    \includegraphics[width=0.49\linewidth]{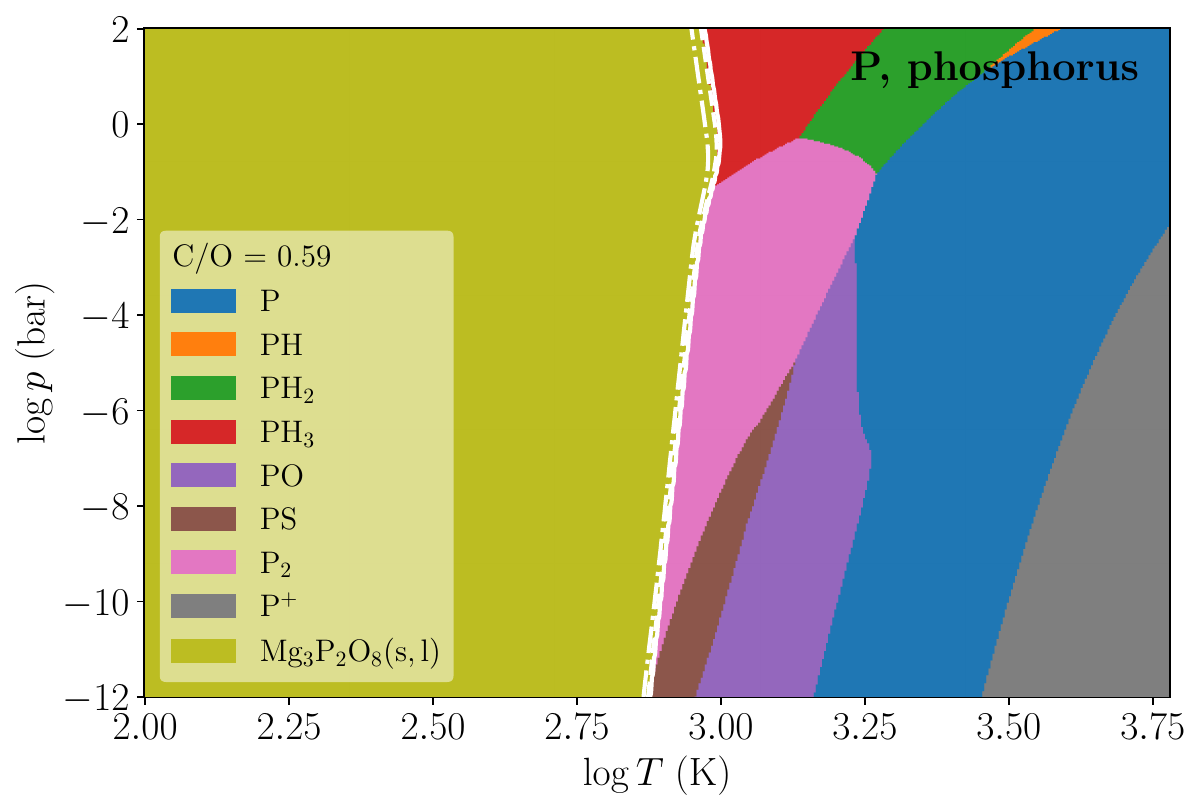}\hfill
    \includegraphics[width=0.49\linewidth]{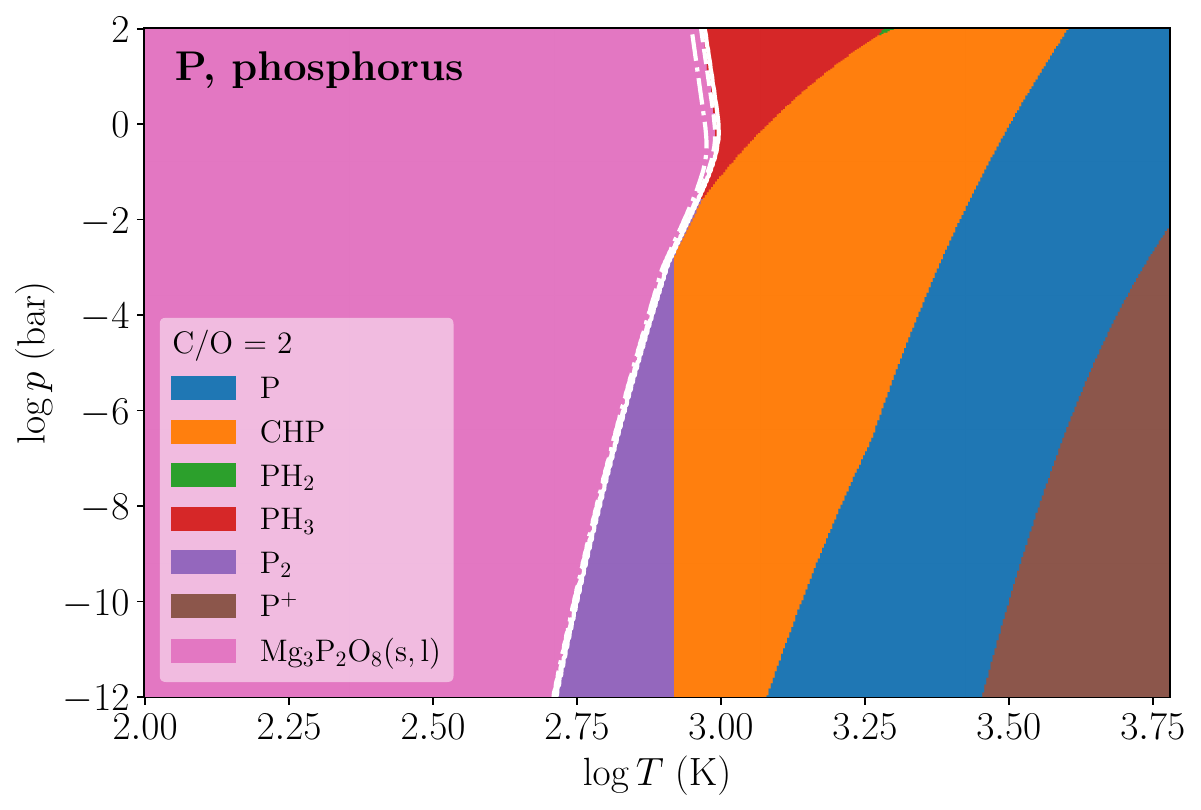}\\
    \includegraphics[width=0.49\linewidth]{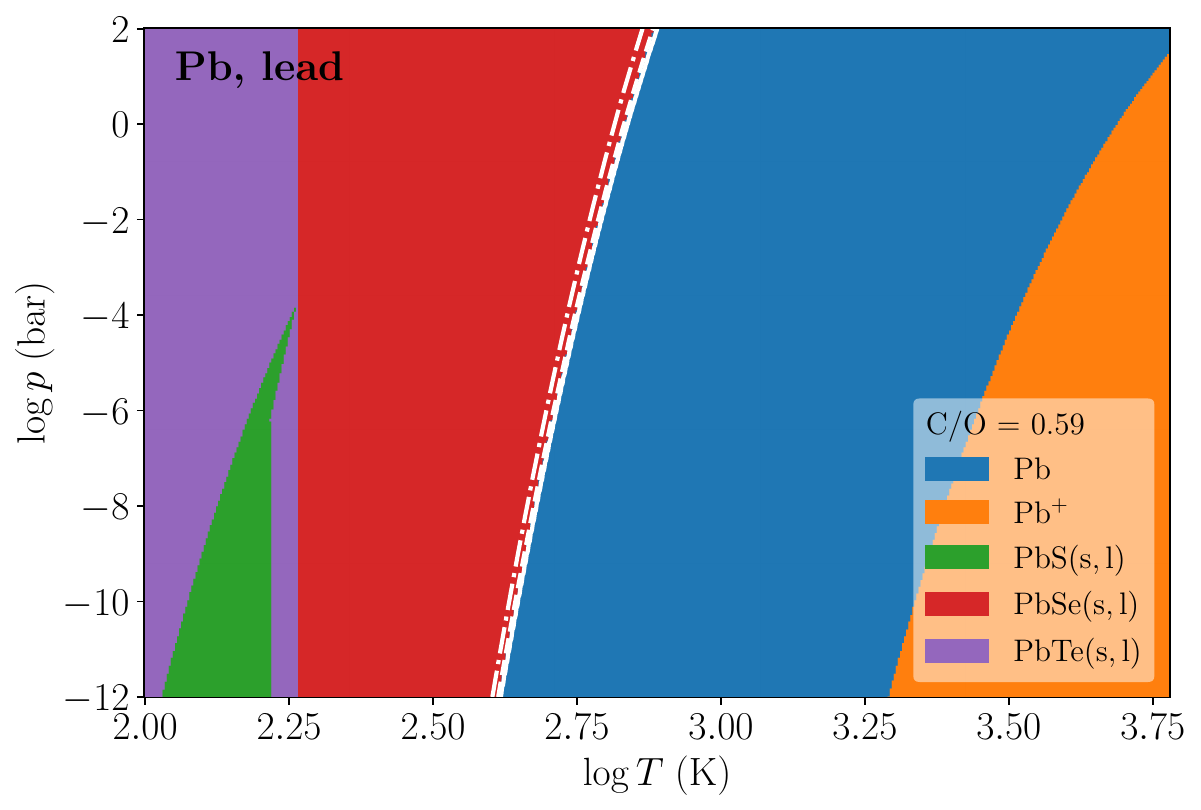}\hfill
    \includegraphics[width=0.49\linewidth]{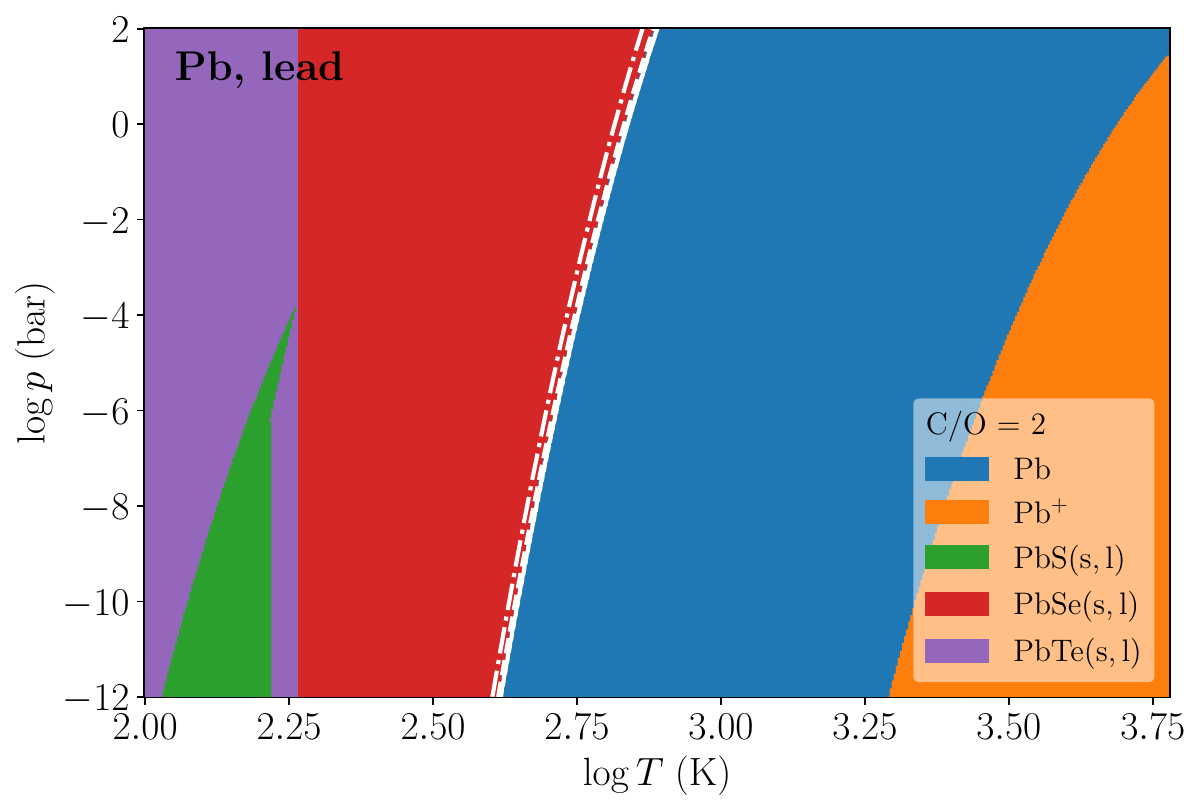}\\
    \includegraphics[width=0.49\linewidth]{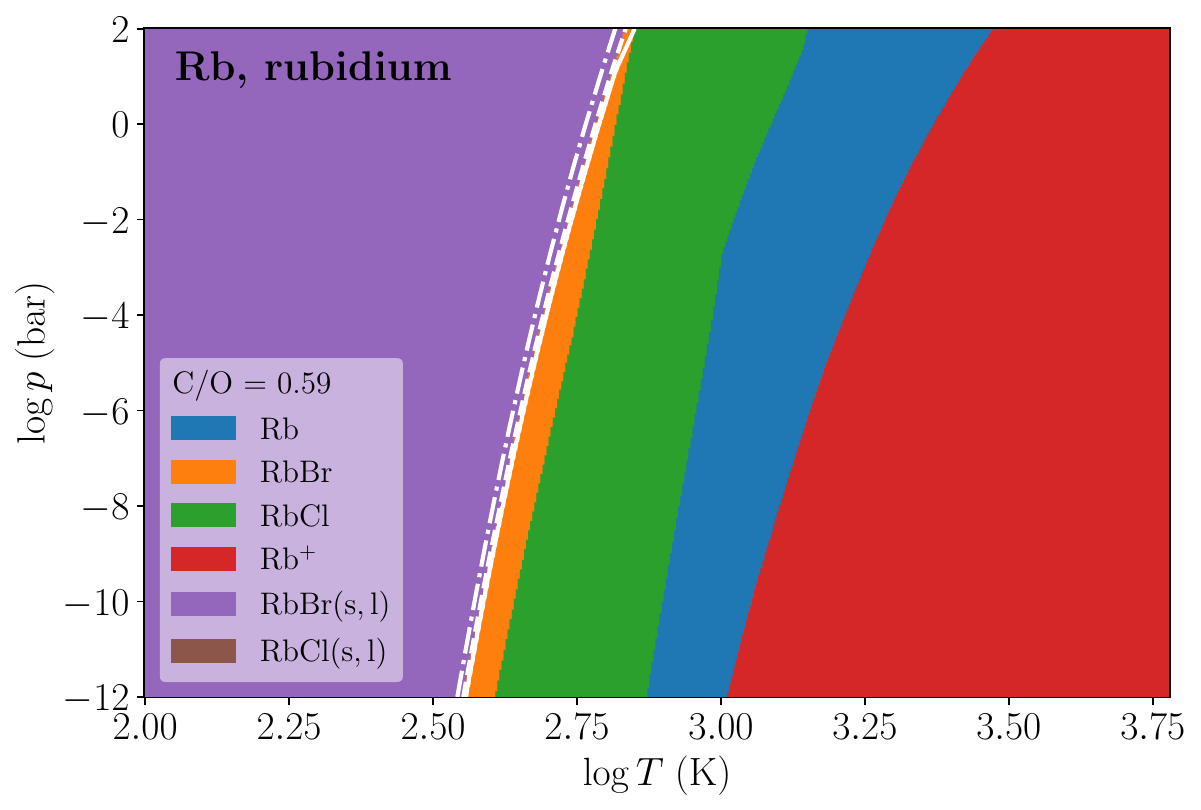}\hfill
    \includegraphics[width=0.49\linewidth]{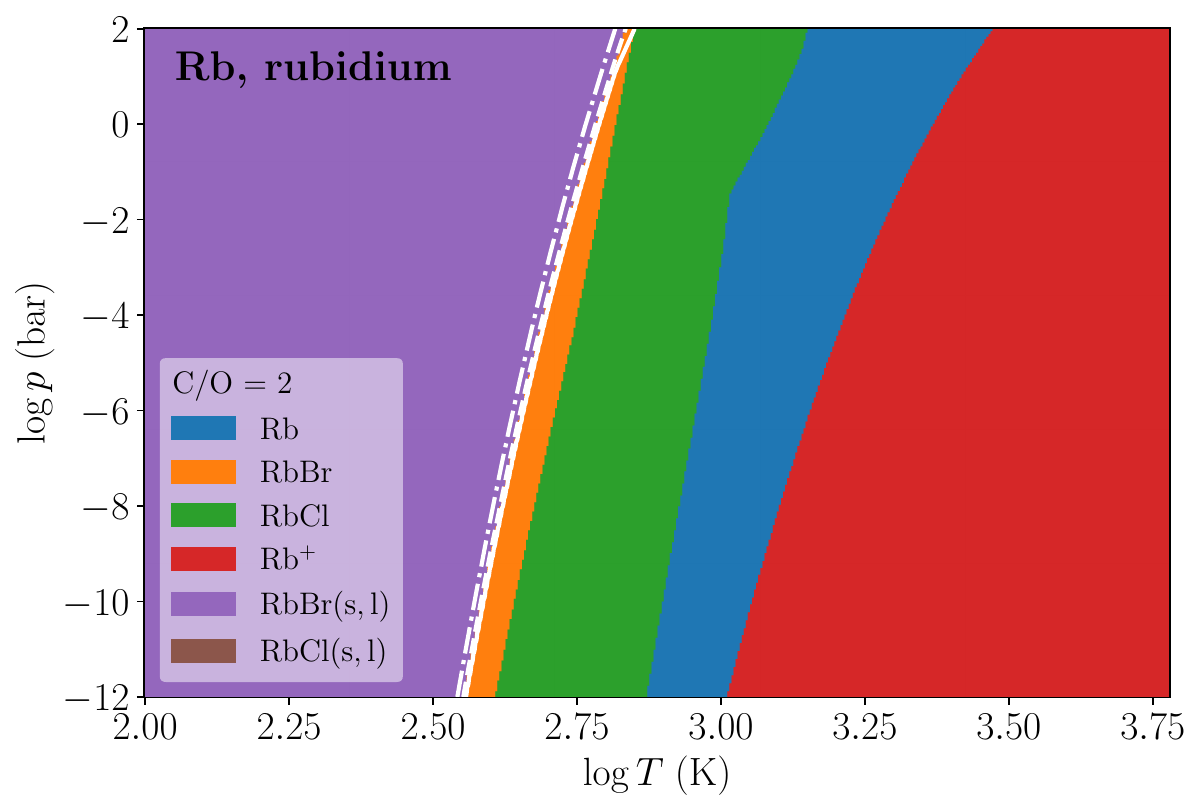}\\
    \includegraphics[width=0.49\linewidth]{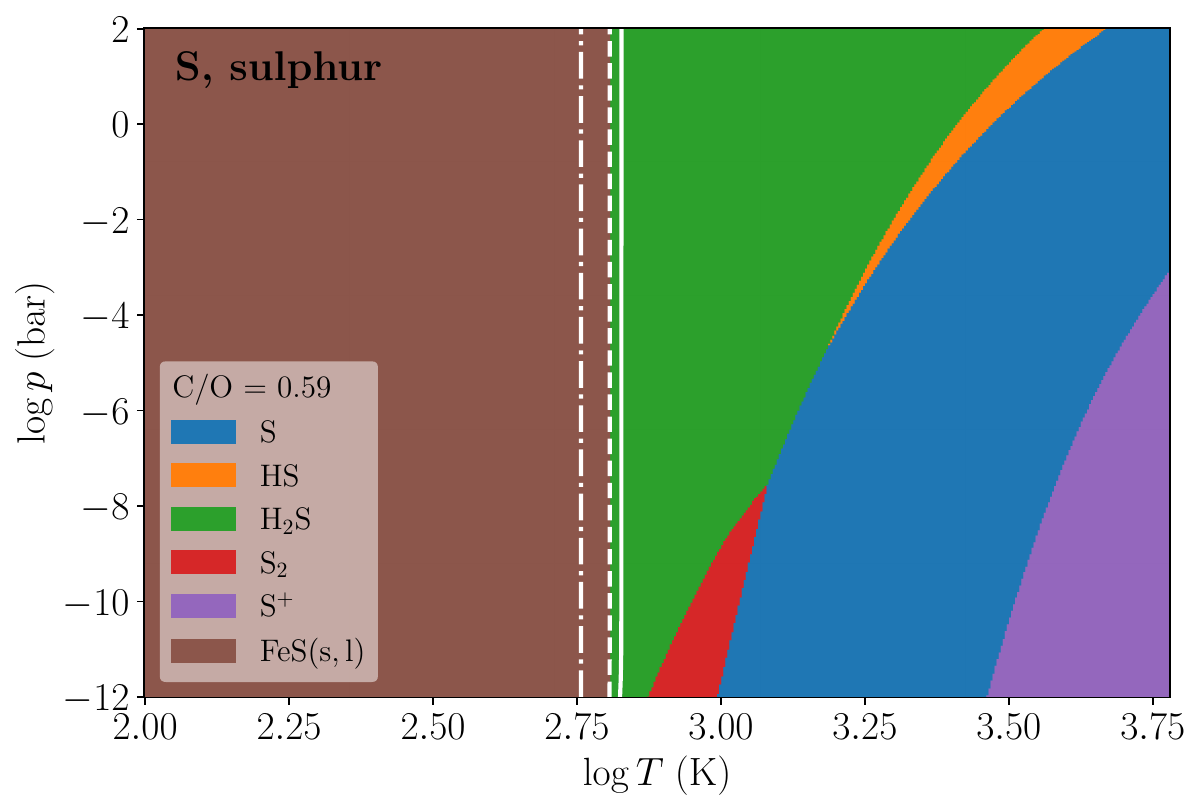}\hfill
    \includegraphics[width=0.49\linewidth]{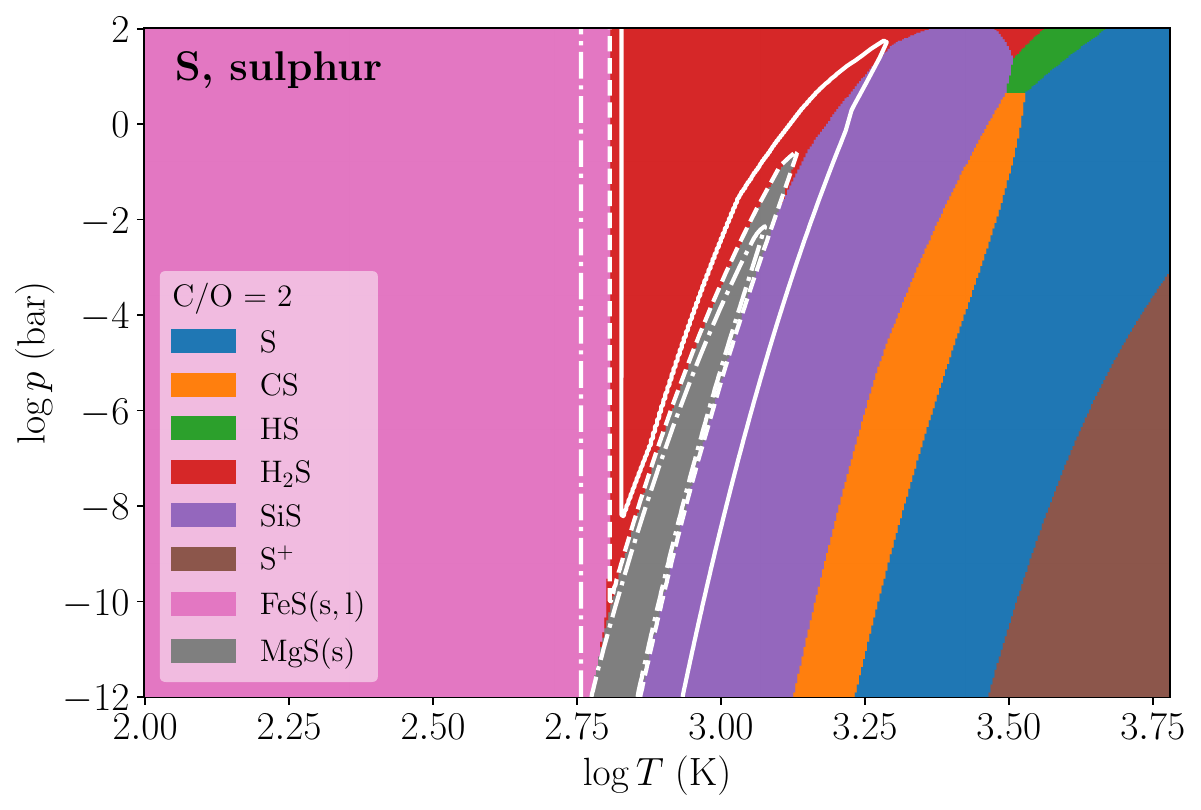}%
  }
  \caption{continued.}
\end{figure}

\begin{figure}[h!]
  \ContinuedFloat
  \sidecaption
  \parbox[b]{12cm}{%
    \includegraphics[width=0.49\linewidth]{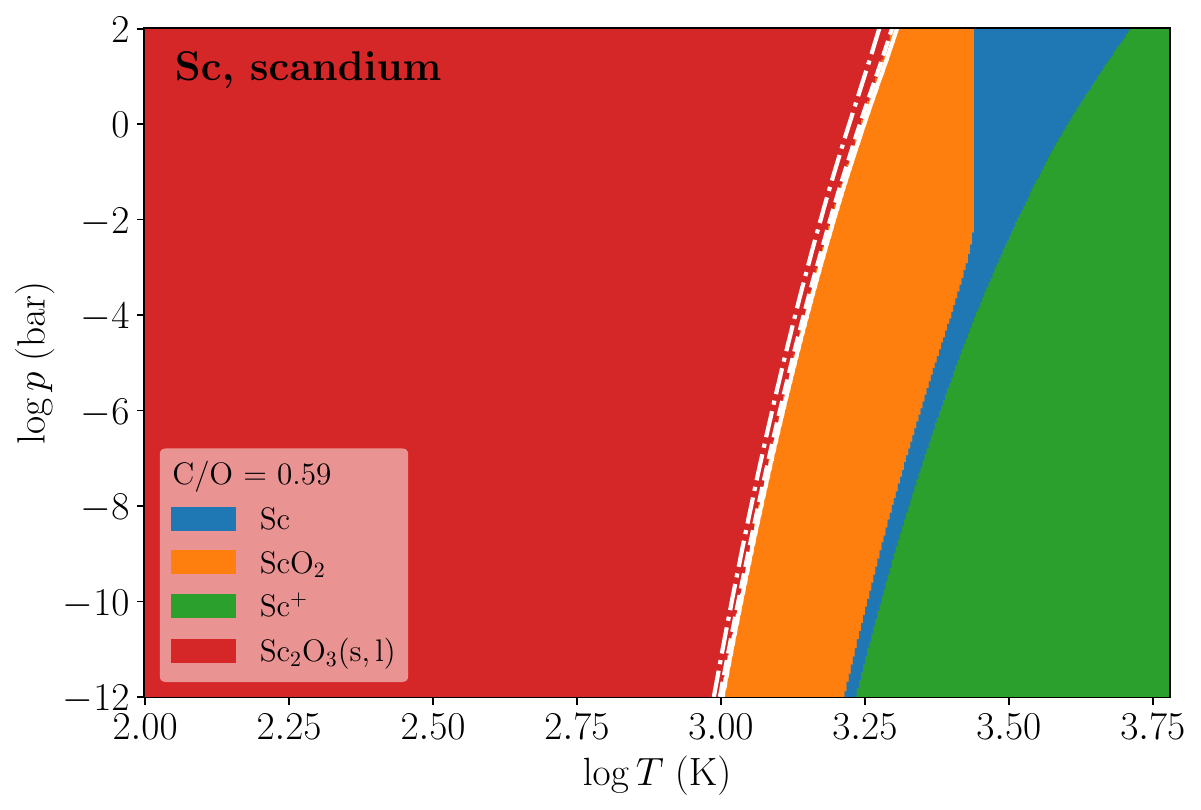}\hfill
    \includegraphics[width=0.49\linewidth]{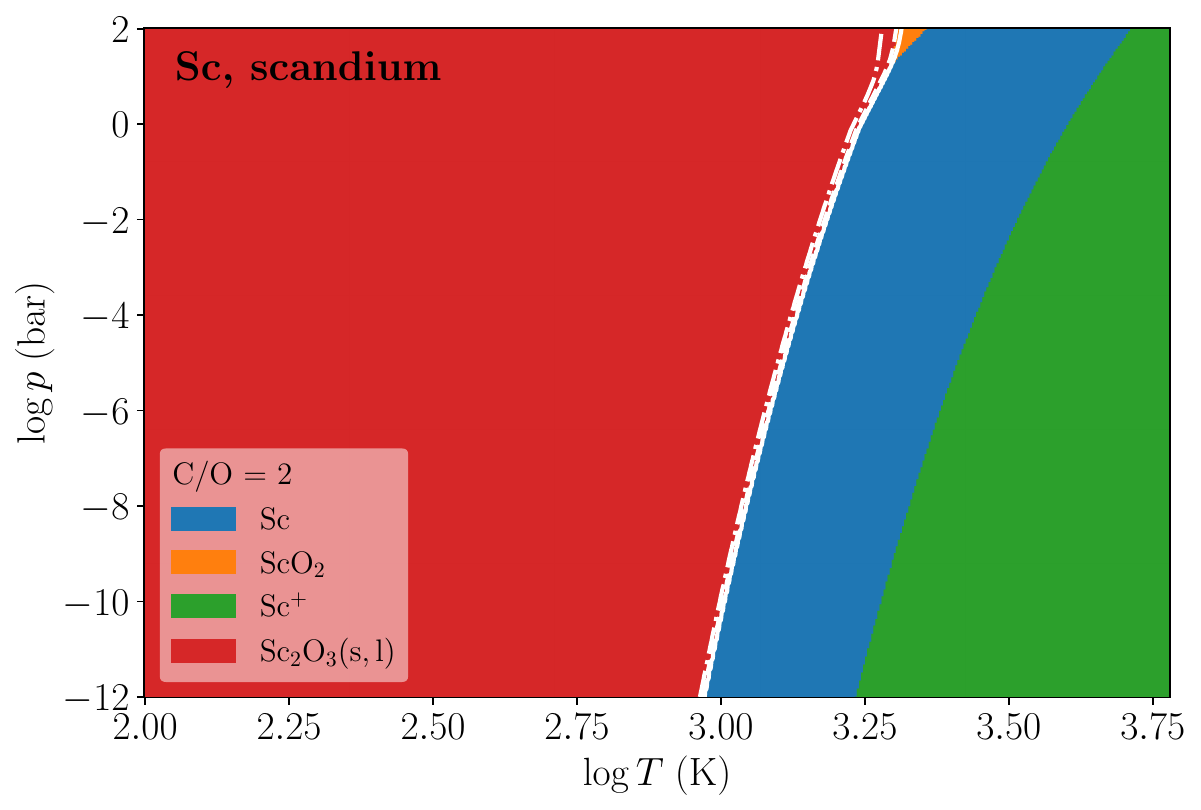}\\
    \includegraphics[width=0.49\linewidth]{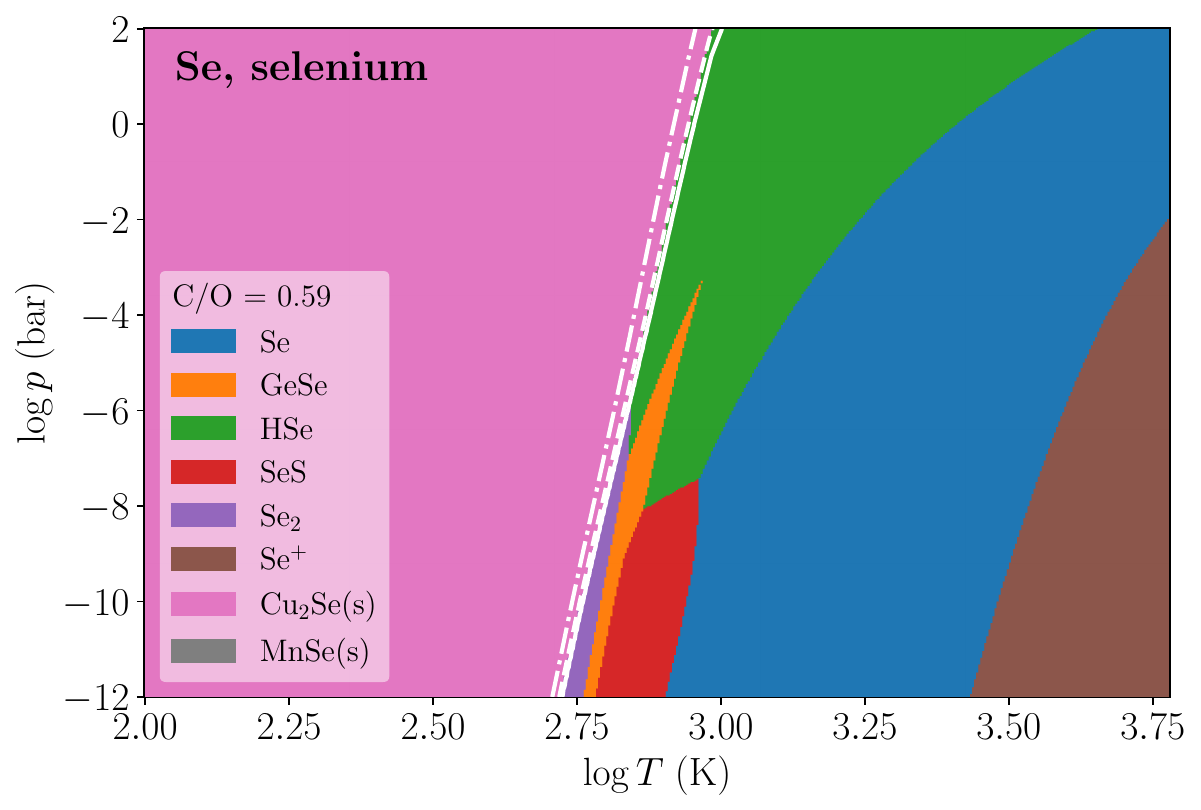}\hfill
    \includegraphics[width=0.49\linewidth]{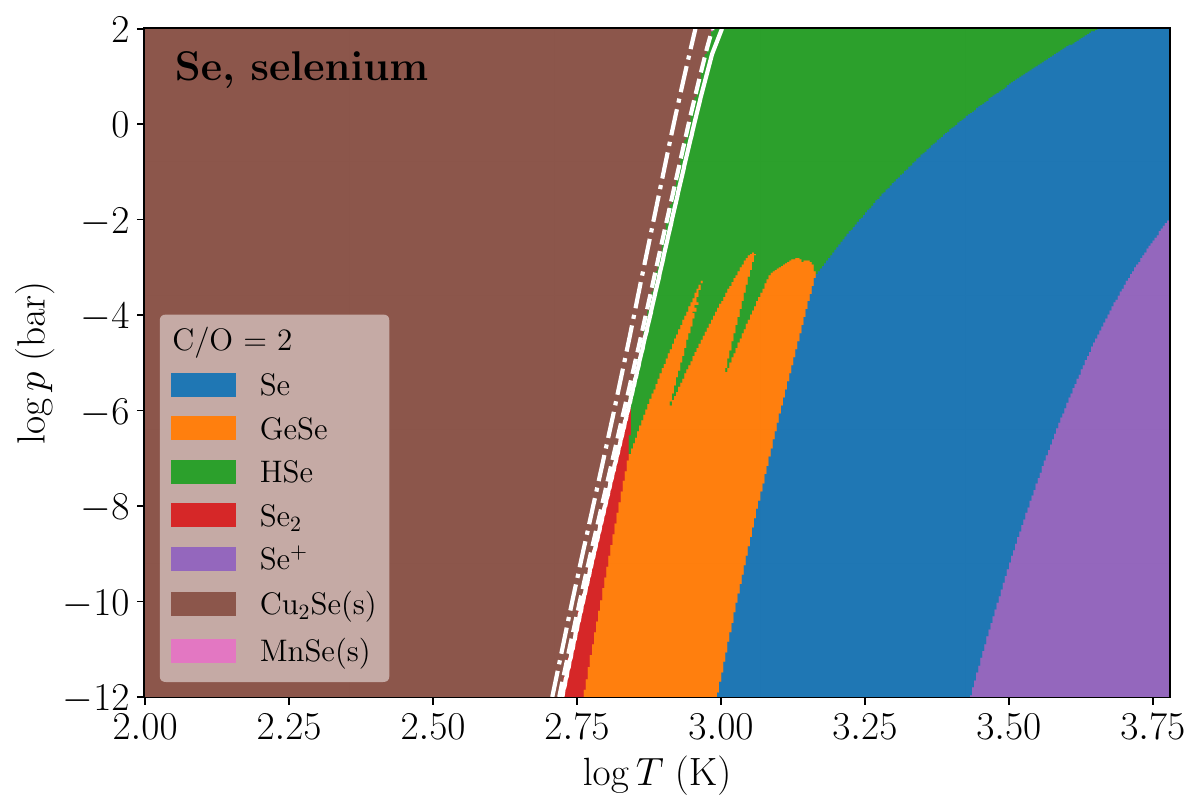}\\
    \includegraphics[width=0.49\linewidth]{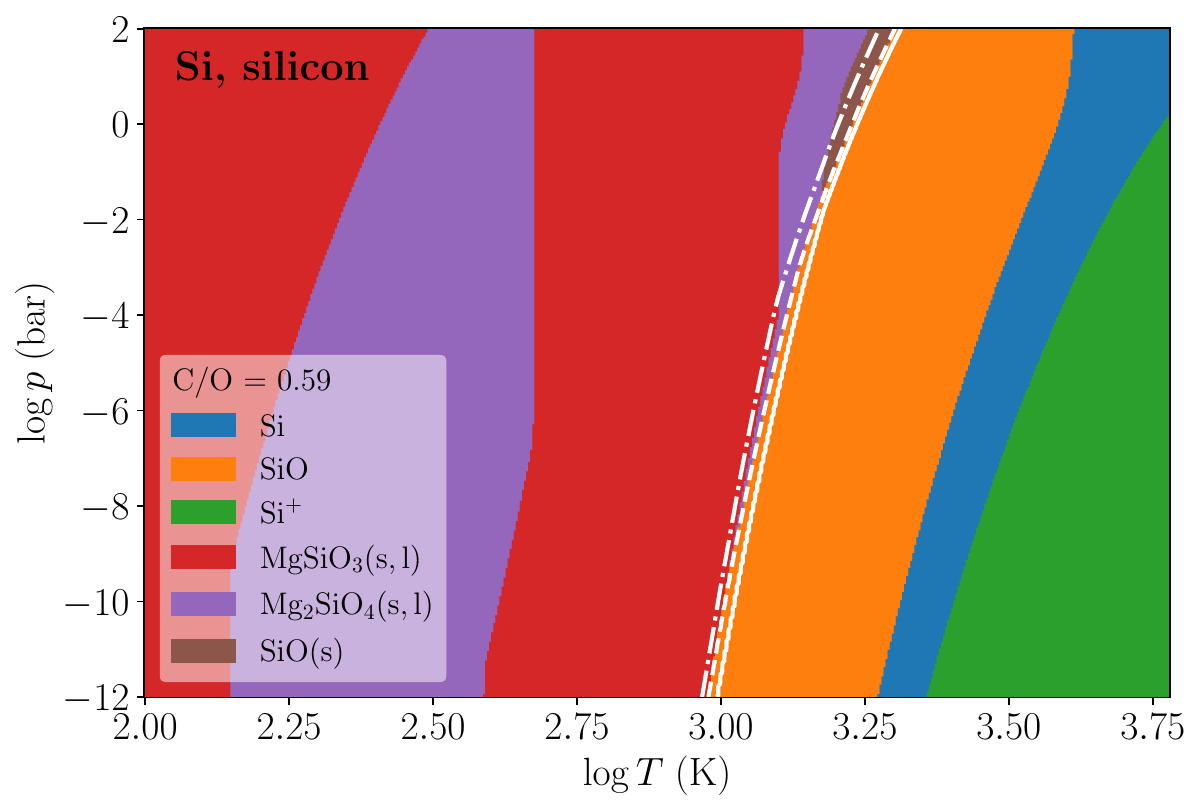}\hfill
    \includegraphics[width=0.49\linewidth]{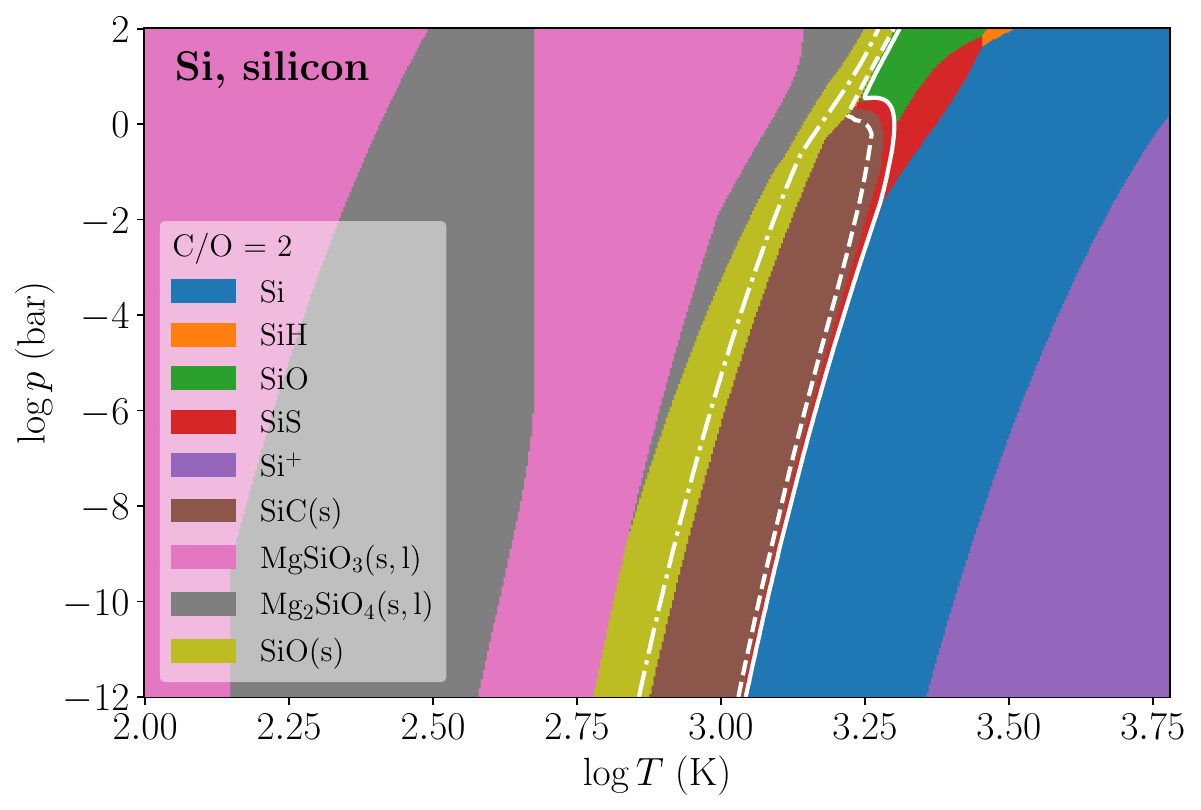}\\
    \includegraphics[width=0.49\linewidth]{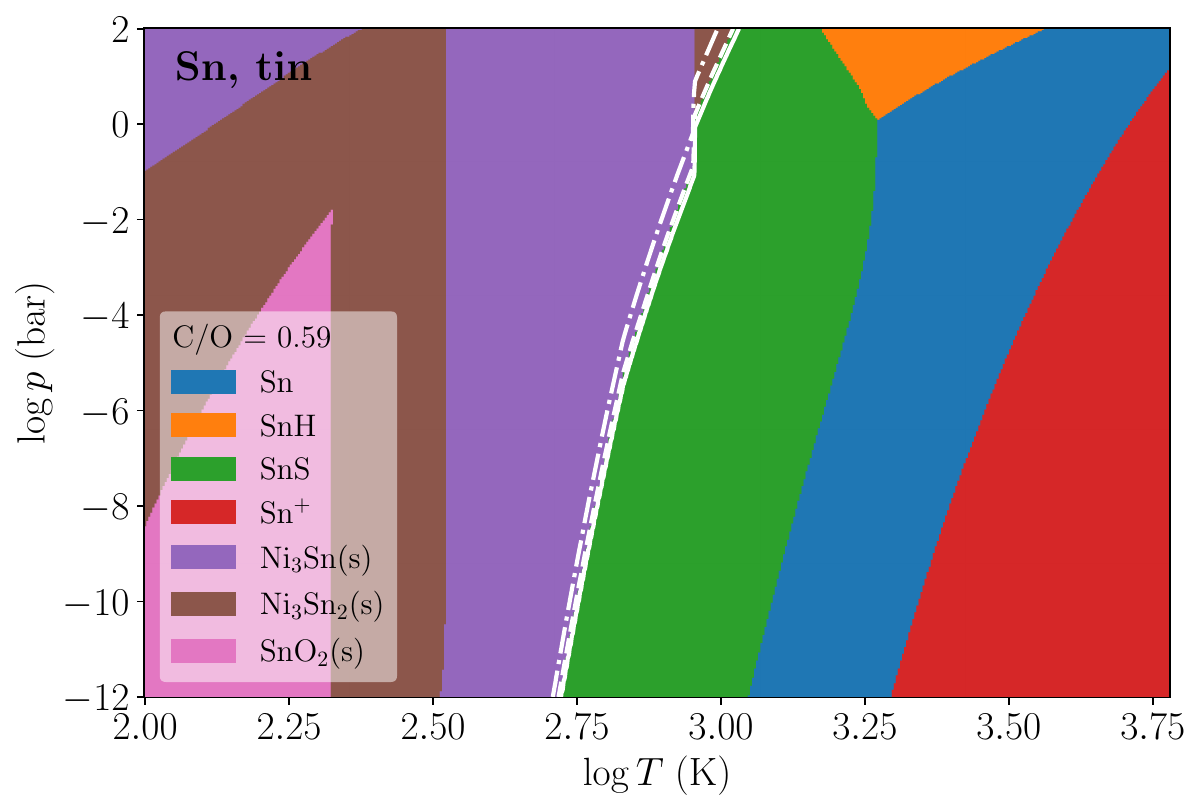}\hfill
    \includegraphics[width=0.49\linewidth]{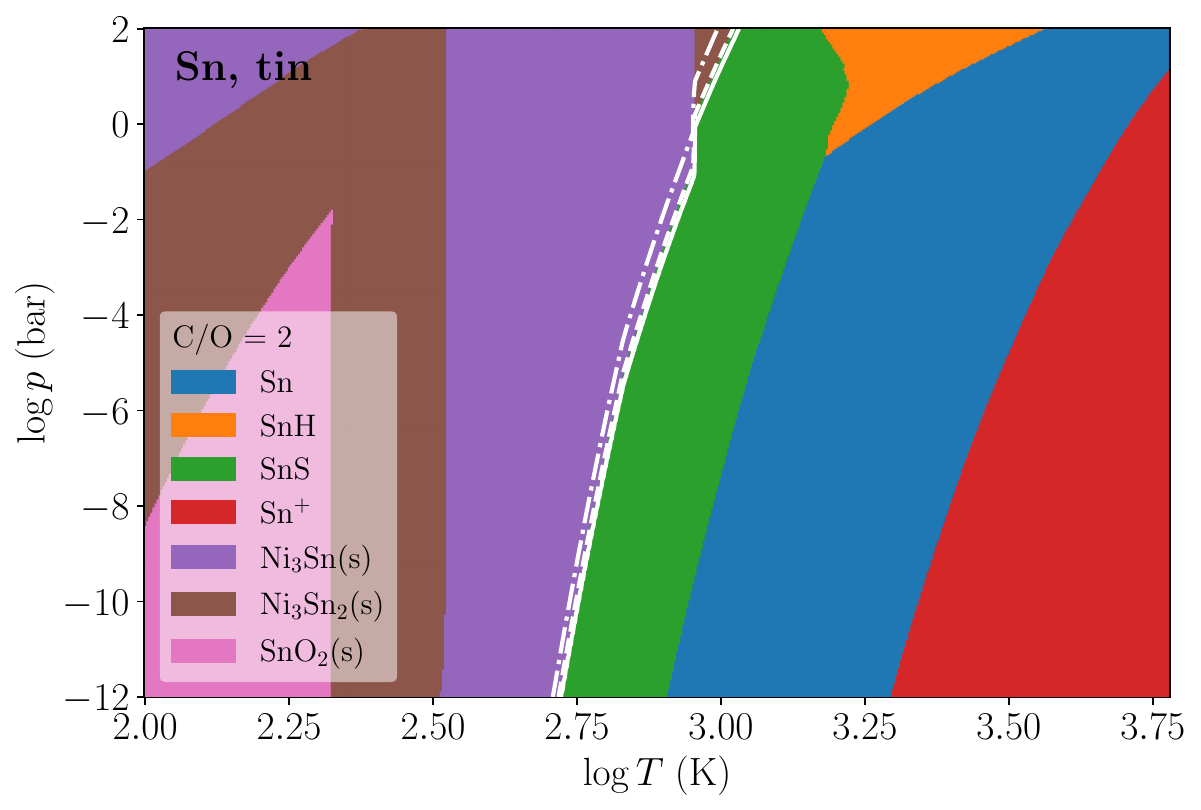}\\
    \includegraphics[width=0.49\linewidth]{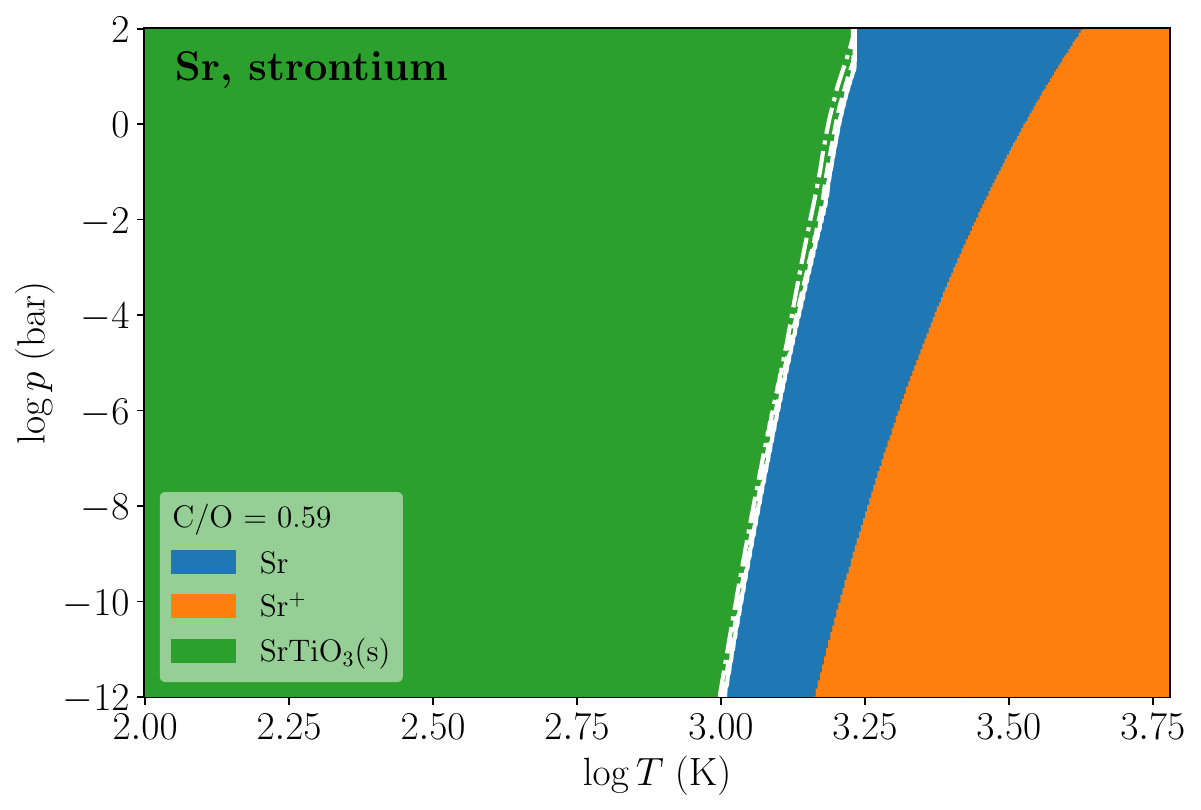}\hfill
    \includegraphics[width=0.49\linewidth]{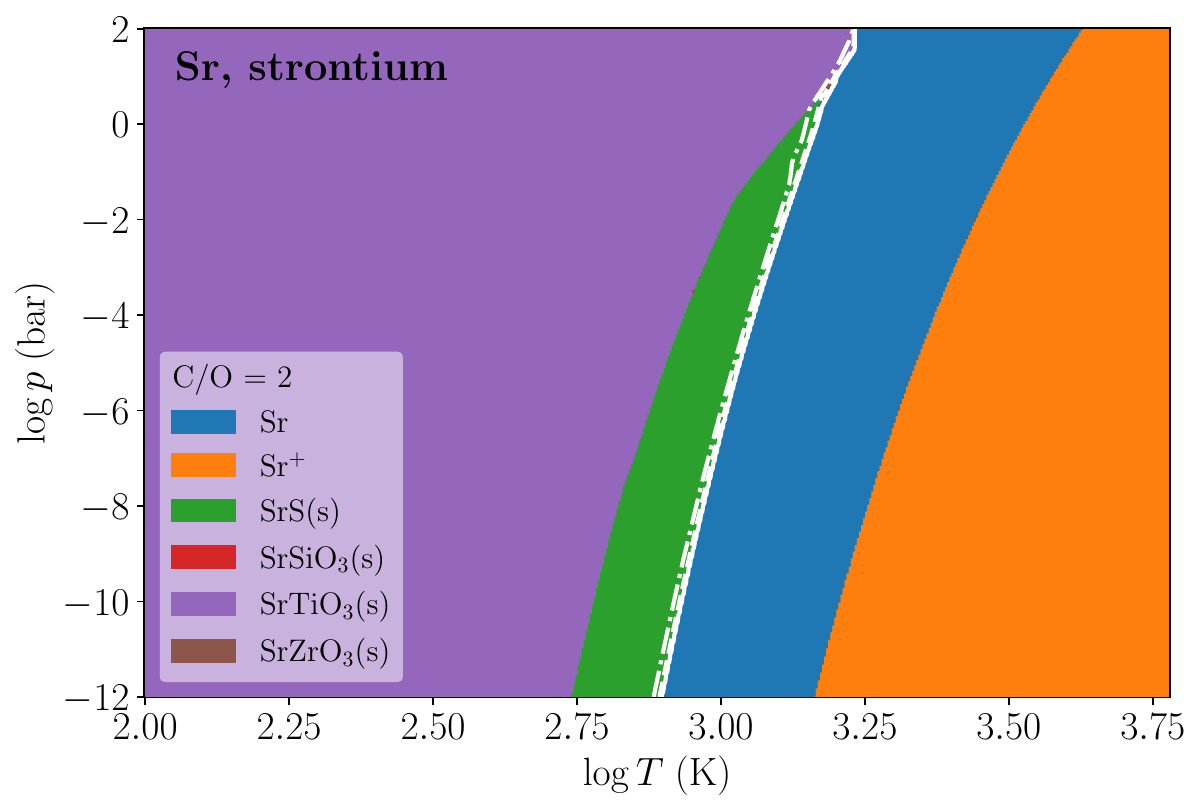}\\
    \includegraphics[width=0.49\linewidth]{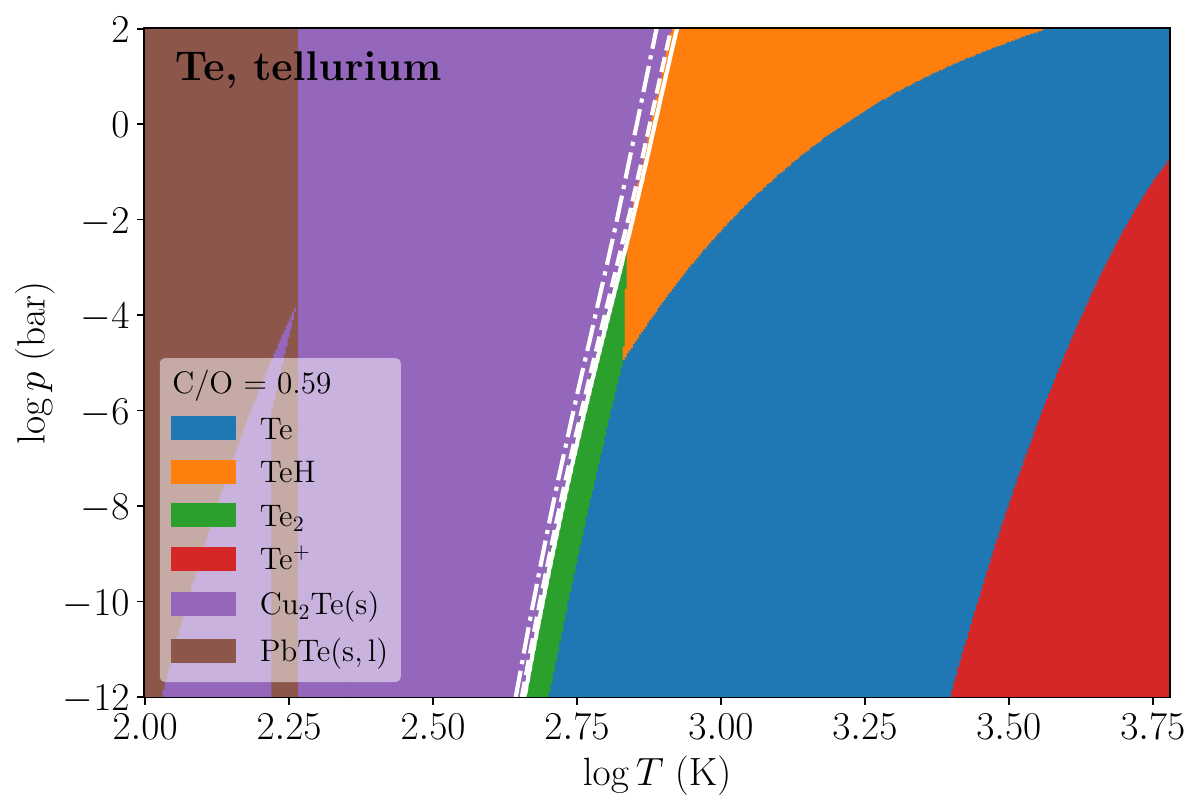}\hfill
    \includegraphics[width=0.49\linewidth]{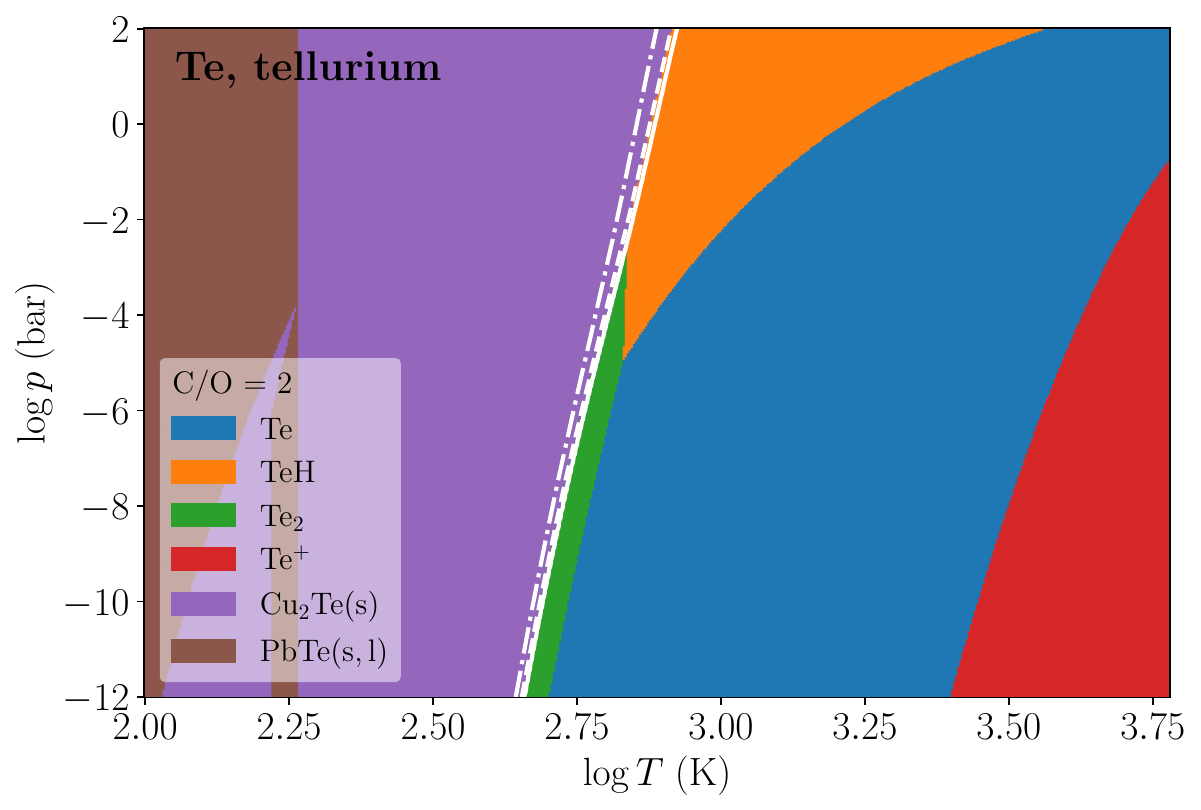}%
  }
  \caption{continued.}
\end{figure}

\begin{figure}[h!]
  \ContinuedFloat
  \sidecaption
  \parbox[b]{12cm}{%
    \includegraphics[width=0.49\linewidth]{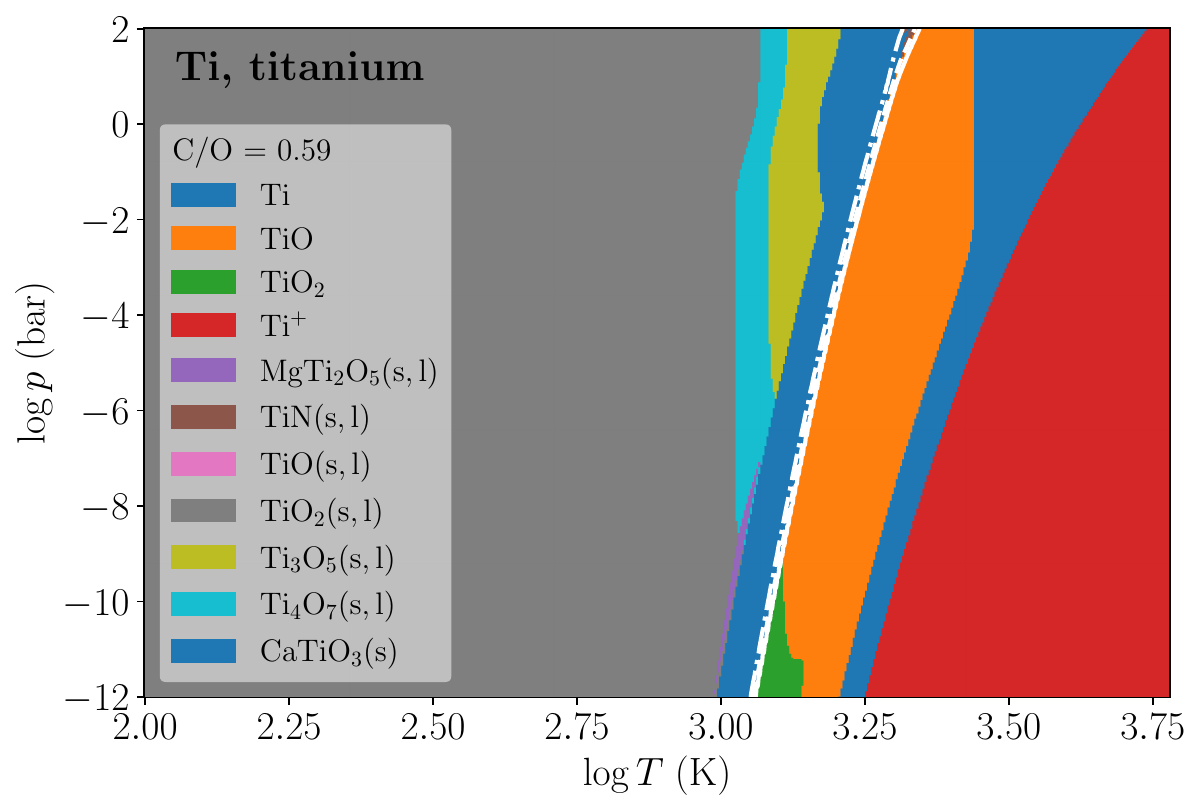}\hfill
    \includegraphics[width=0.49\linewidth]{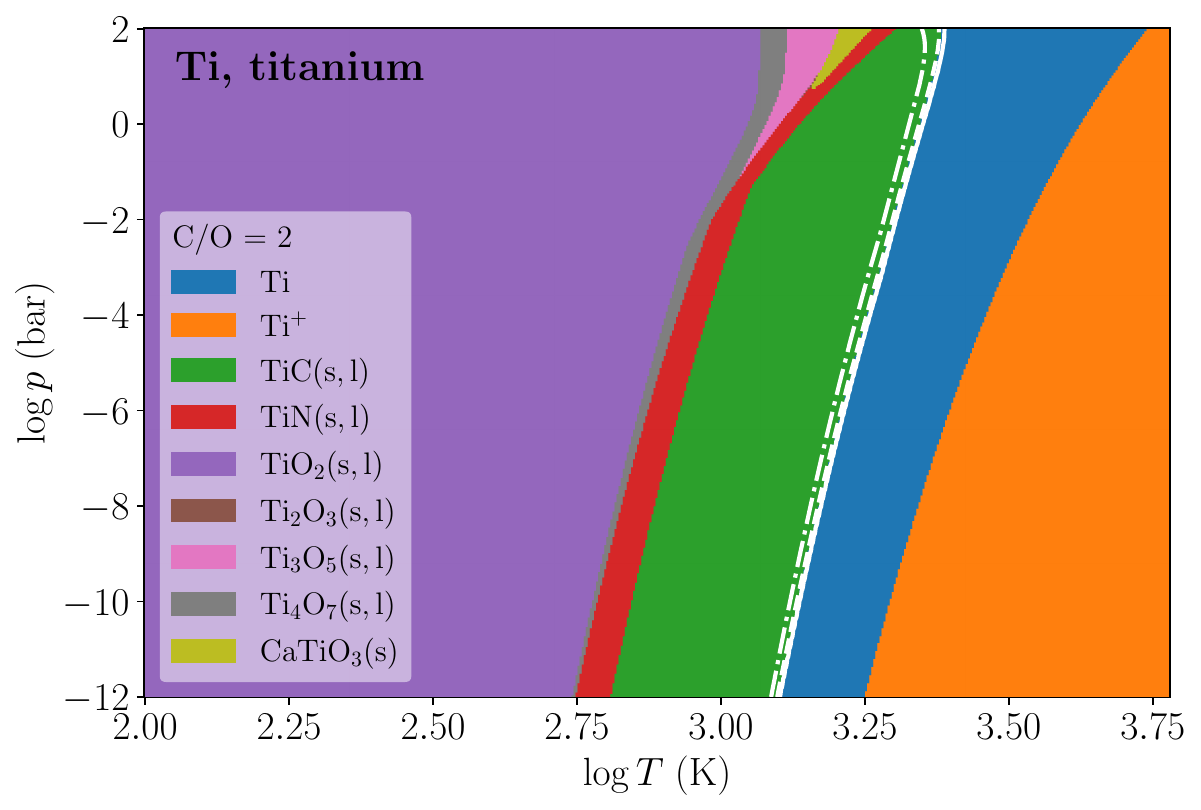}\\
    \includegraphics[width=0.49\linewidth]{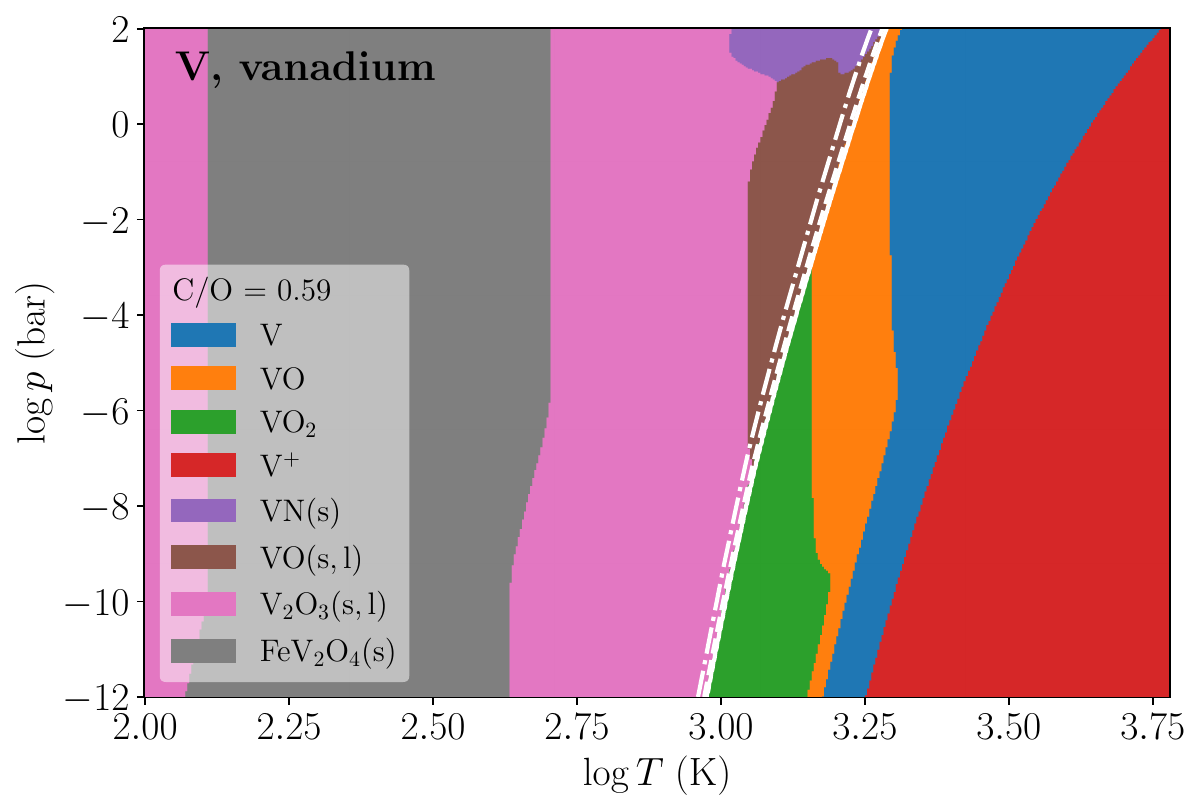}\hfill
    \includegraphics[width=0.49\linewidth]{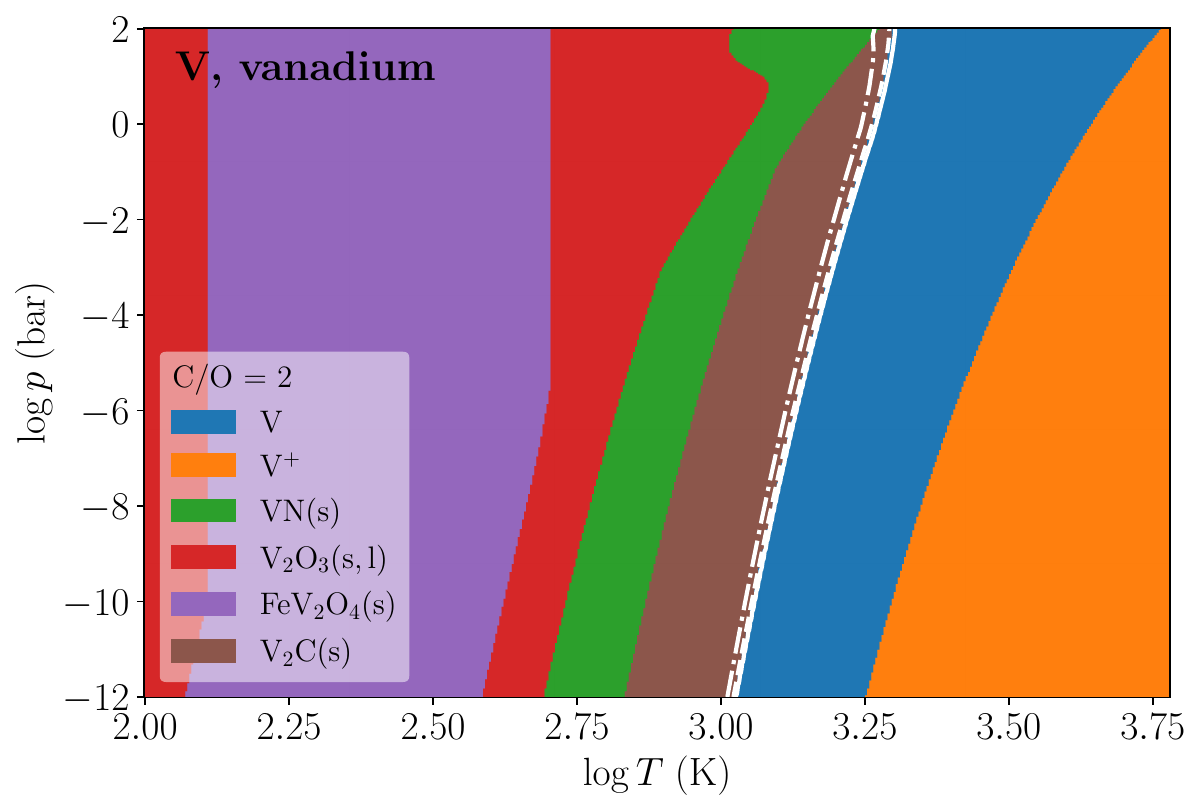}\\
    \includegraphics[width=0.49\linewidth]{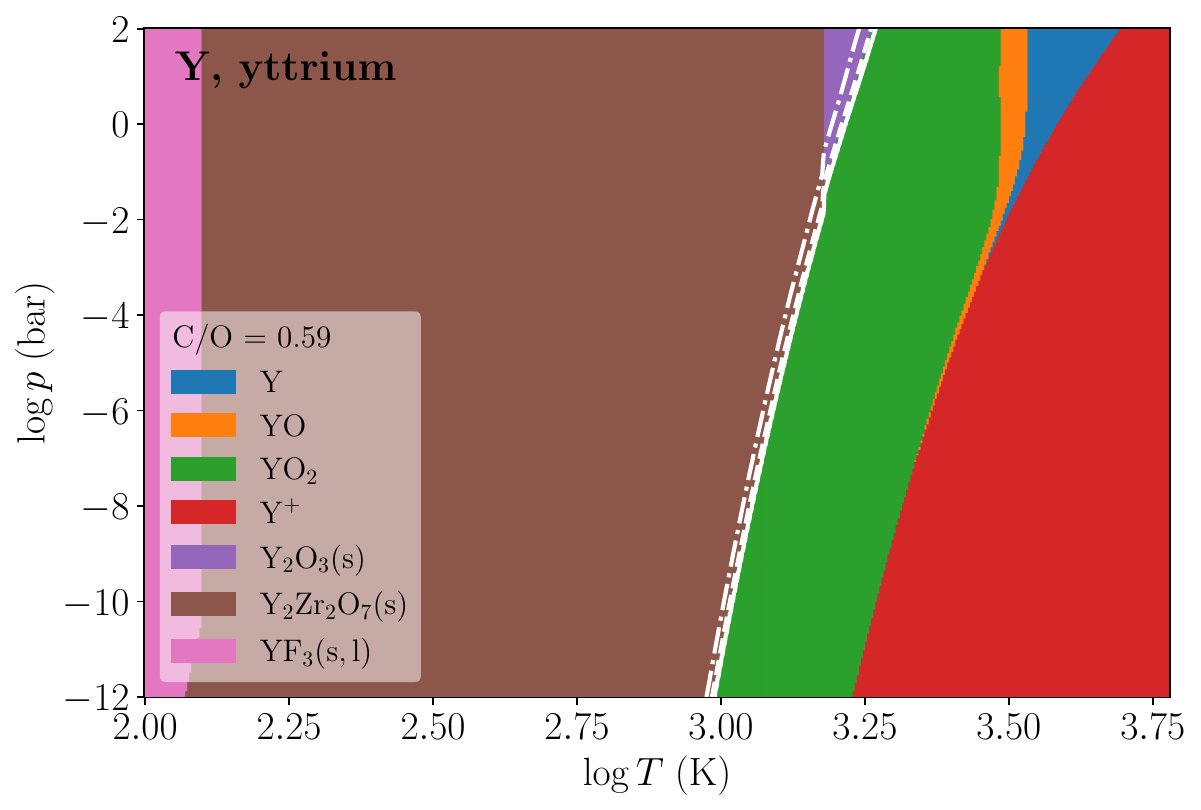}\hfill
    \includegraphics[width=0.49\linewidth]{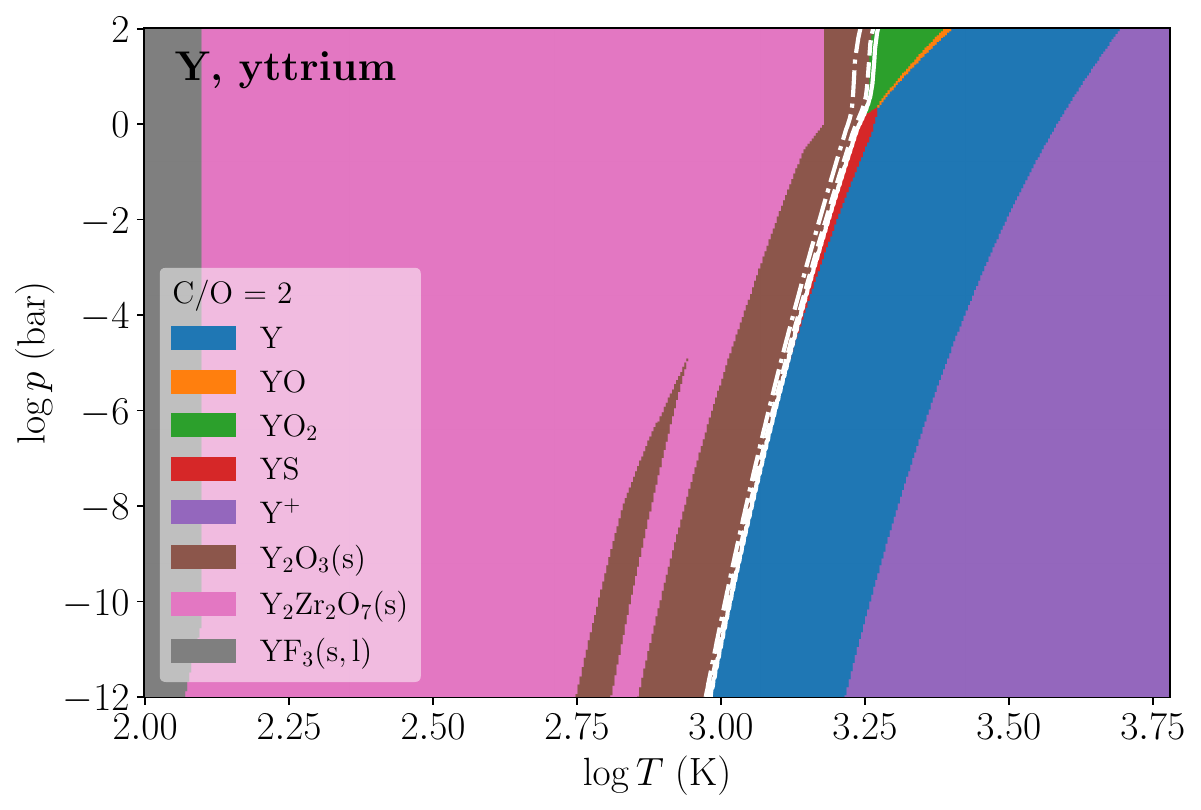}\\
    \includegraphics[width=0.49\linewidth]{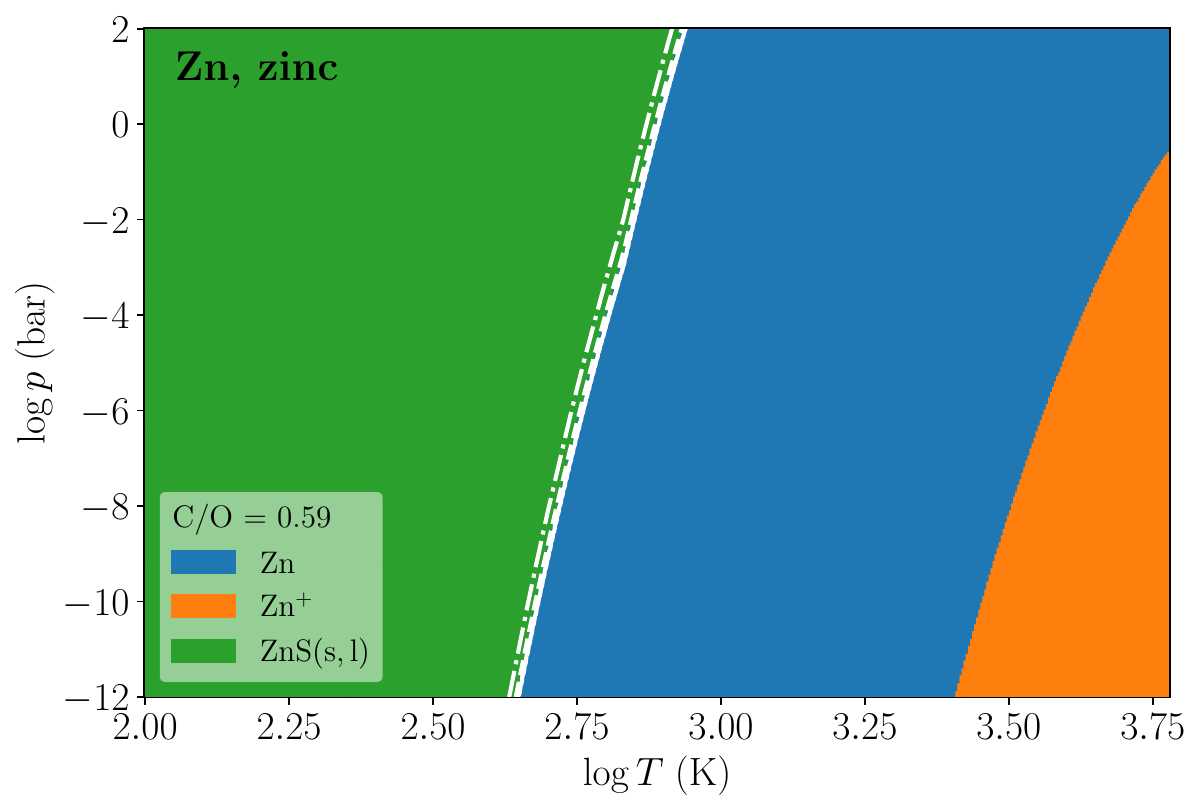}\hfill
    \includegraphics[width=0.49\linewidth]{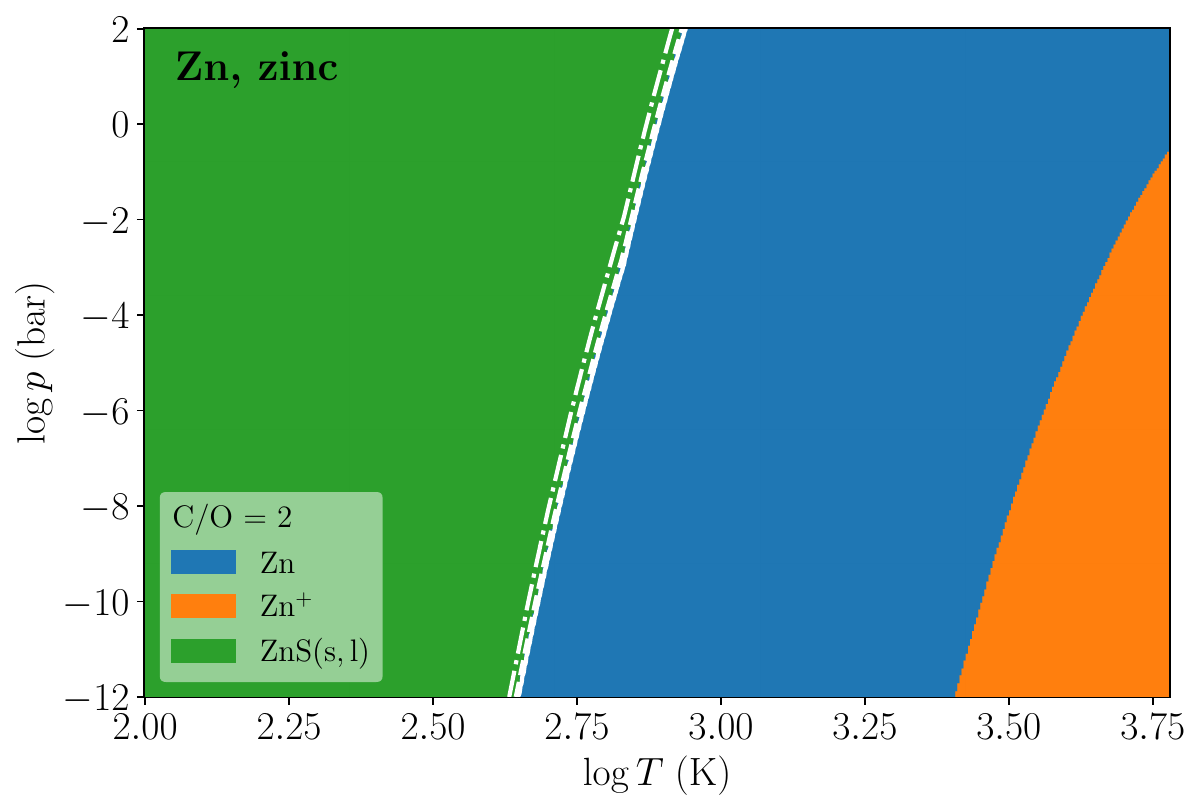}\\
    \includegraphics[width=0.49\linewidth]{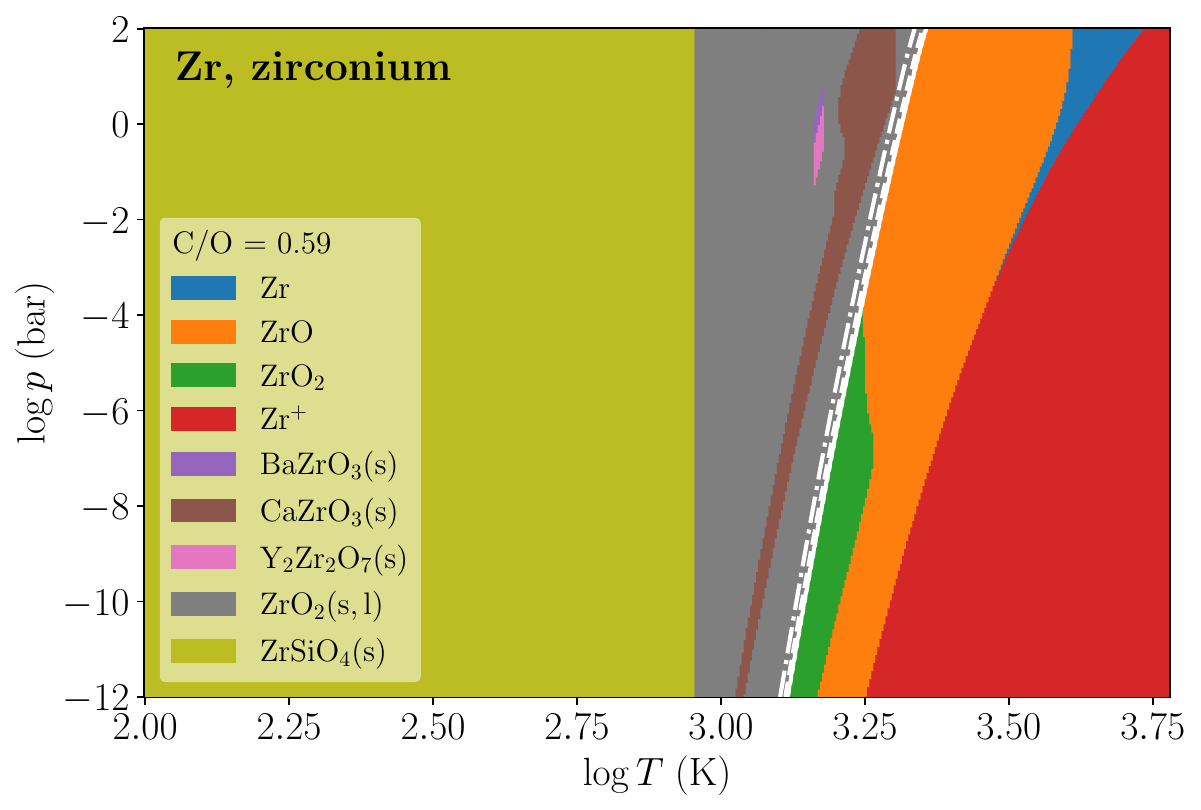}\hfill
    \includegraphics[width=0.49\linewidth]{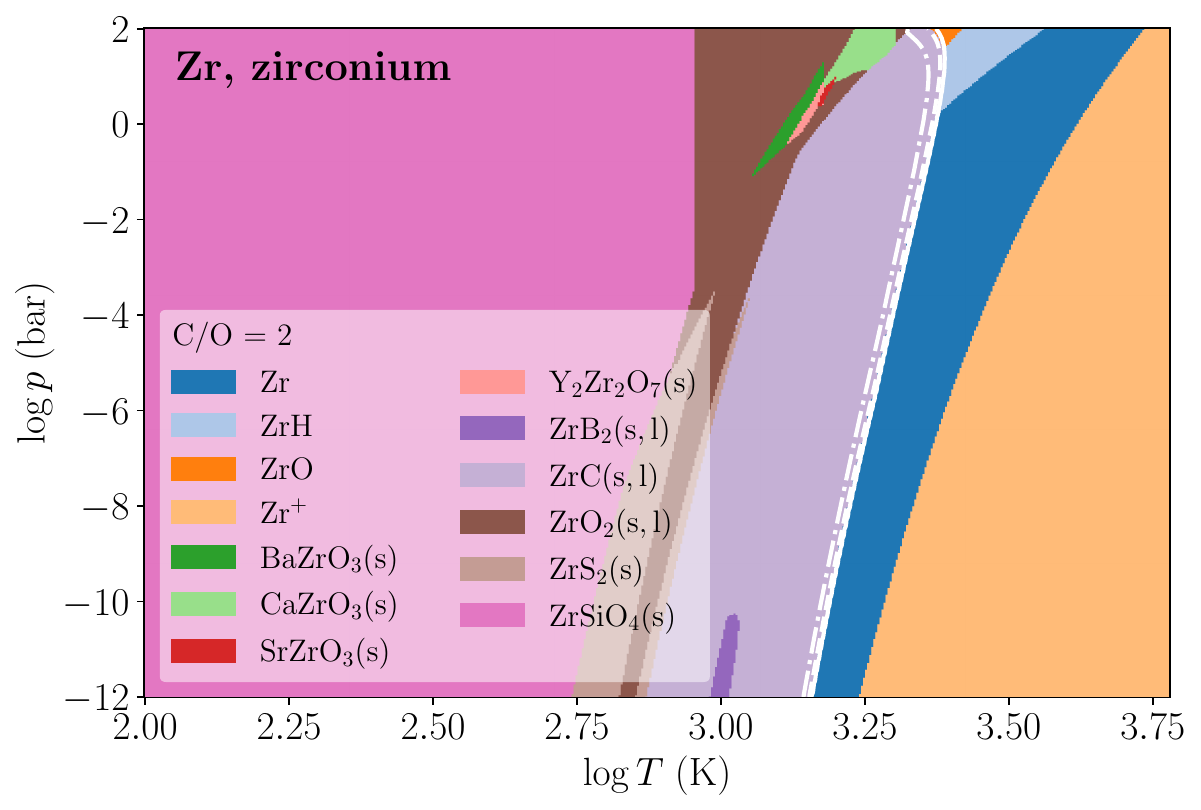}%
  }
  \caption{continued.}
\end{figure}

\end{appendix}

\end{document}